\documentclass{article}

\usepackage{fullpage}

\usepackage{amssymb}                                    
\usepackage{mathrsfs}                    
\usepackage{amsmath}                    
\usepackage{amsthm}                     
\usepackage{makeidx}
\usepackage{graphics}
\usepackage{hyperref}

\usepackage{color}

\usepackage[all]{xy}

\theoremstyle{plain}

\newtheorem{theorem}{Theorem}[section]
\newtheorem{lemma}[theorem]{Lemma}

\newtheorem{claim}[theorem]{Claim}
\newtheorem{proposition}[theorem]{Proposition}
\newtheorem{corollary}[theorem]{Corollary}

\theoremstyle{definition}
\newtheorem{definition}[theorem]{Definition}
\newtheorem{example}[theorem]{Example}

\newtheorem{remark}[theorem]{Remark}

\renewcommand{\(}{\begin{equation}}
\renewcommand{\)}{\end{equation}}
\newcommand{\bea}{\begin{eqnarray}}
\newcommand{\eea}{\end{eqnarray}}

\def\proof {{Proof.}\hspace{7pt}}

\def\endofproof {\hfill{$\Box$}\vspace{5pt}\\}

\begin{document}

\title{Quantization via Linear Homotopy Types}
\author{Urs Schreiber}

\maketitle

\begin{abstract}
 In the foundational logical framework of homotopy-type theory
 we discuss
 a natural formalization of secondary integral transforms in stable geometric homotopy theory.
 We observe that this yields a process of non-perturbative cohomological quantization
 of local pre-quantum field theory; and show that quantum anomaly cancellation amounts to
 realizing this as the boundary of a field theory that is given by genuine (primary) integral transforms,
 hence by linear polynomial functors.

 Recalling that traditional linear logic has semantics
 in symmetric monoidal categories and serves to formalize quantum mechanics \cite{Yetter, Pratt, AbDu05, Du06},
 what we consider is its refinement to
 \emph{linear homotopy-type theory} with semantics in stable $\infty$-categories of bundles of
  stable homotopy types (generalized cohomology theories)
  formalizing  Lagrangian quantum field theory, following \cite{dcct, Nuiten13}
  and closely related to \cite{Haugseng, HopkinsLurie}.

  For the reader interested in technical problems of quantization we
  provide non-perturbative quantization of Poisson manifolds and of the superstring; and find
  insight into quantum anomaly cancellation, the holographic principle and motivic structures in quantization.
  For the reader inclined to the interpretation of quantum mechanics we
  exhibit quantum superposition and interference as existential quantification in
  linear homotopy-type theory. For the reader inclined to foundations we provide
  a refinement of the proposal
  in \cite{Lawvere91, Lawvere94} for a formal foundation of physics
  \cite{Lawvere86, Lawvere97}, lifted from classical continuum mechanics to
  local Lagrangian quantum gauge field theory.
\end{abstract}

\medskip
\medskip

\begin{center}
  This here are expanded notes for talks given at:
  \\
  \medskip
  \medskip

  {\it Philosophy of Mechanics: Mathematical Foundations}
  \\
  Workshop at Paris Diderot
  \\
  February 12-14, 2014
  \\
  \href{http://phil.physico-math.gie.im/indexPMMF}{phil.physico-math.gie.im/indexPMMF}

  \medskip
  \medskip

  {\it Modern Trends in Topological Quantum Field Theory}
  \\
  Workshop at ESI Vienna
  \\
  February 17-21, 2014
  \\
  \href{http://www.ingvet.kau.se/juerfuch/conf/esi14/esi14_33.html}{www.ingvet.kau.se/juerfuch/conf/esi14/esi14\_33.html}

  \medskip
  \medskip

  {\it String Geometry Network Meeting}
  \\
  Workshop at ESI Vienna
  \\
  February 24-28, 2014
  \\
  \href{http://www.ingvet.kau.se/juerfuch/conf/esi14/esi14_34.html}{www.ingvet.kau.se/juerfuch/conf/esi14/esi14\_34.html}

\end{center}

\vfill

\begin{center}
  This document and related material is kept online at
  \\
  \href{http://ncatlab.org/schreiber/show/Quantization+via+Linear+homotopy+types}{ncatlab.org/schreiber/show/Quantization+via+Linear+homotopy+types}.
\end{center}

\newpage

\tableofcontents

\newpage

\section{Introduction and Summary}

\subsection{Introduction}

The modern theory of fundamental physics is local quantum field theory (local QFT).
The axiomatic characterization of local QFT on spacetimes/worldvolumes of arbitrary topology
(see \cite{SatiSchreiber} for survey and review)
had been completed in \cite{LurieQFT} via a universal construction in monoidal higher category theory.
Here the ``higher'' categorical structures are a direct reflection of the locality of the QFT.
While in the ``bulk'' of the worldvolume this describes topological QFT (TQFT, depending on the topology of the
worldvolume but not on any geometric structure on it), the axiomatization also captures defects (domain walls)
and boundaries (branes) and the corresponding boundary field theories may be geometric (non-topological)
\cite{Freed, dcct}.
In particular quantum mechanics is a boundary field theory of the non-perturbative
2d Poisson-Chern-Simons TQFT in this way
\cite{Bongers, Nuiten13}.

But modern quantum physics is more than quantum mechanics, and the local quantum field theories
of interest both in nature and in theory are not random examples of the axioms
of local QFT, instead they are supposed to arise via a process of ``quantization'' from geometric data,
namely from higher pre-quantum geometry \cite{hgp}. Here now the ``higher'' geometry
(hence geometric homotopy theory, higher topos theory) is the reflection of
the locality of the pre-quantum field theory.

Hence the process of quantization is central to modern fundamental physics and the areas of
mathematics influenced by it -- but the mysteries involved in making
full formal sense of it are proverbial. At least the perturbative approximation
(power series expansion) to quantization of
Poisson manifolds, hence of systems of phase spaces\footnote{see \href{http://ncatlab.org/nlab/show/deformation+quantization}{ncatlab.org/nlab/show/deformation+quantization} for
survey and references} has been understood in \cite{CattaneoFelder}
as being the boundary field theory of the perturbative 2d Poisson field theory. The refinement of
this from perturbative mechanics to perturbative field theory is currently being investigated
in terms of ``factorization algebras'' by Costello et. al.
On the other hand, the full (non-perturbative)
quantization of symplectic manifolds (phase spaces) via Kostant-Souriau geometric quantization
had been observed by Bott to be given by index theory in K-theory\footnote{see \href{http://ncatlab.org/nlab/show/geometric+quantization}{ncatlab.org/nlab/show/geometric+quantization}
for survey and references}. A refinement of this
from mechanics to field theory has been proposed and studied in \cite{dcct, Nuiten13} in terms
of index theory in the twisted generalized cohomology of cohesive homotopy types.

This note here describes how this cohomological
(or ``motivic'', see \ref{Motives} below)
non-perturbative quantization of local boundary pre-quantum field theory is naturally and usefully formalized
in the logical framework of type theory --
in its dependent and intensional flavor called \emph{homotopy type theory} \cite{HoTT}.
The axiomatization is meant to provide clarity where proverbial mysteries have
prevailed and seems to justify itself by its results and implications.

The process is naturally divided into two stages: first, pre-quantum geometry
(involving local Lagrangians and phase spaces)
is naturally axiomatized in \emph{cohesive homotopy type theory} \cite{ScSh}; second, quantization
(geometric quantization and path integral quantization, in fact we find a subtle mix of both)
is naturally axiomatized in \emph{linear homotopy-type theory}
(see \ref{DependentLinearTypeTheory} below). In fact we find that linear homotopy-type theory
provides an improved \emph{quantum logic} that, contrary to the common perception of
traditional quantum logic, indeed serves as a powerful tool for reasoning about what is
just as commonly perceived as the more subtle aspects of quantum theory,
 such as the path integral, quantum anomalies, holography, motivic structure.

The first step has been discussed in
detail in \cite{dcct, ScSh}, see \cite{Schreiber13} for a quick exposition.
Aspects of the second step have been worked out in examples in \cite{Nuiten13}.
There however a more fundamental type-theoretic formalization
is not made explicit yet. The present note essentially points out how the quantization process in
section 4 of \cite{Nuiten13} has a natural abstract formalization in linear homotopy-type theory
which provides deeper insights.

\begin{center}
\begin{tabular}{|c|c|}
  \hline
  pre-quantum geometry & quantization
  \\
  \hline
  \hline
  cohesive homotopy-type theory & linear homotopy-type theory
  \\
  \hline
\end{tabular}
\end{center}

\medskip

A brief word on the role of ``homotopy'': This is often regarded as a fancy new ingredient,
but the truth is that it is at the very heart of modern physics. The \emph{gauge principle} in physics
says that it is wrong to identify any two field configurations, that instead one is to ask if there
is a specified ``gauge'' equivalence relating them. Any two gauge equivalences in turn may be related
themselves by higher
order gauge equivalences. This state of affairs concerning identity/equivalence is exactly
what makes ``homotopy type theory'' be about homotopy types. That homotopy theory can be at the very
heart of modern fundamental physics and still be largely under-appreciated among researchers has
a simple reason: since quantum field theory is demanding, it is mostly considered
only in perturbative (infinitesimal) approximation, a limit in which much of its genuine structure
can be ignored. (What physicist call the ``BRST complex'' is the infinitesimal approximation
to the geometric homotopy-type of gauge field configurations.)
For instance most (if not all) mathematical physics textbooks agree that a field
in physics is to be formalized as a section of some fiber bundle called the ``field bundle''.
But for gauge field theories, and hence for the majority of all field theories of interest,
this statement makes sense only perturbatively, while non-perturbatively it is plain wrong:
a gauge field is itself a fiber bundle equipped with connection, and there is in general no
bundle such that its space of sections is equivalenty the space of
gauge fields. (This is true only perturbatively, when one fixes one background gauge field
and considers only a neighbourhood of that.) Instead, the space of gauge fields is the space of
sections of what is called a ``2-bundle'' \cite{NSS},  hence a \emph{stack}, a higher geometric homotopy type.
Non-perturbatively one can only avoid the need for homotopy types if one breaks locality
(a stack is a \emph{local} assignment of homotopy types).\footnote{Recently when the
description of Yang-Mills gauge theory was undertaken, using the non-perturbative
methods of ``algebraic quantum field theory'', just this breaking of locality was found to
be a consequence of assuming an ordinary field bundle instead of a higher homotopy type of fields, in \cite{BDS}.}
In more refined situations this state of affairs becomes only more pronounced. For instance
the full non-perturbative description of higher gauge fields, such as the ``RR-field'' \cite{DFM},
without homotopy theory is simply out of the question, see \ref{Gauge} below. Indeed,
claim \ref{UnityOfShapeFlatIsGauge} below says that gauge field theory is identified with
homotopy type theory qualified by a certain dual pair of modalities.

\medskip

The formalization presented here emerged in the course of attacking a class of
open problems in modern quantum field theory:
Non-perturbative local quantum physics has subtle consistency conditions on well-definedness of
its action functionals called ``quantum anomaly cancellation'' conditions, see \cite{Freed86} for a clean account
of the traditional story and see \ref{CoboundingTheory} below for our general abstract formulation.
By the discussion there, these anomalies are obstructions to defining action functional data
globally on moduli stacks of physical fields, hence on geometric homotopy types.
A famous example in 2-dimensional QFT is known as the Freed-Witten-Kapustin anomaly
(see example \ref{DBraneChargeAndTDuality} below);
in its full form this requires a field called the RR-field
to be given by cocycles in twisted differential K-theory \cite{DFM}. This anomaly is at the heart
of all discussion of Yang-Mills gauge field theory \cite{JaffeWitten} from open strings on D-branes
(see e.g. \cite{Moore}), hence of AdS/CFT duality (see. e.g. \cite{Nastase} for the traditional story
and see \ref{Holography} below for our abstract formulation)
and it affects the consistency of points in the
moduli space of string theory vacua, the ``landscape of string vacua'' \cite{Schellekens}.
Despite this importance, the definition and construction of cocycles and moduli stacks of twisted differential K-theory
used to be elusive. A resolution was only recently given, in terms of
cohesive homotopy theory see \cite{BNV} and section 4.1.2 of \cite{dcct}.
Quantum anomalies in higher dimensional field theory are expected to be similarly expressed
in terms of twisted differential generalized cohomology of higher chromatic degree
\cite{Sati}. For more discussion of these applications see section 5 of \cite{dcct}.

It is for such purposes of problem solving in QFT that we develop the
present formalism; but the more philosophically inclined reader (and all others should skip ahead)
might appreciate that our formalization has something to offer beyond just
technical problem solving. It may also be regarded as resolving some of the infamous ``problems of interpretation''
of quantum physics, by translating them from inadequate common language to a genuinely well adapted formal logic.
For instance in the discussion below, the path integral turns out to be given essentially by the linear homotopy theoretic refinement of the existential quantifier $\exists$, the linear dependent sum $\sum$. Therefore
our discussion may be regarded
as providing a formal and useful sense in which the statement ``There is a physical trajectory.'' of classical logic first turns into ``The space of all trajectories.'' in homotopy type theory and then into ``The linear space of quantum interfering trajectories.'' in linear homotopy-type theory.
This is the content of remark \ref{SuperpositionAsLinearDependentSum} below.
This might be thought of as a mathematical resolution of the interpretation of quantum interference via the non-classical formal meaning of ``existence'' in linear homotopy-type
theory.

For readers with even more tolerance for metaphysics (and all others are urged to skip this)
we indicate below how the axioms in type theory
that we base our discussion on have a striking resemblance to core parts of the metaphysics laid out in
\cite{Hegel}, once one posits, following \cite{Lawvere91, Lawvere94}, that the ``unity of opposites''
in which this is formulated is to be formalized by adjoint pairs of higher modalities \cite{Shulman12a} in
type theory, see \ref{AxiomaticMetaphysics} below.

\medskip

\subsection{Summary (point of view of Foundations)}
\label{SummaryFoundations}

This document is an expanded set of notes for some talks I gave in February 2014.
This section here is the script for what I actually said in a talk at 
the meeting ``Philosophy of Mechanics: Mathematical Foundations'' at Paris Diderot, to an audience
of mathematical physicists interested in the mathematical and philosophical foundations
of quantum physics. (Another summary aimed instead at TQFT theoreticians is below in
\ref{SummaryForTQFT}. A summary of more technical details is below in \ref{SummaryTechnicalDetails}.)

\subsubsection{Categories of homotopy-types and The gauge principle}

We will be considering a formal system that refines
traditional quantum logic such as to allow it to speak about quantization of
field theories. Since the proverb has it   that quantization is a mystery, such
a formalization inevitably tends to raise the question of whether it helps with
``interpreting'' the theory (in the physicist's vague sense, not the logician's
precise one).
That is not our actual concern, but let it here serve
as a lead-in.

Ever since Galileo, mathematical formalization alone serves to clarify and demystify physics\footnote{
See also J. Butterfield's talk \cite{Butterfield} at the same meeting.}).
The ancients found the daily rising of the sun a mystery, some found it
non-evident enough to sacrifice their own kin in the hope to propel the process.
Later people were still mystified by
the epicyclic intransparency of heavenly motions.
That people tend to not be mystified by any of this anymore is not because we now
have some deep "interpretation" of the concept of moving point masses acted on by
forces-at-a-distance; instead, we just found a formal system
(Newton's equations of motion, to start with)
that naturally and transparently allows to deduce these processes; and after staring at that for a while
and finding all the useful facts it implies,
it began to look very much self-evident.

The same ought to be true for a working formal quantum physics, a
``quantum logic''. A working quantum logic should be a formalism that is more than
the ``QM 101 made difficult'' as which traditional quantum logic must appear (see \ref{LinearLogic} below
for review);
it should instead be a formalism that empowers us to think
useful thoughts that were previously hard to think and that inform us about the genuine deep
aspects of quantum physics. To recall some of these:

\noindent{\bf Deep structural aspects of quantum physics.}
\begin{enumerate}
  \item  the path integral;

  \item  quantum anomaly cancellation;

  \item non-perturbative effects;

  \item  holography;

  \item  motivic structures.
\end{enumerate}

This points to what is a well-kep secret in much of the literature on quantum logic
and related issues:
Quantum mechanics is not actually our most fundamental theory of nature.
Instead, modern physics says that reality is fundamentally governed by
{\it quantum field theory}. More specifically, modern physics is based on
{\it local Lagrangian gauge quantum field theory}:

\newpage
{\noindent{\bf Characteristics of fundamental quantum physics.}
\begin{enumerate}

\item {\bf fields} -- types of configurations depending on $n$-dimensional spaces;

\item {\bf gauge} -- types of fields are really  {\it moduli stacks}, geometric homotopy types;

\item {\bf Lagrangian}  -- obtained via quantization from Lagrangian $n$-cocycle data;

\item {\bf  local} -- $n$-dimensional theory is an $n$-categorical construction.
\end{enumerate}
}
Quantum mechanics itself is just one limiting case of that.
But if there is interesting formal structure in the foundations of physics at all,
then it seems plausible that these are most purely exhibited by the very foundations, and not so much by some
special limiting case.

\medskip

The other proverb, the one referring to the effectiveness of mathematics in the
physical sciences, suggests that a foundations of fundamental physics should
go along well with the very foundations of mathematics and logic themselves.
These have seen some considerable advances since, say, Hilbert, and we are going to take these into full account.

\medskip

\noindent{\bf Logic sits inside type theory.}
A deep insight (attributed to Brouwer-Heyting-Kolmogorov, also to Howard) says that
propositional or first order logic may naturally be regarded as a subsystem of what more fundamentally is
\emph{type theory} or \emph{dependent type theory}, respectively 
(see \ref{NotionsJudgementDeduction} below for a little bit more on this).
Under this \emph{BHK correspondence} a proposition $\phi$ about terms $x$ of type $X$
is identified as the sub-type
$$
  \raisebox{20pt}{
  \xymatrix{
    \underset{x \in X}{\sum}\phi(x) =  \{x \in X | \phi(x) \, \mathrm{true}\}
    \ar@{^{(}->}[d]
    \\
    X
  }
  }
  \;\;\;
  \in
  \mathbf{H}_{/X}
$$
of $X$ of all those terms $x$ which validate $\phi$. Here $\mathbf{H}$ denotes the
\emph{category} of all types -- and that is the insight of \emph{categorical logic},
that the syntax of type theories has semantics in suitable categories.
Moreover, $\mathbf{H}_{/X}$ denotes the slice category over the type $X$, and this
is what interprets $X$-\emph{dependent} types.

It should be plausible that using types instead of (just) propositons as the
fundamental logical substrate suits formalization of fundamental physics,
for here we are clearly concerned with talking not just about propositions, but about ``things'', notably when talking
about some \emph{type of fields}, which we generically write
\begin{itemize}
  \item $\mathbf{Fields} \in \mathbf{H}$ -- a moduli space of fields.
\end{itemize}
And indeed these should depend on other types, such as on spacetimes $X \in \mathbf{H}$,
for instance in order to form the \emph{field bundle}
\begin{itemize}
\item
$\raisebox{20pt}{\xymatrix{\mathbf{Fields}_X \ar[d] \\ X}} \in \mathbf{H}_{/X}$
-- a bundle of moduli of fields parameterized over $X$
\end{itemize}
of which the actual field configurations of a field theory on $X$ would be sections.

\medskip

\noindent {\bf Dependent types and Existence.}
The only basic operations on (dependent) types are these:
for any morphism $f \;:\; X \longrightarrow Y$ in $\mathbf{H}$ (hence a function
sending terms of type $X$ to terms of type $Y$) there is an adjoint triple
$$
  (\underset{f}{\sum} \dashv f^\ast \dashv \underset{f}{\prod})
  \;:\;
  \xymatrix{
    \mathbf{H}_{/X}
    \ar@<+8pt>@{->}[rr]^{\sum_f}
    \ar@<0pt>@{<-}[rr]|{f^\ast}
    \ar@<-8pt>@{->}[rr]_{\prod_f}
    &&
    \mathbf{H}_{/Y}
  }
$$
whose operations are called, in type theory and in order of appearance, the dependent sum, the context extension and the
dependent product along $f$. (In geometry and topos theory this is instead known as base change, or similar.)
Here sum and product are to be understood fiberwise over the fibers of $f$, and so if we think of a bundle
$E$ over $X$ as a collection of types $E(x)$ for $x \in X$, then the dependent sum reads
$$
  (\underset{f}{\sum}E)(y) = \underset{x \in f^{-1}(y)}{\sum} E(x)
$$
and is hence manifestly a form of fiber integration along $f$. This is going to play a role in the
path integral below (see \ref{IntegralTransformations} for more).

We already used the $\sum$-notation above for indicating how
propositions appear as types, and indeed the combined restriction and co-restriction of
$\underset{X}{\sum} : \mathbf{H}_{/X} \to X$ to propositions is existential quantification:
$\underset{x \in f^{-1}(y)}{\exists} \phi(x)$. So where propositional logic has the
proposition ``There exists an $x$ such that $\phi(x)$ is true.'' its embedding into
type theory replaces that with ``The collection of all $x$ such that $\phi(x)$ is true.''
Something to keep in mind when we get to the path integral below.

\medskip

\noindent {\bf Constructive} $\leftrightarrow$ {\bf Physically realizable}.
Closely related is the fact that type theory embodies \emph{constructive mathematics},
where nothing is regarded as true unless its proof may be constructed in
a way that yields an algorithm. For instance in type theory to prove
that there exists $x$ such that $\phi(x)$ is true is to actually construct a term
$t \in \underset{x\in X}{\sum}\phi(X)$ from the deductive rules of the theory. This explains the fundamental
relevance of type theory in computer science, where these proofs are
the very programs -- which shows that the constructive
concept of existence is in some way closely related to the physical concept of existence.

Taking this constructivism fully seriously lead to a  breakthrough
convergence of formerly disparate concepts:

\medskip

\noindent{\bf Constructive identity types $\leftrightarrow$ Gauge principle}.
Constructivism demands that given a (dependent) type $\mathbf{Fields}_X$ (of fields, for instance) and given
two terms $\phi_1$, $\phi_2$ of that type (so: two field configurations of given type over
a spacetime $X$) then
it is misguided to ask whether these are equal or not, instead
we have to construct a witness $\alpha$ exhibiting their equivalence,
hence produce a \emph{gauge transformation} making them gauge equivalent.
It is in turn
wrong to assert that two such gauge equivalences $\alpha_1$  and $\alpha_2$ are equal,
instead we have to
exhibit a gauge-of-gauge transformation between those:
$$
  
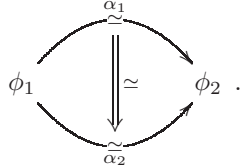
\xymatrix{
    \phi_1
    \ar@/^2pc/[rr]^{\alpha_1}|\simeq_{\ }="s"
    \ar@/_2pc/[rr]_{\alpha_2}|\simeq^{\ }="t"
    &&
    \phi_2
    \ar@{=>}^\simeq "s"; "t"
  }
  \,.
$$
And so on. This is really the \emph{gauge principle} in physics. At the same time, this
is how constructive type theory is \emph{automatically} a theory of homotopy types
(of $\infty$-groupoids).

\medskip

\noindent{\bf Example} \cite{dcct}. For the purpose of the following
exposition, a running example for $\mathbf{H}$ to keep in mind is
$$
  \mathbf{H} = \mathrm{SynthDiff}\infty\mathrm{Grpd} := \mathrm{Sh}_\infty(\mathrm{FormMfd})
$$
the homotopy topos of sheaves of homotopy types
on formal smooth manifolds. This is a homotopy-theoretic version of a topos that
interprets synthetic differential geometry \cite{Lawvere97}.

Noteworthy geometric homotopy types that we encounter below are
$$
  \mathbf{B}^n U(1) \in \mathbf{H}
$$
which are obtained by delooping the abelian Lie group $U(1) = \mathbb{R}/\mathbb{Z} \in \mathrm{Grp}(\mathbf{H})$
$n$ times.
These are the moduli for instanton sectors of $n$-form $U(1)$-gauge fields.

Notice a basic fact of homotopy theory, a first little hint of holography:
Giving a homotopy as on the left of
$$
  \raisebox{35pt}
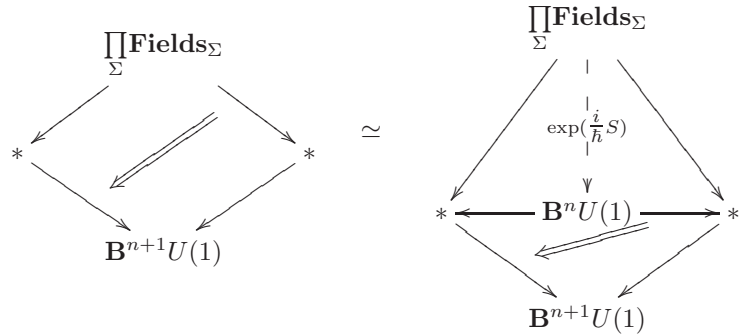
{
  \xymatrix{
    & \underset{\Sigma}{\prod}\mathbf{Fields}_{\Sigma}
    \ar[dl]
    \ar[dr]_{\ }="s"
    \\
    \ast \ar[dr]^{\ }="t" && \ast \ar[dl]
    \\
    & \mathbf{B}^{n+1}U(1)
    \ar@{=>} "s"; "t"
  }
  }
  \;\;\;\;\simeq\;\;\;\;
  \raisebox{45pt}{
  \xymatrix{
    & \underset{\Sigma}{\prod}\mathbf{Fields}_{\Sigma}
    \ar[ddl]
    \ar[ddr]
    \ar@{-->}[dd]|-{\exp(\tfrac{i}{\hbar}S)}
    \\
    \\
    \ast \ar[dr]^{\ }="t" & \mathbf{B}^n U(1) \ar[r]_{\ }="s" \ar[l] & \ast \ar[dl]
        \\
    & \mathbf{B}^{n+1}U(1)
    \ar@{=>} "s"; "t"
  }
  }
$$
is equivalently a dashed map as shown on the right.
If here we think of the tip as a type of fields over a closed manifold
$\Sigma$, and if we furthermore restrict to $n = 0$ in which case $\mathbf{B}^0 U(1) \simeq U(1)$,
then on the right the map denoted $\exp(\tfrac{i}{\hbar}S)$ may be regarded as
an \emph{action functional} on these fields, as indicated. More generally
$\Sigma$ may have boundaries, in which case the situation is more interesting,
we come to this below.

\medskip

\noindent{\bf Infinitesimal identity types $\leftrightarrow$ homotopy Lie algebroids / BRST complexes.}
Homotopy types of gauge equivalences are best known in physics in
the approximation of perturbation theory,
where they appear as homotopy Lie algebroids known as BV-BRST complexes \cite{HenneauxTeitelboim}.
For instance if $\mathbf{Fields} \in \mathbf{H}$ is a type of fields
and $G$ is a gauge group acting on that, then there is the homotopy quotient
$\mathbf{Fields}/\!/G$. Infinitesimally this is the \emph{BRST-complex}.

\subsubsection{Categories of being and Prequantum geometry}

To make formal sense of what we mean by ``infinitesimally'' here,
and generally by ``differential'' etc.
we need to equip the types with \emph{geometric quality}.
In logic and type theory such ``quality'' is called ``modality''.
For propositions a modality is ``a way of being true'' -- for instance 
and traditionally: ``possibly being true''
or ``necessarily being true''.
But for types a modality is simply ``a way of being''.

A modality is formalized (see \ref{ModalitiesMomentsOpposites} below for more) 
as \emph{monad} or \emph{comonad} on the type system
$\mathbf{H}$. Lawvere observed \cite{Lawvere91} that adding an \emph{adjoint idempotent modality}
$$
  \xymatrix{
    \emptyset
    \ar@{}[r]|{\subset}
    \ar@{-|}[d]
    & \flat
    \ar@{-|}[d]
    \\
    \ast
    \ar@{}[r]|{\subset}
    &
    \sharp
  }
$$
naturally has the following meaning:
\begin{itemize}
 \item $\flat$ is the modality of ``being geometrically discrete'';
 \item $\sharp$ is the modality of ``being geometrically codiscrete''.
\end{itemize}
Adding one more makes it a \emph{category of cohesion}
$$
  \xymatrix{
    \\
    &\int \ar@{-|}[d]
    \\
    \emptyset
    \ar@{}[r]|{\subset}
    \ar@{-|}[d]
    & \flat
    \ar@{-|}[d]
    \\
    \ast
    \ar@{}[r]|{\subset}
    &
    \sharp
  }
$$
In homotopy-type theory we have that
\begin{itemize}
  \item $\int$ is the modality of ``being homotopy invariant''.
\end{itemize}
In \cite{dcct} we add three more
to obtain what we called a \emph{category of differential cohesion}
$$
  \xymatrix{
    && \Re
    \ar@{-|}[d]
    \\
    &\int \ar@{-|}[d] \ar@{}[r]|{\subset} & \oint \ar@{-|}[d]
    \\
    \emptyset
    \ar@{}[r]|{\subset}
    \ar@{-|}[d]
    & \flat
    \ar@{}[r]|{\subset}
    \ar@{-|}[d]
    &
    \Im
    \\
    \ast
    \ar@{}[r]|{\subset}
    &
    \sharp
  }
$$
We find that
\begin{itemize}
  \item $\oint$ is the modality of ``being formally {\'e}tale'',
  orthogonal to ``being infinitesimal''.
\end{itemize}
\medskip

\noindent {\bf Claim (Theorem) a)} (\cite{dcct}): In cohesive homotopy-type theory there is a
natural construction of  differential cohomology = higher gauge theory fields.

For instance there is the type $\mathbf{B}U(1)_{\mathrm{conn}} \in \mathbf{H}$
characterized by the fact that for $X$ a smooth manifold, then functions
$$
  \nabla \;\colon\; X \longrightarrow \mathbf{B}U(1)_{\mathrm{conn}}
$$
are equivalently $U(1)$-principal connections on $X$ -- for instance electromagnetic field configurations.

\medskip

\noindent {\bf Improved Claim a)} (\cite{BNV}) In cohesive homotopy-type theory \emph{every}
stable homotopy-type represents a generalized differential cohomology theory
(such as differential K-theory, differential elliptic cohomology, etc.)
hence a higher gauge theory.

\medskip

\noindent {\bf Claim (Theorem)  b)} (\cite{hgp}, \cite{dcct}): Cohesive homotopy-type theory naturally encodes
local pre-quantum geometry, hence Lagrangian cocycle data on higher moduli stacks of fields.

\medskip

We continue to provide some examples of this.

\medskip

\noindent {\bf running Example a) (\ref{KUBoundaryQuantization}) -- The particle at the boundary of 2d Poisson-Chern-Simons theory.}
Consider a Poisson manifold $(X,\pi)$, the phase spaces of a foliation by mechanical systems.
It is encoded in its Poisson Lie algebroid $\mathfrak{P}$, which is
equipped with a 2-plectic cocycle $\pi$
$$
  \mathfrak{P} \stackrel{\pi}{\longrightarrow} \mathbf{B}^2 \mathbb{R}
$$

\noindent {\bf Theorem.} \cite{DiffClasses}\cite{Bongers}
Higher Lie integration $\exp(-)$ of $\pi$ yields the Lagrangian $\mathbf{L}_{2d\mathrm{CS}}$ for non-perturbative
2-d Poisson-Chern-Simons theory
$$
  \mathbf{L}_{2d\mathrm{CS}}
  =
  \exp(\pi)
  \;:\;
  \mathrm{SymplGrpd}(\mathfrak{P},\pi)_{\mathrm{conn}}
  \longrightarrow
  \mathbf{B}^2 U(1)_{\mathrm{conn}}
$$
whose moduli stack of fields is differential cohomology refinement of the ``symplectic groupoid''.
Moreover, the original Poisson manifold is a boundary condition of this 2-d theory
exhibited by a correspondence diagram in the slice $\mathbf{H}_{\mathbf{B} (\mathbf{B}U(1)_{\mathrm{conn}})}$
$$
  \xymatrix{
    & X
    \ar[dr]_-{\ }="s"
    \ar[dl]
    \\
    \ast \ar[dr]^{\ }="t" && \mathrm{SymplGrpd}(X,\pi)_{\mathrm{conn}}
    \ar[dl]^{\mathbf{L}}
    \\
    & \mathbf{B} \mathbf{B}U(1)_{\mathrm{conn}}
    \ar@{=>} "s"; "t"
  }
$$
which describes the points of $X$ as ``trajectories'' along which the fields of
the 2d theory may approach the boundary.

\medskip

\noindent {\bf running Example b) (\ref{stringattheboundary}) -- The string at the boundary of 3d Chern-Simons theory.}

\noindent {\bf Theorem} \cite{DiffClasses}, exposition in \cite{StackyPerspective}.
The Lie integration of the canonical Lie algebra 3-cocycle on the orthogonal Lie algebra
$$
  \xymatrix{
    \mathfrak{so}
    \ar[rr]^-{\langle -,[-,-]\rangle}
    &&
    \mathbf{B}^2 \mathbb{R}
  }
$$
is a differential cohomology refinement of the first fractional
Pontryagin class $\tfrac{1}{2}p_1$
$$
  \mathbf{L}_{3d\mathrm{CS}}
  =
  \tfrac{1}{2}\mathbf{p}
  \;:\;
  \mathbf{B}\mathrm{Spin}_{\mathrm{conn}}
  \longrightarrow
  \mathbf{B}^3 U(1)_{\mathrm{conn}}
$$
and this is the local Lagrangian for 3d $\mathrm{Spin}$ Chern-Simons theory.

Moreover, the \emph{universal} boundary condition for this is the
delooped String 2-group $\mathbf{B}\mathrm{String}$ (see section 5.1 of \cite{dcct}), and hence
a manifold $X$ via a $\mathrm{Spin}$-structure $\nabla_{\mathrm{Spin}}$
yields a boundary for the 3d Chern-Simons theory here precisely if it lifts to a String-structure
$\nabla_{\mathrm{String}}$
$$
  
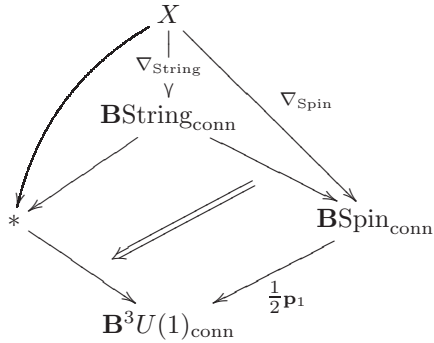
\xymatrix{
    & X \ar[ddr]^{\nabla_{\mathrm{Spin}}}
    \ar@{-->}[d]|{\mathrm{\nabla}_{\mathrm{String}}}
    \ar@/_1pc/[ddl]
    \\
    & \mathbf{B}\mathrm{String}_{\mathrm{conn}}
    \ar[dl]
    \ar[dr]_{\ }="s"
    \\
    \ast
    \ar[dr]^{\ }="t"
     && \mathbf{B}\mathrm{Spin}_{\mathrm{conn}}
     \ar[dl]^{\tfrac{1}{2}\mathbf{p}_1}
    \\
    & \mathbf{B}^3 U(1)_{\mathrm{conn}}
    \ar@{=>} "s"; "t"
  }
$$

Below we see the holographic quantization of these two examples.
In general, a space of field trajectories $\mathbf{Fields}_{\mathrm{traj}}$ equipped with
action functional data for an $n$-dimensional field theory is a correspondence in
$\mathbf{B}_{/\mathbf{B}^n U(1)}$ of the form
$$
  
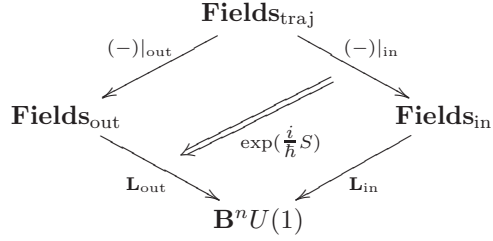
\xymatrix{
    & \mathbf{Fields}_{\mathrm{traj}}
    \ar[dl]_{(-)|_{\mathrm{out}}}
    \ar[dr]^{(-)|_{\mathrm{in}}}_{\ }="s"
    \\
    \mathbf{Fields}_{\mathrm{out}}
    \ar[dr]_{\mathbf{L}_{\mathrm{out}}}^{\ }="t"
    &&
    \mathbf{Fields}_{\mathrm{in}}
    \ar[dl]^{\mathbf{L}_{\mathrm{in}}}
    \\
    & \mathbf{B}^n U(1)
    \ar@{=>}^{\exp(\tfrac{i}{\hbar} S)} "s"; "t"
  }
$$
Now to quantize all this higher prequantum geometry.

\subsubsection{Categories of linear homotopy-types and Quantization}

There is a simple idea: quantization is \emph{linearization} of the above pre-quantum
geometry, analogous to how motivic geometry is a linearization of algebraic geometry.
To formalize this, think of ``linear'' as being ``affine with basepoint''.

\medskip

\noindent {\bf quantum $\leftrightarrow$ linear = base-point + affine.}
The ``modality of being pointed''
is the \emph{maybe monad}
$$
  \ast/ \;:\; X \mapsto X \sqcup \ast
  \,.
$$
Indeed, the $\ast/$-modal types are equivalently the pointed types
canonically equipped with the
smash tensor product as ``linear conjunction'' (\ref{DependentLinearTypeTheory} below). Notice that this is
a non-cartesian tensor product, the hallmark of quantum theory in category theory
(see \ref{LinearLogic} for more on this).

Moreover, according to deformation theory a pointed space is affine if
it is infinitesimally extended, which by the above means it is orthogonal
to $\oint$-modal types.

\noindent {\bf Definition}:
For $X\in \mathbf{H}$ then  $\mathrm{Mod}(X)$ are the $\ast/$-modal types in $\mathbf{H}_{/X}$
which are left orthogonal to $\oint$-modal types.

\noindent{\bf Proposition} (\ref{DependentLinearTypeTheory}): this forms a \emph{linear} homotopy-type theory
$$
  \xymatrix{
    \mathrm{Mod}(X)
    \ar@<+8pt>@{->}[rr]^{\sum_f}
    \ar@<0pt>@{<-}[rr]|{f^\ast}
    \ar@<-8pt>@{->}[rr]_{\prod_f}
    &&
    \mathrm{Mod}(Y)
  }
$$

Under this extension the meaning of the existential quantifier $\sum$ changes drastically
(for the following assume for simplicity of notation that $\mathrm{Line}(\ast)$
has a reflection as $\mathrm{Line} \in \mathbf{H}$ that classifies
invertible linear types):

\medskip

\noindent {\bf Definition} (the path integral, \ref{IntegralTransformations}):
First turn an action functional on trajectories as above into an
\emph{integral kernel} by associating linear coefficients
$$
  
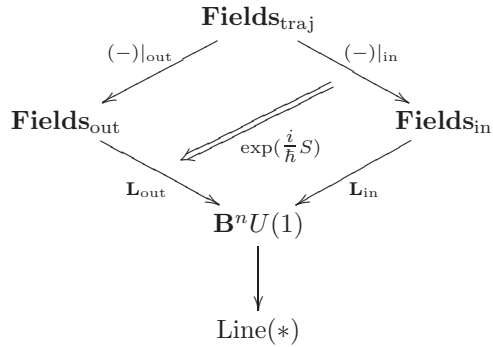
\xymatrix{
    & \mathbf{Fields}_{\mathrm{traj}}
    \ar[dl]_{(-)|_{\mathrm{out}}}
    \ar[dr]^{(-)|_{\mathrm{in}}}_{\ }="s"
    \\
    \mathbf{Fields}_{\mathrm{out}}
    \ar[dr]_{\mathbf{L}_{\mathrm{out}}}^{\ }="t"
    &&
    \mathbf{Fields}_{\mathrm{in}}
    \ar[dl]^{\mathbf{L}_{\mathrm{in}}}
    \\
    & \mathbf{B}^n U(1)
    \ar[d]
    \\
    & \mathrm{Line}(\ast)
    \ar@{=>}^{\exp(\tfrac{i}{\hbar} S)} "s"; "t"
  }
$$
The path integral is then effectively the linear sum over this:
$\underset{\mathbf{Fields}_{\mathrm{traj}}}{\sum} \exp(\tfrac{i}{\hbar}S)$.

The full expression involves pre-composing this with the $(\underset{\mathrm{in}}{\prod} \dashv \mathrm{in}^\ast)$-unit
followed by a ``twisted ambidexterity'' measure $d\mu$, and postcomposing with the
$(\underset{\mathrm{out}}{\sum} \dashv \mathrm{out}^\ast)$-counit:
$$
  \hspace{-2.3cm}
  \xymatrix{
    & \mathbb{D} \underset{{\mathbf{Fields}_{\mathrm{traj}}}}{\int} \exp(\tfrac{i}{\hbar}S) d\mu =
   \\
    \underset{\mathbf{Fields}_{\mathrm{out}}}{\sum} \mathbf{L}_{\mathrm{out}}
    \ar@{<-}[r]^-{\underset{\mathbf{Fields}_{\mathrm{out}}}{\sum} \!\!\!\epsilon_{\mathbf{L}_{\mathrm{out}}}}
    &
    \underset{\mathbf{Fields}_{\mathrm{out}}}{\sum} \mathrm{out}_! \mathrm{out}^\ast
      \mathbf{L}_{\mathrm{out}}
    \ar@{<-}[r]^-{\simeq}
    &
    \underset{\mathbf{Fields}_{\mathrm{traj}}}{\sum} \mathrm{out}^\ast \mathbf{L}_{\mathrm{out}}
    \ar@{<-}[rr]^-{\underset{\mathbf{Fields}_{\mathrm{traj}}}{\sum} \!\!\!\!\!\exp(\tfrac{i}{\hbar}S)}
    &&
    \underset{\mathbf{Fields}_{\mathrm{traj}}}{\sum} \mathrm{in}^\ast \mathbf{L}_{\mathrm{in}}
    \ar@{<-}[r]^-{\simeq}
    &
    \underset{\mathbf{Fields}_{\mathrm{in}}}{\sum} \mathrm{in}_! \mathrm{in}^\ast \mathbf{L}_{\mathrm{in}}
    \ar@{<-}[r]^-{\underset{\mathbf{Fields}_{\mathrm{in}}}{\sum} \!\!\![\mathrm{in}]}
    &
    \underset{\mathbf{Fields}_{\mathrm{in}}}{\sum} \mathbf{L}_{\mathrm{in}} \otimes \tau
  }
$$

\medskip

Notice how here what used to be existential quantification becomes the path integral
with its superposition and quantum interference:

\noindent {\bf The quantum incarnation of existential quantification.}
\begin{enumerate}

\item in logic: That there exists a path.

\item in type theory: The collection of all paths.

\item in homotopy-type theory: The collection of all paths with gauge equivalences between them.

\item in  linear type theory: The linear addition (superposition) of amplitudes of all paths.

\item in linear homotopy-type theory: The linear addition (superposition) of amplitudes of all paths
 with gauge equivalences taken into account.

\end{enumerate}

\noindent {\bf Theorem} \ref{CoboundingTheory}: This is quantum anomaly free if there is cobounding theory.

\medskip

\noindent {\bf running example a) -- The particle at the boundary of 2d Poisson-Chern-Simons theory}:

\noindent {\bf Proposition} \cite{Nuiten13}, \ref{KUBoundaryQuantization}:
The above path integral yields the geometric quantization of symplectic manifolds
in its K-theoretic incarnation (following Bott). It generalizes it to a
geometric quantization of Poisson manifolds and reproduces there for instance the
universal orbit method of \cite{FHT05}.

\medskip

\noindent{\bf running example b)  -- The string at the boundary of 3d Chern-Simons theory}

\medskip

\noindent {\bf Proposition} \ref{stringattheboundary} Quantization of the boundary of $\mathrm{Spin}$ Chern-Simons
with coefficients in the universal elliptic cohomology ring
$\mathrm{tmf}$
  $$
    \raisebox{20pt}
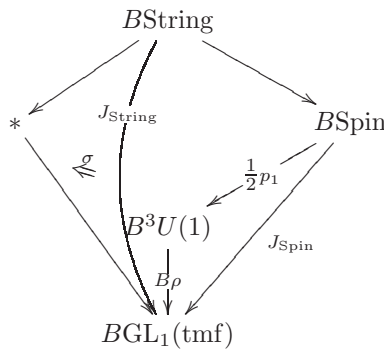
{
    \xymatrix{
      & B \mathrm{String}
      \ar[dl]
      \ar[dr]
      \ar@/_1.52pc/[ddd]|<<<<<<<<<<{J_{\mathrm{String}}}_-{\ }="s"
      \\
      \ast \ar[ddr]^<<<<<<{\ }="t"
      && B \mathrm{Spin} \ar[dl]|-{\tfrac{1}{2}p_1}
      \ar[ddl]^{J_{\mathrm{Spin}}}
      \\
      & B^3 U(1)
      \ar[d]|{B\rho}
      \\
      & B \mathrm{GL}_1(\mathrm{tmf})
      \ar@{=>}_\sigma "s"; "t"
    }
    }
    \,.
  $$
yields the
non-perturbative refinement of the Witten genus \cite{WittenGenus}, the
partition function of the string, to the String-orientation of $\mathrm{tmf}$ \cite{AHR}
$$
  \xymatrix{
    \mathrm{tmf}
    =
    \underset{\ast}{\sum} 1_\ast
    \longleftarrow
    \underset{B \mathrm{String}}{\sum} J_{\mathrm{String}}
    =
    M \mathrm{String}\wedge \mathrm{tmf}
  }
  \,.
$$
(This follows using section 8 of \cite{ABG11}.)

\subsubsection{Outlook}

We are seeing here the pattern of the holographic principle \cite{Maldacena, Witten98}.
The next example of interest of this form is induced from a local Lagrangian
for 7-dimensional Chern-Simons theory \cite{FSS}.
\begin{center}
\begin{tabular}{|l|c|c||c|c|c|}
  \hline
  {\bf field theory} & {\bf spaces of states} & {\bf propagator} & particle/2dCS & string/3dCS & 6dSCFT/7dCS
  \\
  \hline
  $\mathrm{TQFT}_{d+1}$ & $\mathrm{Mod}_2 \in \mathrm{Cat}_2$ & integral transform
  & $\mathrm{TQFT}_3$ & $\mathrm{TQFT}_4$ & $\mathrm{TQFT}_8$
  \\
  \hline
  $\mathrm{TQFT}_{d}^\tau$ & $\mathrm{Mod}(\ast) \in \mathrm{Mod}_2$ & secondary integral transform
  &
  $\mathrm{CS}_{2}$ & $\mathrm{CS}_3$ & $\mathrm{CS}_7$
  \\
  \hline
  $\mathrm{QFT}_{d-1}$ & $\underset{X}{\sum} A_X \in \mathrm{Mod}(\ast)$ 
  &
  \begin{tabular}{l}
    equivariance under
    \\
    Hamiltonian group action
  \end{tabular}
  &
  $\mathrm{QM}_1$ & $\mathrm{WZW}_2$ & $\mathrm{WZW}_6$
  \\
  \hline
\end{tabular}
\end{center}
See also \cite{Freed}.
Notice that according to \cite{WittenSix} the Kaluza-Klein compactification of
$\mathrm{WZW}_6$ on a torus is 4-dimensional (super-)Yang-Mills theory. This way we find
at least a sketch of a plausible path here for how to approach the quantization of Yang-Mills
theory \cite{JaffeWitten} in linear homotopy-type theory. A central open question from
this perspective is which brave new ring would serve as the right coefficient for quantization of
the 7d Chern-Simons theory.

\subsection{Summary (point of view of TQFT)}
 \label{SummaryForTQFT}

This document is an expanded set of notes for some talks I gave in February 2014.
What follows here is the script for what I actually said in a talk at the
``Modern Trends in TQFT'' meeting at ESI in Vienna, to an audience
of TQFT theoreticians.
(Another summary aimed instead at mathematical physicists interested in
foundations is above in \ref{SummaryFoundations}. A summary of more technical details
is below in \ref{SummaryTechnicalDetails}.)

\subsubsection{The need for Lagrangian TQFT and Quantization}

Since the (very detailed outline of) the proof of the cobordism hypothesis \cite{LurieQFT},
there is a full mathematical classification
of \emph{local} quantum field theory in the form of monoidal $(\infty,n)$-functors
$$
  Z \;\colon\; \mathcal{S}\mathrm{Bord}_n^{\sqcup} \longrightarrow \mathrm{Mod}_n^\otimes
  \,.
$$
(Here we denote by $S \mathrm{Bord}_n$ the symmetric monoidal $(\infty,n)$-category of cobordisms
(spacetimes, worldvolumes)
equipped with some given structure $\mathcal{S}$ -- which we take to include lifts of structure groups
but also singularity data -- while $\mathrm{Mod}_n$ is some symmetric monoidal $(\infty,n)$-category of
$n$-categorical modules. Of course this is \emph{topological} field theory, but one point of relevance
in the following is that once singularity data is allowed, such $Z$ contains boundary and defect data
which corresponds to non-topological/geometric field theory in the sense of physics.)

\medskip

While this result is a celebrated and fundamental result, it is noteworthy that the
field theories appearing in nature and in physical theory are not random examples of
this classification. Instead they are instances of
\emph{Lagrangian quantum field theory}: they arise from a process of ``quantization''
from ``Lagrangian'' geometric $n$-cocycle data.

\medskip

\noindent{\bf Local Lagrangian field theory.}
\\
\begin{tabular}{|ccc|}
  \hline
  {\bf local pre-quantum field theory}& $\stackrel{\mbox{\tiny quantization}}{\longrightarrow}$ &
  {\bf local quantum field theory}
  \\
  \hline
  \hline
  \begin{tabular}{l}Lagrangian, \\action functional
  \end{tabular}
   &&
  \begin{tabular}{l}
     propagator, \\ correlator, \\ S-matrix
  \end{tabular}
  \\
  \hline
  \begin{tabular}{l}
    geometric $n$-cocycle:
    \\
    $\mathbf{Fields} \stackrel{\mathbf{L}}{\longrightarrow} \mathbf{B}^n U(1)_{\mathrm{conn}}$
  \end{tabular}
  &&
  \begin{tabular}{l}
    monoidal $n$-functor:
    \\
    $S\mathrm{Bord}_n^{\sqcup} \stackrel{Z}{\longrightarrow} \mathrm{Mod}_n$
  \end{tabular}
  \\
  \hline
\end{tabular}

\medskip

\noindent Hence the next {\bf open question}
is: \\
\noindent Which higher categorical refinement of the process of quantization
produces such functors $Z$ from Lagrangian data $\mathbf{L}$
so that we may eventually write
$$
  \mbox{
  ``
  $Z \;\simeq\; \underset{\phi \in \mathbf{Fields}_{\mathrm{traj}}}{\int} \!\!\!\!\!\exp(\tfrac{i}{\hbar}S(\phi)) \,d\mu$''
  }
  \;\;\;\;\;
$$
with $\exp(\tfrac{i}{\hbar}S)$ the action functional obtained by transgression from
the Lagrangian $\mathbf{L}$?

\medskip

\noindent {\bf Relevance in physics.}
To see that this is an issue relevant for mathematical/theoretical physics, the most famous
example is 3-dimensional Chern-Simons gauge field theory \cite{WittenCS}
for a compact Lie group $G$. The theory is defined to be the quantization of the
Chern-Simons Lagrangian $\mathbf{L}_{\mathrm{CS}}$
(whose action functional sends fields given by $G$-principal connections to
the volume holonomy of the secondary differential characteristic of their second Chern class
\cite{StackyPerspective}).
It has always been conjectured that the result of this is equivalently the
functor $Z_{\mathrm{MTC}}$ on 3-bordisms which is constructed algebraically
(via Reshetikhin-Turaev, Turaev-Viro, skein relations, etc., see for instance \cite{Rowell} for a review) from the
modular tensor category of $G$.
This conjectured equivalence is crucial for
various results that build on it, for instance the construction and classification \cite{FuchsRunkelSchweigert}
of full (defined on all cobordisms) rational 2d conformal field theory $\mathrm{CFT}_{\mathrm{rat}}$
as the holographic boundary theory of Chern-Simons .
$$
  \xymatrix{
    \mathbf{L}_{\mathrm{CS}}
    \ar@{|->}[rr]^{\mbox{\tiny ? quantization ?}}
    &&
    Z_{\mathrm{MTC}}
    \\
    && \mathrm{CFT}_{\mathrm{rat}}
    \ar@{<-|}[u]_{\mbox{\tiny holography}}
  }
$$
(To put this in perspective: this is the full classification of a (tiny) fraction of the
the moduli space of 2d CFTs, the ``landscape'' of string theory \cite{Schellekens}.)
However, an actual proof of $\mathbf{L}_{\mathrm{CS}} \mapsto Z_{\mathrm{MTC}}$ had been elusive.
One is claimed only rather recently\footnote{See also J. Andersen's talk at the
\emph{String Geometry Network} meeting.} in \cite{Andersen}.

For non-compact gauge groups, the situation is even more interesting and even more
subtle. In \cite{Witten3d} it was famously argued that
3d quantum gravity with suitable cosmological constant is equivalent to
3d quantum Chern-Simons theory with suitable non-compact gauge group. But later in \cite{Witten3dAgain}
are listed arguments why this cannot quite be after all, and the issue remains open.
On the answer to this it depends whether a whole sequence of mathematical results
informs us about 3d quantum gravity or not, for instance Liouville field theory
\cite{CHvD} and quantum Teichm{\"u}ller theory \cite{Teschner}.
$$
  \xymatrix{
    \mathbf{L}_{\mathrm{3dGrav}}
    \ar@{|->}[rr]^{\mbox{\tiny ?}}
    &&
    \mathbf{L}_{\mathrm{CS}}
    \ar@{|->}[d]
    \\
    && \mathrm{CFT}_{\mathrm{Liouv.}}
  }
$$

And of course for the case of actual physical interest, quantum gravity in dimensions $\geq 4$ the
problem is, infamously, more interesting and wide open.

\medskip

\noindent Here we won't solve this problem, clearly, but we do want to
start to systematically investigate it with suitable mathematical formalism.

\medskip

\noindent {\bf Relation to existing literature.} Proposals along the lines to follow in available literature include:
\begin{itemize}

\item There is a long tradition, going back to \cite{Weinstein} in the context of mechanics
and then particularly picked up in \cite{Freed92} in the context of field theory, to propose that
 the correct domain for quantization are correspondences of phases spaces (or configuration
 spaces) of fields, and that (geometric/path integral) quantization is a kind of linearization
 of such correspondences of spaces given by pull-push, much as in the theory of motives. (More on this relation
 below in \ref{Motives}).

\item For the comparatively simple case of higher gauge theories of Dijkgraaf-Witten type
(with finite gauge group)
a method for such pull-push quantization is sketched in sections 3 and 8 of \cite{FHLT}.
Partial details on how to flesh out that proposal have now been made available in
\cite{HopkinsLurie}.

\item Non-perturbative quantization of non-finite and non-topological theories
 as boundary field theories of local topological field theories
 has been studied in \cite{Nuiten13}.
 The two main examples there we will re-consider below: the quantum particle at the boundary
 of the 2d Poisson-Chern-Simons theory and the quantum superstring at the boundary of the
 3d $\mathrm{Spin}$-Chern-Simons theory.
\end{itemize}
\noindent These proposals focus on different sub-aspects of the general issue and should
be subsumed in a comprehensive theory.
Here we mean to further push in this direction. We give a formalization of the path integral that
subsumes the notion of integration in  \cite{HopkinsLurie} as a special case.
It generalizes it in particular by allowing certain ``twists'' to appear on
boundary data (twists that would be anomalous in the bulk) where we find them to
give the boundary field theory of a TQFT its non-topological ``geometric'' character
as in \cite{Nuiten13}. Finally we show that the condition that the TQFT in the
bulk is quantum anomaly free means equivalently that it is itself the boundary field
theory of a yet higher dimensional theory, making the non-topological field theory
not just a boundary field theory but  \emph{corner field theory}. This is a pattern
that has been expected in \cite{Sati}, see also \cite{Freed}. For Lagrangian pre-quantum
geometry we had discussed this in sections 3.9.14 and 5.7 of  \cite{dcct}, based on \cite{FiorenzaValentino}.

\medskip

\noindent {\bf What we do.} Here we consider quantization for local (extended) TQFT
in dimensions $(d-1, d, d+1)$, localized to the point in a way that it produces
some of the corner and boundary data for a monoidal 2-functor of the form\footnote{An detailed discussion of
$\mathrm{Corr}_2(-)$ is for instance in section 10 of \cite{DyckerhoffKapranov}.}
$$
  \xymatrix{
    \mathcal{S} \mathrm{Bord}_{2}^{\sqcup} \ar[rr]^-{\exp(\tfrac{i}{\hbar}S)}_-{\mbox{\tiny loc. Lagrangian}}
    \ar@/_2pc/[rrrr]_{Z := \underset{\phi \in \mathbf{Fields}}{\int} \!\!\!\!\exp(\tfrac{i}{\hbar}S(\phi)) d\mu}|{\mbox{\tiny local Lagrangian TQFT}}
    &&
    \mathrm{Corr}_2(\mathbf{H})^{\otimes}
    \ar[rr]^-{\int (-)d\mu}_-{\mbox{\tiny quantization}}
   &&\mathrm{Mod}_2^{\otimes}
   }
  \,,
$$
where $\mathcal{S} = \{ \mathrm{corner} \}$ is the singularity datum of a  \emph{corner} (a codim-$(n-1)$ boundary ending itself on a codim-$(n-2)$ boundary) and find an interpretation as:
\begin{enumerate}
  \item in the middle dimension this is a $\mathrm{TQFT}_d^\tau$  such as 2d Poisson-Chern-Simons or 3d $\mathrm{Spin}$-Chern-Simons;
  \item whose boundary field theory is a non-topological $\mathrm{QFT}_{d-1}$  such as the quantum particle
  and the quantum string;
  \item whose cobounding theory $\mathrm{TQFT}_{d+1}$ exhibits its quantum anomaly cancellation.
\end{enumerate}

\noindent {\bf What we do not consider.}
We discuss the data needed for such a functor essentially only on the generating corner 2-cell
without discussing all the conditions needed to lift this to a complete monoidal 2-functor or rather,
in the end, to a monoidal $n$-functor. In a somewhat simpler version, the problem of extending the bulk field theory
to a full $(\infty,2)$-functor is the topic remark 4.2.5 of \cite{HopkinsLurie}. Using this one may
begin to see what the necessary conditions are to lift the data that we consider here to a full
monoidal $n$-functor, but we will not look into this here.

\subsubsection{Local prequantum field theory with Boundaries and Corners}

The idea of local pre-quantum field theory\footnote{This here is a lightning review of section 3.9.14 in \cite{dcct}  (inspired from \cite{FiorenzaValentino}).}
is that where a genuine quantum field theory
assigns a space of quantum states, its pre-quantum version assigns
a moduli stack $\mathbf{Fields}$ of field configurations ($\sim$ a phase space, but see below)
equipped with a Lagrangian and a prequantum line bundle.
And where a genuine quantum field theory assigns a propagator or S-matrix, the pre-quantum field theory
assigns a space of field trajectories and a transformation between the incoming/outgoing
prequantum bundles.

\medskip

\noindent Let $\mathbf{H}$ be a differentially cohesive $\infty$-topos \cite{dcct} such as
$\infty$-stacks over the site of formal manifolds
$\mathbf{H}= \mathrm{Sh}_\infty(\mathrm{FormMfds})$ as discussed in section 4.5 of \cite{dcct}. 
It being differentially cohesive
means that we have moduli for differential cohomology\footnote{
See also U. Bunke's talk at the ``Modern Trends in TQFT'' meeting.} in $\mathbf{H}$
in particular that we know how to find moduli $n$-stacks
of $(n-1)$-bundle gerbes with connection, which we write
$$
  \mathbf{B}^n U(1)_{\mathrm{conn}} \in \mathbf{H}
  \,.
$$

\medskip

\noindent {\bf Example.} For $(X_1, \omega_1)$ and $(X_2, \omega_2)$ two symplectic manifolds
(phase spaces) and $L \stackrel{i}{\hookrightarrow} (X_1 \times X_2, p_1^\ast \omega_1 - p_2^\ast \omega_2)$
a Lagrangian subspace of their product (with one of the symplectic structures taken with opposite sign),
hence a \emph{Lagrangian correspondence} \cite{Weinstein},
defines a correspondence in $\mathbf{H}_{/\Omega^2_{\mathrm{cl}}}$
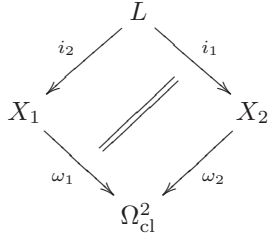
$$
  \raisebox{20pt}{
  \xymatrix{
    & L
    \ar[dl]_{i_2}
    \ar[dr]^{i_1}_{\ }="s"
    \\
    X_1
    \ar[dr]_{\omega_1}^{\ }="t"
     && X_2
     \ar[dl]^{\omega_2}
    \\
    & \Omega^2_{\mathrm{cl}}
    \ar@{=} "s"; "t"
  }
  }
  \,.
$$
A \emph{pre-quantization} of this is a lift through the universal curvature map
$F_{(-)} : \mathbf{B}U(1)_{\mathrm{conn}} \to \Omega^2_{\mathrm{cl}}$
to a correspondence of the form \cite{Schreiber13}.
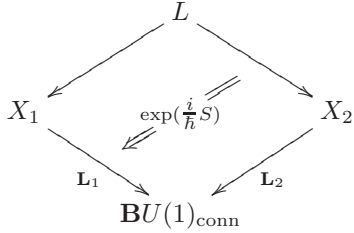
$$
  \raisebox{20pt}{
  \xymatrix{
    & L
    \ar[dl]
    \ar[dr]_{\ }="s"
    \\
    X_1
    \ar[dr]_{\mathbf{L}_1}^{\ }="t"
     && X_2
     \ar[dl]^{\mathbf{L}_2}
    \\
    & \mathbf{B}U(1)_{\mathrm{conn}}
    \ar@{=>}|{\exp(\tfrac{i}{\hbar}S)} "s"; "t"
  }
  }
  \,.
$$

\medskip

\noindent {\bf Proposition.} \cite{hgp} Given a prequantum line bundle
$\mathbf{L} : \mathbb{R}^{2n} \to \mathbf{B}U(1)_{\mathrm{conn}}$ prequantizing
the canonical phase space, then morphisms in $\mathbf{H}$ of the form
$$
  \mathbf{B}\mathbb{R}
  \longrightarrow
  \mathbf{B}\mathrm{Aut}_{\mathbf{B}U(1)_{\mathrm{conn}}}(\mathbf{L})
$$
which are ``concrete'' (something expressible using the fourth adjoint in the definition of cohesion)
are equivalent to choices of Hamiltonians $H \in C^\infty(\mathbb{R}^n)$ and send
$t \in \mathbb{R}$ to the prequantized Lagrangian correspondence as above with
$S = \int_{0^t} L \, dt$ the respective Hamilton-Jacobi action,
i.e. the integral over the Legendre transform $L$ (the Lagrangian)
of $H$.

\medskip
So this expresses the core ingredients of classical mechanics.
Passing here from $\mathbf{B}U(1)_{\mathrm{conn}}$ to $\mathbf{B}^n U(1)_{\mathrm{conn}}$
produces the Hamilton-De Donder-Weyl formulation of classical local field theory
(see section 1.2.11 of \cite{dcct} or \cite{Schreiber13}).

\medskip

\noindent {\bf Example.} The gauge coupling action functional $S_{\mathrm{int}}$ for an electromagnetically charged particle propagating on a (spacetime) manifold $X$ with electromagnetic field given by a $U(1)$-principal
connection $\nabla : X \to \mathbf{B}U(1)_{\mathrm{conn}}$
on a $U(1)$-principal bundle $\mathbf{L} : X\to\mathbf{B}U(1)$ is a
correspondence in $\mathbf{H}_{/\mathbf{B}U(1)}$ with tip the smooth path space
of $X$:
$$
  \raisebox{20pt}{
  \xymatrix{
    & X^{[0,1]}
    \ar[dl]_{(-)|_{0}}
    \ar[dr]^{(-)|_{1}}_{\ }="s"
    \\
    X \ar[dr]_{\mathbf{L}}^{\ }="t" && X \ar[dl]^{\mathbf{L}}
    \\
    & \mathbf{B}U(1)
    \ar@{=>}|{\exp(\tfrac{i}{\hbar}S_{\mathrm{int}} )} "s"; "t"
  }
  }
$$
(the components of the homotopy filling this diagram is over each path an assignment which
sends a trivialization of $\mathbf{L}$ over the endpoints of these paths to a value in $U(1)$:
the \emph{parallel transport} of along with path with respect to $\nabla$).

\medskip

\noindent {\bf Definition.} For $\mathbf{H}$ an $\infty$-topos, write $\mathrm{Corr}_n(\mathbf{H})$
for the $(\infty,n)$-category of $n$-fold correspondences in $\mathbf{H}$.

\medskip

\noindent {\bf Fact.} Every object in $\mathrm{Corr}_n(\mathbf{H})$ is fully dualizable,
and in fact is fully self-dual.

\medskip

\noindent Hence for an $n$-dimensional prequantum field theory the
action functional looks like
$$
  \xymatrix{
    & \mathbf{Fields}_{\mathrm{traj}}
    \ar[dl]_{(-)|_{\mathrm{out}}}
    \ar[dr]^{(-)|_{\mathrm{in}}}_{\ }="s"
    \\
    \mathbf{Fields}_{\mathrm{out}}
    \ar[dr]_{\mathbf{L}_{\mathrm{out}}}^{\ }="t"
    &&
    \mathbf{Fields}_{\mathrm{in}}
    \ar[dl]^{\mathbf{L}_{\mathrm{in}}}
    \\
    & \mathbf{B}^n U(1)
    \ar@{=>}|{\exp(\tfrac{i}{\hbar} S)} "s"; "t"
  }
$$
and we may axiomatize such a pre-quantum bulk field theory as
$$
  \exp(\tfrac{i}{\hbar}S) \;:\; S\mathrm{Bord}_n^{\sqcup} \longrightarrow \mathrm{Corr}_n(\mathbf{H}_{/\mathbf{B}^n U(1)})^\otimes
  \,.
$$
Then the idea would be that to formalize the path integral is to produce a
monoidal $n$-functor
$$
  \int (-) d\mu : \mathrm{Corr}(\mathbf{H}_{/\mathbf{B}^n U(1)})^{\otimes} \longrightarrow \mathrm{Mod}_n^\otimes
$$
then the composite $\int \exp(\tfrac{i}{\hbar}S)d\mu$ would be the quantized field theory.

This is essentially what we will consider, only that it turns out to be natural to
first realize the action functional itself as a boundary effect of an auxiliary $(d+1)$-dimensional
theory.
To that end notice that on the last few pages of \cite{LurieQFT} there is
a vast generalization of the cobordism theorem to the case of cobordisms with singularities:

\medskip

\noindent {\bf Cobordism hypothesis with singularities} (theorem 4.3.11 in \cite{LurieQFT}, rough paraphrase):
{\it If $\mathcal{S}$ is a collection of ``catastrophy diagrams'' characterizing types of $k$-dimensional singularities
for $0 \leq k \leq n$, then the $(\infty,n)$-category $\mathcal{S} \mathrm{Bod}_n^\sqcup$ of
cobordisms with singularities of these types is the free symmetric monoidal $(\infty,n)$-category
with duals generated from the elements of $S$ regarded as $k$-morphisms.}

\medskip

\noindent {\bf Example.} $\mathcal{S} = \{(\emptyset \to \ast)\}$ encodes a single bulk theory $\ast$
with a single type of codimension $(n-1)$-boundary. A monoidal $n$-functor
$\mathbf{Fields} :  S \mathrm{Bord}_n^\sqcup \to \mathrm{Corr}_n(\mathbf{H})^\otimes$ is this case is
equivalently a choice $\mathbf{Fields}(\ast)\in \mathbf{H}$ of bulk fields together with a
morphism from $\ast$ to $\mathbf{Fields}(\ast)$ in $\mathrm{Corr}_n(\mathbf{H})$, hence
a correspondence in $\mathbf{H}$
$$
  \xymatrix{
     \ast & \mathbf{Fields}^\partial
     \ar[r]
     \ar[l]
    & \mathbf{Fields}(\ast)
  }
  \,,
$$
hence a choice of moduli $\mathbf{Fields}^\partial$ of boundary fields and a choice of map from these into
the bulk fields.

\medskip

\noindent {\bf Example.} The singularity datum that we consider in the following is an elementary \emph{corner}
(formalizing corner field theories as envisioned in \cite{Sati})
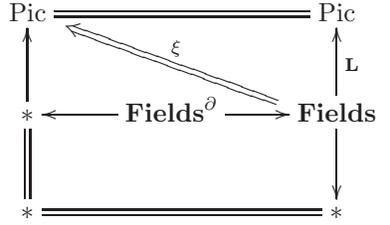
$$
  \raisebox{20pt}{
  \xymatrix{
    \mathrm{Pic} \ar@{=}[rr]_<{\ }="t" && \mathrm{Pic}
    \\
    \ast \ar[u] \ar@{=}[d] & \mathbf{Fields}^\partial
    \ar[l] \ar[r]^>{\ }="s"
    & \mathbf{Fields} \ar[u]_{\mathbf{L}} \ar[d]
    \\
    \ast \ar@{=}[rr] && \ast
    \ar@{=>}_\xi "s"; "t"
  }
  }
  \,,
$$
This means that the $d$-dimensional bulk $\mathbf{Fields}$ are realized themselves as
the boundary fields of a $(d+1)$-dimensional theory with fields $\mathrm{Pic} \in \mathbf{H}$,
so that the boundary $\mathbf{Fields}^\partial$ of the $d$-dimensional theory are now
corner fields of the $(d+1)$-dimensional theory.

A diagram showing trajectories of the $d$-dimensional bulk theory ``approaching and then hitting the boundary''
is hence of the following form
$$
  
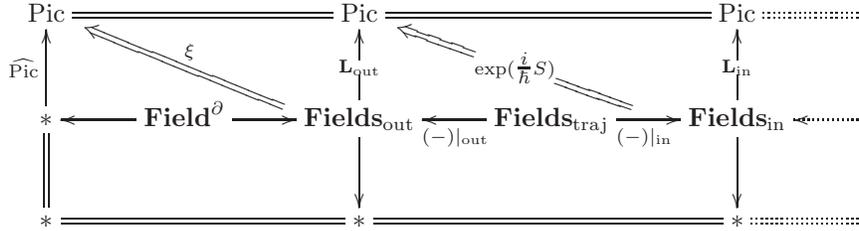
\xymatrix{
    \mathrm{Pic} \ar@{=}[rr]_<{\ }="t1" && \mathrm{Pic} \ar@{=}[rr]_<{\ }="t" && \mathrm{Pic} \ar@{::}[r] &
    \\
    \ast \ar[u]^{\widehat{\mathrm{Pic}}} \ar@{=}[d]
    & \mathbf{Field}^\partial
    \ar[l] \ar[r]^>{\ }="s1" & \mathbf{Fields}_{\mathrm{out}} \ar[u]|{\mathbf{L}_{\mathrm{out}}}
     \ar[d]
    &
    \mathbf{Fields}_{\mathrm{traj}}
    \ar[l]^{(-)|_{\mathrm{out}}}
    \ar[r]_{(-)|_{\mathrm{in}}}^{\ }="s"
    &
    \mathbf{Fields}_{\mathrm{in}}
    \ar[u]|{\mathbf{L}_{\mathrm{in}}}
    \ar[d]
    \ar@{<..}[r]
    &
    \\
    \ast \ar@{=}[rr] && \ast \ar@{=}[rr] && \ast \ar@{::}[r] &
    \ar@{=>}|{\exp(\tfrac{i}{\hbar}S)} "s"; "t"
    \ar@{=>}_\xi "s1"; "t1"
  }
$$
(We will see that composition to the right here imposes strong ``anomaly cancellation'' constraints on the
quantization to make the composites functorial. But at the boundary itself, where no further
composition is possible (to the left) these conditions are relaxed and the available choices
are part of what makes the boundary theory itself ``geometric'' (non-topological).)

\subsubsection{Linear homotopy-types and Spaces of states}

Traditional geometric quantization produces vector spaces of quantum states as sections
of bundles of vector spaces (``prequantum bundles'') over configuration spaces/phase spaces
of fields.\footnote{Of course geometric quantization produces \emph{polarized} sections.
We take \emph{all} sections for the cobounding TQFT and find -- Example a) below -- that this automatically
makes polarized sections appear in the boundary theory.}
$$
 \left(
  \mbox{
    \begin{tabular}{l}
      prequantum  line bundle
      \\
      $\mathbf{L}$
      \\
      on phase space $X$
    \end{tabular}
  }
  \right)
  \;\;
  \mapsto
  \;\;
  \left(
  \mbox{
    \begin{tabular}{l}
      vector space of states
      \\
      $\Gamma(\mathbf{L})$
      \\
      over the point
    \end{tabular}
  }
  \right)
  \,.
$$

\noindent{\bf Example.} If $X$ here is just a finite set, then $\mathbf{L} \in \mathrm{Vect}(X)$
is just an $X$-parameterized collection of vector spaces, and then
$\Gamma_X(\mathbf{L}) \simeq \underset{\underset{x \in X}{\longrightarrow}}{\lim} \mathbf{L}_x$, which is
left Kan extension along the terminal map $X \to \ast$
$$
  \xymatrix{
    \mathrm{Vect}(X)
    \ar@<+8pt>@{->}[rr]^-{\underset{\rightarrow}{\lim}}
    \ar@{<-}[rr]|-{\mathrm{const}}
    \ar@<-8pt>@{->}[rr]_-{\underset{\leftarrow}{\lim}}
    &&
    \mathrm{Vect}(\ast) = \mathrm{Vect}
  }
  \,.
$$

\noindent Plain vector spaces over the complex or real numbers like this will be
too inflexible for a decent formalization of the path integral. But more general
``linear homotopy-types'' will do. We axiomatize the minimum structure that we need:

\medskip

\noindent {\bf Definition} A (2-monoidal, Beck-Chevalley) model for \emph{linear homotopy-type theory}
is a Cartesian fibration of closed symmetric monoidal $(\infty,1)$-categories
$$
  \xymatrix{
    \mathrm{Mod}(-)
    \ar[d]
    \\
    \mathbf{H}
  }
$$
over a Cartesian monoidal $\infty$-category $\mathbf{H}$,
that satisfies good base change in that for every map $f : X \longrightarrow Y$ in $\mathbf{H}$
\begin{enumerate}
  \item
    pullback $f^\ast$ is a strong monoidal $\infty$-functor;
  \item
     which
    has a left and a right adjoint
    $(f_! \dashv f^\ast \dashv f_\ast)
    :
    \xymatrix{
      \mathrm{Mod}(X)
        \ar@<+8pt>@{->}[rr]|{f_!}
        \ar@{<-}[rr]|{f^\ast}
        \ar@<-8pt>@{->}[rr]|{f_\ast}
        &&
      \mathrm{Mod}(Y)
    }
    $;
  \item
    and satisfies the ``projection formula''
    $
      f_!((f^\ast A)\otimes B)
      \stackrel{\simeq}{\longrightarrow}
      A \otimes f_! B
    $
  \item
    such that
    \begin{itemize}
      \item  the Beck-Chevalley condition holds for the left adjoints, meaning that
      $$
        X (\stackrel{f}{\leftarrow} Z \stackrel{g}{\rightarrow}) Y
        \mapsto
        g_! f^\ast  : \mathrm{Mod}(X) \longrightarrow \mathrm{Mod}(Y)
      $$
      constitutes an $(\infty,1)$-functor\footnote{The BC condition says that we get an $\infty$-functor
      to $\infty$-categories, but with the projection formula it follows that this indeed takes values in
      $\mathrm{Mod}(\ast)$-linear $\infty$-categories.}
      $$
        \mathrm{Corr}_1(\mathbf{H}) \longrightarrow (\mathrm{Mod}(\ast))\mathrm{Mod}
        \,,
      $$
      \item  there are natural equivalences
$\mathrm{Mod}(X \times_Z Y)\simeq \mathrm{Mod}(X) \underset{\mathrm{Mod}(Z)}{\otimes}
  \mathrm{Mod}(Y)$.
    \end{itemize}
\end{enumerate}

\medskip

\noindent This is one incarnation of the ``yoga of six operations'', the one called a ``Wirthm{\"u}ller context''
in \cite{May05}.
In the context of (linear) homotopy type theory \cite{HoTT} the left adjoint
$f_!$ is also written $\underset{f}{\sum}$, expressing the fact that this operation ``sums up''
data over the (homotopy-)fibers of $f$. This is a very suggestive notation for the path integral
that we are after, and we will interchangeably use it below.

\medskip

\noindent {\bf Proposition.} For $E$ an $E_\infty$-ring, and $\mathbf{H} = \infty \mathrm{Grpd}$,
then the assignment
$$
  X \mapsto \mathrm{Mod}(X) := E \mathrm{Mod}(X) := \mathrm{Func}(X, E \mathrm{Mod})
$$
which sends an $\infty$-groupoid (fundamental $\infty$-groupoid of a topological space)
to the $\infty$-category of $E$-module spectrum bundles ($E$-linear ``local systems'' )
over $X$
is a (2-monoidal, Beck-Chevalley) model for linear homotopy-type theory.

\medskip

\noindent {\bf Definition.} Given $\mathrm{Mod}(-)$, the $(\infty,2)$-category $\mathrm{Mod}_2$ of 2-modules is
$\mathrm{Mod}_2 := (\mathrm{Mod}(\ast), \otimes)\mathrm{Mod}$.

\medskip

Now given $\mathbf{L}\in \mathrm{Mod}(\mathbf{Fields})$ regarded
as a prequantum bundle, then the linear homotopy-type of (dual)
prequantum states is
$$
  \underset{\mathbf{Fields}}{\sum} \mathbf{L}
  \in
  \mathrm{Mod}(\ast)
  \,.
$$

We want that a correspondence between field moduli induces a linear map
between these spaces given by integrating over spaces of trajectories.
For this path integral we first need a notion of measure to integrate against.

\subsubsection{Fundamental classes, measures and Path integral quantization via Pull-push}

We now formulate in the general context of a linear homotopy-type theory as above
what it means to produce (``secondary'') integral transformations by ``secondary'' pull-push
of linear homotopy types and discuss how this subsumes as a special cases pull-push in
generalized twisted cohomology.

The operation is ``secondary'' as the pull-push is
implemented by the units and counits of the $(f_! \dashv f^\ast \dashv f_\ast)$-operations,
instead of by these operations themselves as in familiar categorified integral transforms.
Below we discuss that the ``secondary'' operation discussed here is precisely a boundary condition
for the ``primary'' operation and as such in direct analogy to the traditional
terminology by which the Chern-Simons functional
is a ``secondary characteristic'' where the topological Yang-Mills functional is ``primary''.

\begin{center}
\begin{tabular}{|l|c|l|}
  \hline
  {\bf field theory} & {\bf spaces of states} & {\bf propagator}
  \\
  \hline
  $\mathrm{TQFT}_{d+1}$ & $\mathrm{Mod}_2 \in \mathrm{Cat}_2$ & integral transform
  \\
  \hline
  $\mathrm{TQFT}_{d}^\tau$ & $\mathrm{Mod}(\ast) \in \mathrm{Mod}_2$ &
   \begin{tabular}{l} secondary integral transform, \\ path integral \end{tabular}
  \\
  \hline
  $\mathrm{QFT}_{d-1}$ & $\underset{X}{\sum} A_X \in \mathrm{Mod}(\ast)$ &
  \begin{tabular}{l}
    equivariance under
    \\
    Hamiltonian $G$-action
  \end{tabular}
  \\
  \hline
\end{tabular}
\end{center}

The crucial ingredient to make sense of integration is a concept of
measure to integrate against. We formalize this as follows.

\noindent {\bf Definition.} A choice of \emph{fiberwise fundamental class} $[f]$
on a morphism
$$
  \xymatrix{
    \widehat{\mathbf{Fields}}
    \ar[d]^f
    \\
    \mathbf{Fields}
  }
$$
in $\mathbf{H}$ is
(if it exists) a choice of dualizable $\tau \in \mathrm{Mod}(\mathbf{Fields})$
together with a choice of equivalence
$$
  f_! f^\ast (1) \simeq f_\ast f^\ast (\tau)
  \,.
$$

\noindent {\bf Example.} If we happen to have even a natural equivalence $f_! \simeq f_\ast$
then this induces in particular a fundamental class $[f]$ with vanishing twist, $\tau = 1$.
This case of a single but two-sided (``ambidextrous'') adjoint to $f^\ast$ is what is used to
formulate integration in \cite{HopkinsLurie}. For our purposes we need the more general
concept where only the above equivalence with possibly non-trivial $\tau$ is required.

\medskip

\noindent{\bf Proposition.} A choice of fiberwise fundamental class, induces a natural morphism
of the form
$$
  [f]_A : A \otimes \tau \longrightarrow f_! f^\ast A
  \,.
$$
Notice that this goes reverse to the $(f_! \dashv f^\ast)$-unit.

\medskip

\noindent {\bf Definition (path integral).}
Given an integral kernel
$$
  \xymatrix{
    & \mathbf{Fields}_{\mathrm{traj}}
    \ar[dl]_{(-)|_{\mathrm{out}}}
    \ar[dr]^{(-)|_{\mathrm{in}}}_{\ }="s"
    \\
    \mathbf{Fields}_{\mathrm{out}}
    \ar[dr]_{\mathbf{L}_{\mathrm{out}}}^{\ }="t"
    &&
    \mathbf{Fields}_{\mathrm{in}}
    \ar[dl]^{\mathbf{L}_{\mathrm{in}}}
    \\
    & \mathbf{B}^n U(1)
    \ar[d]
    \\
    & \mathrm{Line}(\ast)
    \ar@{=>}|{\exp(\tfrac{i}{\hbar} S)} "s"; "t"
  }
$$
and a fiberwise fundamental class $[\mathrm{in}]$, then
the path integral is the composite
$$
  \hspace{-2.4cm}
  \xymatrix{
    & \mathbb{D} \underset{{\mathbf{Fields}_{\mathrm{traj}}}}{\int} \exp(\tfrac{i}{\hbar}S) d\mu :=
   \\
    \underset{\mathbf{Fields}_{\mathrm{out}}}{\sum} \mathbf{L}_{\mathrm{out}}
    \ar@{<-}[r]^-{\underset{\mathbf{Fields}_{\mathrm{out}}}{\sum} \!\!\!\epsilon_{\mathbf{L}_{\mathrm{out}}}}
    &
    \underset{\mathbf{Fields}_{\mathrm{out}}}{\sum} \mathrm{out}_! \mathrm{out}^\ast
      \mathbf{L}_{\mathrm{out}}
    \ar@{<-}[r]^-{\simeq}
    &
    \underset{\mathbf{Fields}_{\mathrm{traj}}}{\sum} \mathrm{out}^\ast \mathbf{L}_{\mathrm{out}}
    \ar@{<-}[rr]^-{\underset{\mathbf{Fields}_{\mathrm{traj}}}{\sum} \!\!\!\!\!\exp(\tfrac{i}{\hbar}S)}
    &&
    \underset{\mathbf{Fields}_{\mathrm{traj}}}{\sum} \mathrm{in}^\ast \mathbf{L}_{\mathrm{in}}
    \ar@{<-}[r]^-{\simeq}
    &
    \underset{\mathbf{Fields}_{\mathrm{in}}}{\sum} \mathrm{in}_! \mathrm{in}^\ast \mathbf{L}_{\mathrm{in}}
    \ar@{<-}[r]^-{\underset{\mathbf{Fields}_{\mathrm{in}}}{\sum} \!\!\![\mathrm{in}]}
    &
    \underset{\mathbf{Fields}_{\mathrm{in}}}{\sum} \mathbf{L}_{\mathrm{in}} \otimes \tau
  }
$$

\medskip \noindent {\bf Example.} If the linear homotopy-type theory is that given by
vector spaces parameterized over (finite) sets, then this formula reproduces
traditional matrix calculus in linear algebra.

\medskip

\noindent {\bf Proposition.} If the linear homotopy-type theory is that given by
$E$-module bundles over (small) $\infty$-groupoids for $E \in \mathrm{CRing}_\infty$ an $E_\infty$-ring,
then this formula reproduces the pull-push in twisted generalized $E$ cohomology by
twisted Umkehr maps as described in \cite{ABG11}.

\subsubsection{Examples}

\noindent {\bf Example a) particle at the boundary of 2d Poisson-Chern-Simons theory}

(see example \ref{KUBoundaryQuantization})

\medskip

\noindent {\bf Example b) superstring at the boundary of the 3d $\mathrm{Spin}$-Chern-Simons theory}

(see example \ref{stringattheboundary})

\medskip

\noindent {\bf Example c) D-brane charge and T-duality}

(see example \ref{DBraneChargeAndTDuality})

\subsubsection{Quantum anomaly cancellation via Cobounding field theory}

Above we gave a definition of path integral that turns a correspondence given by field
trajectories into a linear map between the spaces of quantum states of the incoming and
the outgoing prequantum field theories. There are two obstructions to this construction being
functorial
\begin{enumerate}
  \item For measures with nontrivial $\tau$ the incoming space of quantum states
  picks up a twist and hence will not match the outgoing space of quantum states
  for two consecutive correspondences.
  \item Even if the twist vanishes, then the choices of fundamental classes/measures
  might not be consistent in that doing the path integral first through one
  correspondence and then through the second is not equivalent to doing one single
  path integral through the composite correspondence.
\end{enumerate}
These obstructions to functoriality prevent the path integral from actually constituting
a quantum field theory. Such obstructions are known as quantum anomalies.
Lifting them is ``quantum anomaly cancellation''.

We close here by observing that the condition that the quantum anomalies in the path
integral cancel is precisely the condition that there is a cobounding field theory
in one dimension higher of which the given field theory is a boundary.

\medskip

\noindent Take the functor on $\{\mathrm{corner}\}\mathrm{Bord}_2$ to be given as follows:

\begin{itemize}
\item on objects: $X \mapsto \mathrm{Mod}(X) \in \mathrm{Mod}_2$

\item on morphisms $(\stackrel{f}{\longleftarrow}\stackrel{g}{\longrightarrow}) \mapsto g_! f^\ast$
(integral transform, as in Fourier-Mukai transf., Penrose transf., Harish-Chandra transf. etc.)

\item on 2-morphisms
 \begin{itemize}
   \item
     of the form
     $\xymatrix{
       \mathrm{Pic} \ar@{=}[rr]_<{\ }="t" && \mathrm{Pic}
       \\
       \mathbf{Fields}_{\mathrm{out}}
       \ar[u]|{\mathbf{L}_{\mathrm{out}}}
       &
       \mathbf{Fields}_{\mathrm{traj}}
       \ar[l]^{(-)|_{\mathrm{out}}}
       \ar[r]_{(-)|_{\mathrm{in}}}^>{\ }="s"
       &
       \mathbf{Fields}_{\mathrm{in}}
       \ar[u]|{\mathbf{L}_{\mathrm{in}}}
       \ar@{=>}|{\exp(\tfrac{i}{\hbar}S)} "s"; "t"
     }$
 \end{itemize}
 it assigns the induced
 $$
   \mathrm{out}^\ast\mathbf{L}_{\mathrm{in}}^\ast
   \longrightarrow
   \mathrm{in}^\ast\mathbf{L}_{\mathrm{out}}^\ast
   \;\;\;
   \mbox{in $\mathrm{Mod}(\mathbf{Fields}_{\mathrm{traj}})$}
 $$
 \item
 of the form
 $$
   \xymatrix{
       \mathbf{Fields}_{\mathrm{out}}
       \ar[d]
       &
       \mathbf{Fields}_{\mathrm{traj}}
       \ar[l]^{(-)|_{\mathrm{out}}}_>{\ }="t"
       \ar[r]_{(-)|_{\mathrm{in}}}
       &
       \mathbf{Fields}_{\mathrm{in}}
       \ar[d]
       \\
       \ast
       \ar@{=>}[rr]
       &&
       \ast
   }
 $$
 it assigns
 $$
   \xymatrix{
     \mathrm{Mod}(\mathbf{Fields}_{\mathrm{out}})
     \ar@{<-}[rr]^{  \mathrm{out}_! \mathrm{in}^\ast}_<{\ }="t"
     \ar[d]|{\underset{\mathbf{Fields}_{\mathrm{out}}}{\sum}}
     &&
     \mathbf{Mod}(\mathbf{Fields}_{\mathrm{in}})
     \ar[d]|{\underset{\mathbf{Fields}_{\mathrm{in}}}{\sum}}
     \\
     \mathrm{Mod}(\ast)
     \ar@{=}[rr]^>{\ }="s"
     &&
     \mathrm{Mod}(\ast)
     \ar@{=>}|{\underset{\mathbf{Fields}_{\mathrm{in}}}{\sum}\!\!\![\mathrm{in}]} "s"; "t"
   }
 $$
\end{itemize}

\noindent{\bf Proposition.} This assignment induces the above path integral. Therefore the
path integral extends to a functor and is hence anomaly free if it arises as the boundary
theory to the primary integral transform theory.

\newpage

\subsection{Summary (technical fact sheets)}
 \label{SummaryTechnicalDetails}

This document is an expanded set of notes for some talks I gave in February 2014.
What follows here are some notes handed out in a talk that I gave together
with Joost Nuiten in the String Geometry Network meeting at ESI in Vienna.
This talk discussed more of the technical details of how pull-push in generalized
cohomology is realized in the abstract theory and how it expresses cohomological quantization,
following \cite{Nuiten13}.

\medskip
\medskip

\medskip

\subsubsection{Translation between linear homotopy-type theory, generalized cohomology and quantization}

\hspace{-1.5cm}
\begin{tabular}{|c|c|c|}
  \hline
  {\bf linear homotopy-type theory}
  &
  {\bf twisted generalized cohomology}
  &
  {\bf quantum theory}
  \\
  \hline
  \hline
  linear homotopy-type & (module-)spectrum & state space
  \\
  \hline
  multiplicative conjunction & smash product of spectra & composite system
  \\
  \hline
  dependent linear type & module spectrum bundle &
  \\
  \hline
  Frobenius reciprocity & \begin{tabular}{l}six operation yoga \\in Wirthm{\"u}ller context\end{tabular} &
  linearity of integrals
  \\
  \hline
  dual type (linear negation) & Spanier-Whitehead duality &
  dual state space
  \\
  \hline
  invertible type & twist & \begin{tabular}{c}prequantum line bundle,\\ quantum anomaly \end{tabular}
  \\
  \hline
  dependent sum & generalized homology spectrum &
  \begin{tabular}{c} space of compactly supported \\ quantum states \\ ``bra''\end{tabular}
  \\
  \hline
  dual of dependent sum & generalized cohomology spectrum & \begin{tabular}{c} space of quantum states \\ ``ket''\end{tabular}
  \\
  \hline
  linear implication & bivariant cohomology & quantum operators
  \\
  \hline
  exponential modality & Goodwillie exponential & Fock space
  \\
  \hline
  \begin{tabular}{l}
  dependent sum \\over finite homotopy type \end{tabular} & Thom spectrum &
  \\
  \hline
  \begin{tabular}{l}
  dualizable dependent sum \\over finite homotopy type \end{tabular}
  &
  \begin{tabular}{l}
    Atiyah duality between
    \\
    Thom spectrum and
    \\ suspension spectrum
  \end{tabular}
  &
  \\
  \hline
  (twisted) self-dual type & Poincar{\'e} duality &  \begin{tabular}{c} inner product (Hilbert) space\end{tabular}
  \\
  \hline
  \begin{tabular}{l}
    dependent sum coinciding
    \\
    with dependent product
  \end{tabular}
  &
  ambidexterity, semiadditivity
  & \begin{tabular}{l}system of \\inner product state spaces \end{tabular}
  \\
  \hline
  \begin{tabular}{l}
    dependent sum coinciding
    \\
    with dependent product
    \\
    up to invertible type
  \end{tabular}
  &
  \begin{tabular}{l}
  Wirthm{\"uller} isomorphism
   \\
   (twisted ambidexterity)
  \end{tabular}
  &
  \begin{tabular}{l}anomalous system \\of inner product state spaces
  \end{tabular}
  \\
  \hline
  $(\sum_f \dashv f^\ast)$-counit & \begin{tabular}{l}pushforward \\ in generalized homology\end{tabular}
  &
  \\
  \hline
  \begin{tabular}{l}
    (twisted-)self-duality-induced
    \\
    dagger of this counit
  \end{tabular}
  &
  \begin{tabular}{c}
   (twisted-)Umkehr map,
   \\
   fiber integration
  \end{tabular}
  &
  \begin{tabular}{l}
  quantum superposition \\ and interference
  \end{tabular}
  \\
  \hline
  linear polynomial functor
  &
  primary integral transform
  &
  \begin{tabular}{l}
    propagator in cobounding
    \\
    $\mathrm{TQFT}_{d+1}$
  \end{tabular}
  \\
  \hline
  \begin{tabular}{l}
    correspondence
    \\
    with linear implication
  \end{tabular}
  &
  \begin{tabular}{c}
    motive
  \end{tabular}
  &
  \begin{tabular}{c}
    prequantized Lagrangian correspondence,
    \\
    action functional
  \end{tabular}
  \\
  \hline
  \begin{tabular}{l}
    composite of this linear implication
    \\
    with unit and daggered counit
  \end{tabular}
  &
  secondary integral transform
  &
  \begin{tabular}{c} cohomological  path integral, \\
    motivic transfer
  \end{tabular}
  \\
  \hline
  trace & Euler characteristic & partition function
  \\
  \hline
\end{tabular}

\newpage

\subsubsection{The quantization process.}

\noindent{\bf Prequantum $d+1$-dimensional field theory }  in $\mathrm{Corr}_2(\mathbf{H})$.

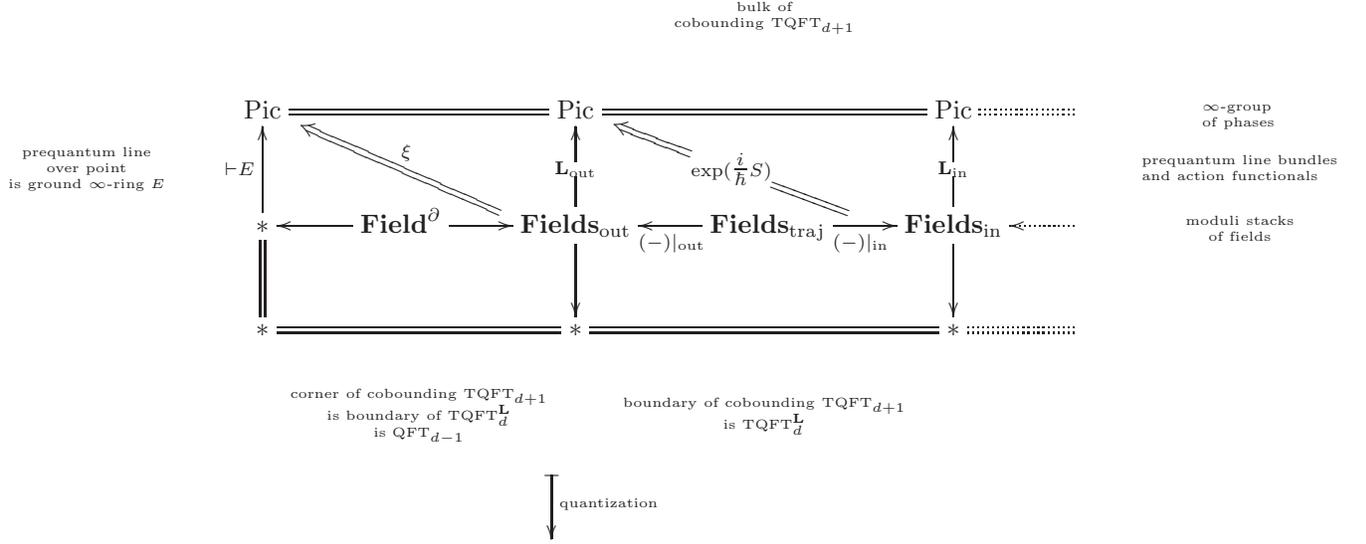
$$
  \xymatrix{
    &&
    && \ar@{}[rr]|{\tiny \begin{tabular}{c} bulk of \\ cobounding $\mathrm{TQFT}_{d+1}$ \end{tabular}}
    &&
    \\
    &&
    \mathrm{Pic} \ar@{=}[rr]_<{\ }="t1" &&
    \mathrm{Pic} \ar@{=}[rr]_<{\ }="t" && \mathrm{Pic} \ar@{::}[r]
    &&
    \mbox{\tiny \begin{tabular}{l} $\infty$-group \\ of phases \end{tabular}}
    \\
    \ar@{}[u]|{\mbox{
      \tiny
      \begin{tabular}{c}
        prequantum line
        \\
        over point
        \\
        is ground $\infty$-ring $E$
      \end{tabular}
    }
    }
    &
    &
    \ast \ar[u]^{\vdash E} \ar@{=}[d]
    & \mathbf{Field}^\partial
    \ar[l] \ar[r]^>{\ }="s1" & \mathbf{Fields}_{\mathrm{out}}
    \ar[u]|{\mathbf{L}_{\mathrm{out}}}
     \ar[d]
    &
    \mathbf{Fields}_{\mathrm{traj}}
    \ar[l]^{(-)|_{\mathrm{out}}}
    \ar[r]_{(-)|_{\mathrm{in}}}^{\ }="s"
    &
    \mathbf{Fields}_{\mathrm{in}}
    \ar[u]|{\mathbf{L}_{\mathrm{in}}}
    \ar[d]
    \ar@{<..}[r]
    &&
    \mbox{
      \tiny
      \begin{tabular}{c}
        moduli stacks
        \\
        of fields
      \end{tabular}
    }
    \ar@{}[u]|{
          \mbox{ \tiny
     \begin{tabular}{l}
       prequantum line bundles
       \\
       and action functionals
     \end{tabular}
    }
    }
    \\
    &&
    \ast \ar@{=}[rr] && \ast \ar@{=}[rr] && \ast \ar@{::}[r] &
    \\
    && \ar@{}[rr]|{\mbox{\tiny \begin{tabular}{c} corner of cobounding $\mathrm{TQFT}_{d+1}$ \\ is boundary of $\mathrm{TQFT}^{\mathbf{L}}_{d}$  \\ is $\mathrm{QFT}_{d-1}$ \end{tabular}}}
    && \ar@{}[rr]|{\mbox{\tiny \begin{tabular}{c} boundary of cobounding $\mathrm{TQFT}_{d+1}$ \\ is $\mathrm{TQFT}^{\mathbf{L}}_{d}$  \end{tabular}}}
    &&
    \ar@{=>}|{\exp(\tfrac{i}{\hbar}S)} "s"; "t"
    \ar@{=>}_\xi "s1"; "t1"
  }
$$

\begin{center}
  $\xymatrix{
    \ar@{|->}[d]^{\mbox{\tiny quantization}}
    \\
    &
  }$
\end{center}

\noindent {\bf Quantum $d+1$-dimensional field theory} in $\mathrm{Mod}_2$.

$$
  
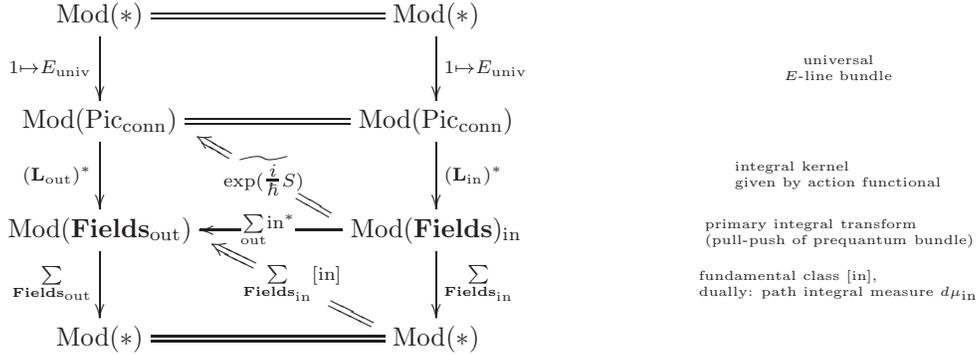
\xymatrix{
    \mathrm{Mod}(\ast)
    \ar@{=}[rr]
    \ar[d]_{1 \mapsto E_{\mathrm{univ}}}
    &&
    \mathrm{Mod}(\ast)
    \ar[d]^{1 \mapsto E_{\mathrm{univ}}}
    &&
    \ar@{}[d]|{\mbox{\tiny \begin{tabular}{c} universal \\ $E$-line bundle\end{tabular}}}
    \\
    \mathrm{Mod}(\mathrm{Pic}_{\mathrm{conn}})
    \ar@{=}[rr]_<{\ }="t"
    \ar[d]_{(\mathbf{L}_{\mathrm{out}})^\ast}
    &&
    \mathrm{Mod}(\mathrm{Pic}_{\mathrm{conn}})
    \ar[d]^{(\mathbf{L}_{\mathrm{in}})^\ast}
    &&
    \\
    \mathrm{Mod}(\mathbf{Fields}_{\mathrm{out}})
    \ar[d]_{\underset{\mathbf{Fields}_{\mathrm{out}}}{\sum}}
    &&
    \mathrm{Mod}(\mathbf{Fields})_{\mathrm{in}}
    \ar[ll]|{\underset{\mathrm{out}}{\sum} \mathrm{in}^\ast}_<{\ }="s"^>{\ }="t1"
    \ar[d]^{\underset{\mathbf{Fields}_{\mathrm{in}}}{\sum}}
    &&
    \mbox{ \tiny
     \begin{tabular}{l}
       primary integral transform \\ (pull-push of prequantum bundle)
     \end{tabular}
    }
    \ar@{}[d]|{\mbox{\tiny  \begin{tabular}{l} fundamental class $[\mathrm{in}]$, \\
     dually: path integral measure $d\mu_{\mathrm{in}}$  \end{tabular}}}
    \ar@{}[u]|{\mbox{\tiny \begin{tabular}{l} integral kernel \\ given by action functional \end{tabular}}}
    \\
    \mathrm{Mod}(\ast)
    \ar@{=}[rr]^>{\ }="s1"
    &&
    \mathrm{Mod}(\ast)
    &&
    \ar@{=>}|{\widetilde{\exp(\tfrac{i}{\hbar}S)}} "s"; "t"
    \ar@{=>}|{\underset{\mathbf{Fields}_{\mathrm{in}}}{\sum} [\mathrm{in}]} "s1"; "t1"
  }
$$

\noindent encodes  \\
\noindent {\bf Quantum $d$-dimensional field theory} \\
as unit-component in $\mathrm{Mod}(\ast)$ of the above transformation: $ \mathbb{D} \!\! \underset{{\mathbf{Fields}_{\mathrm{traj}}}}{\int} \!\!\!\!\exp(\tfrac{i}{\hbar}S) d\mu :=$
$$
  \hspace{-2.4cm}
  \xymatrix{
    \underset{\mathbf{Fields}_{\mathrm{out}}}{\sum} \mathbf{L}_{\mathrm{out}}
    \ar@{<-}[r]^-{\underset{\mathbf{Fields}_{\mathrm{out}}}{\sum} \!\!\!\epsilon_{\mathbf{L}_{\mathrm{out}}}}
    &
    \underset{\mathbf{Fields}_{\mathrm{out}}}{\sum} \mathrm{out}_! \mathrm{out}^\ast
      \mathbf{L}_{\mathrm{out}}
    \ar@{<-}[r]^-{\simeq}
    &
    \underset{\mathbf{Fields}_{\mathrm{traj}}}{\sum} \mathrm{out}^\ast \mathbf{L}_{\mathrm{out}}
    \ar@{<-}[rr]^-{\underset{\mathbf{Fields}_{\mathrm{traj}}}{\sum} \!\!\!\!\!\exp(\tfrac{i}{\hbar}S)}
    &&
    \underset{\mathbf{Fields}_{\mathrm{traj}}}{\sum} \mathrm{in}^\ast \mathbf{L}_{\mathrm{in}}
    \ar@{<-}[r]^-{\simeq}
    &
    \underset{\mathbf{Fields}_{\mathrm{in}}}{\sum} \mathrm{in}_! \mathrm{in}^\ast \mathbf{L}_{\mathrm{in}}
    \ar@{<-}[r]^-{\underset{\mathbf{Fields}_{\mathrm{in}}}{\sum} \!\!\![\mathrm{in}]}
    &
    \underset{\mathbf{Fields}_{\mathrm{in}}}{\sum} \mathbf{L}_{\mathrm{in}} \otimes \tau
  }
$$

\noindent (secondary integral transform: pull-push of states).

\begin{center}
\begin{tabular}{|l|c|c|}
  \hline
  {\bf field theory} & {\bf spaces of states} & {\bf propagator}
  \\
  \hline
  $\mathrm{TQFT}_{d+1}$ & $\mathrm{Mod}_2 \in \mathrm{Cat}_2$ & integral transform
  \\
  \hline
  $\mathrm{TQFT}_{d}^\tau$ & $\mathrm{Mod}(\ast) \in \mathrm{Mod}_2$ &
   \begin{tabular}{l} secondary integral transform, \\ path integral \end{tabular}
  \\
  \hline
  $\mathrm{QFT}_{d-1}$ & $\underset{X}{\sum} \mathbf{L}_X \in \mathrm{Mod}(\ast)$ &
  \begin{tabular}{l}
  equivariance under
  \\
  Hamiltonian group action
  \end{tabular}
  \\
  \hline
\end{tabular}
\end{center}

\newpage

\subsubsection{Translation between linear homotopy-type theory in $E \mathrm{Mod}$ and twisted $E$-cohomology.}

\hspace{-2cm}
\begin{tabular}{|l|l|l|}
\hline
{ special case} & {\bf linear homotopy-type theory} &
\begin{tabular}{l}
  {\bf higher linear algebra}
  \\
  viz.
  \\
  {\bf generalized cohomology theory}
\end{tabular}
\\
\hline
\hline
 & $E \in \mathrm{CRing}_\infty$ & ground ring
 \\
 \hline
 & $X \in \infty \mathrm{Grpd}$ & base homotopy type (base space)
 \\
 \hline
 \hline
 & $\tau : X \stackrel{}{\longrightarrow} \mathrm{Pic}(E)$ & twist
\\
\hline
& $\widehat{\tau} := \tau^\ast \widehat{\mathrm{Pic}}(E) \in \mathrm{Mod}(X)$ & $E$-line bundle
\\
\hline
\begin{tabular}{l}
 canonical twist on moduli
 \\
 for stable vector bundles
\end{tabular}
&
$
  J^E
    :
  \xymatrix@C=11pt{
  \mathbb{Z}\times B O
    \ar[r]^-{J}
    &
  \mathrm{Pic}(\mathbb{S})
    \ar[rr]^-{\mathrm{Pic}(\mathbb{S} \to E)}
    &&
  \mathrm{Pic}(E)
  }
$
&
J-homomorphism
\\
\hline
&
$\underset{X}{\sum} \widehat{\tau} \simeq E_{\bullet+ \tau}(X)$
&
\begin{tabular}{l}
spectrum of \\ $\tau$-twisted $E$-homology cycles
\end{tabular}
\\
\hline
trivial twist
&
$\underset{X}{\sum} 1_X \simeq E_{\bullet}(X) = E\wedge \Sigma_+^\infty X$
&
suspension spectrum
\\
\hline
\begin{tabular}{l}
$X \stackrel{\xi}{\longrightarrow} \mathbb{Z}\times B O$ modulating
\\
 stable vector bundle
\end{tabular}
& $\underset{X}{\sum} \widehat{J^E \circ \xi}  = E \wedge X^\xi $
&
Thom spectrum
\\
\hline
\begin{tabular}{l}
canonical twist on $X := B O\langle n\rangle$
\\
$
 J^E_{B O\langle n\rangle}
   :
 B O \langle n\rangle \to B O \stackrel{J^E}{\longrightarrow} \mathrm{Pic}(E)
$
\\
\end{tabular}
&
$\underset{B O\langle n\rangle}{\sum} \widehat{J^E_{B O\langle n\rangle}}
  \simeq M O\langle n\rangle$
  &
  \begin{tabular}{l}
    universal
    \\
    Thom spectrum
  \end{tabular}
 \\
 \hline
 \\
 low $n$
 &
 \begin{tabular}{ll}
   $n = 0$:
   & $M O$
   \\
     $n = 1$:
  & $M \mathrm{SO}$
  \\
 $n= 2$: &
 $M \mathrm{Spin}$
 \\
 $n = 4$:
 &
 $M \mathrm{String}$
 \end{tabular}
 &
 $
 \left.
 \mbox{
 \begin{tabular}{l}
   Riemannian-
   \\
   oriented-
   \\
   spin-
   \\
   string-
 \end{tabular}
 }
 \right\}
 $
 cobordism spectrum
 \\
 \hline
 \hline
 & $\mathbb{D}$ & Spanier-Whitehead duality
 \\
 \hline
 &
 $\mathbb{D}\underset{X}{\sum} \widehat \tau = E^{\bullet + \tau}(X)$
 &
 \begin{tabular}{l}
 spectrum of
 \\
  $\tau$-twisted $E$-cohomology cocycles
  \end{tabular}
 \\
 \hline
 \begin{tabular}{l}
  $X$ compact smooth manifold
  \\
  with tangent bundle $TX$
  \\
  and stable normal bundle $NX = -  TX$
 \end{tabular}
 &
 $\mathbb{D}(E \wedge \Sigma_+^\infty X) \simeq E_{\bullet +  NX}(X)$
 &
 Atiyah-Whitehead duality
 \\
 \hline
 \hline
 &
 $
   \raisebox{20pt}{
   \xymatrix{
     Z \ar[dr]_{\tau_Z}^{\ }="t" \ar[rr]^-f_{\ }="s" && X \ar[dl]^{\tau_X}
     \\
     & \mathrm{Pic}(E)
     \ar@{=>}^o "s"; "t"
   }
   }
 $
 &
 \begin{tabular}{l}
   fiberwise $E$-orientation
   \\
   of $\tau_Z$ relative to $\tau_X$
 \end{tabular}
 \\
 \hline
 to the point
 &
 $
   \raisebox{20pt}{
   \xymatrix{
     Z \ar[dr]_{\tau_Z}^{\ }="t" \ar[rr]_{\ }="s" && \ast \ar[dl]^0
     \\
     & \mathrm{Pic}(E)
     \ar@{=>}^o "s"; "t"
   }
   }
 $
 &
 \begin{tabular}{l}
   $\tau_Z$-twisted $E$-orientation of $Z$
 \end{tabular}
 \\
 \hline
 vanishing twist on domain
 &
 $
   \raisebox{20pt}{
   \xymatrix{
     Z \ar[dr]_{0}^{\ }="t" \ar[rr]^-f_{\ }="s" && X \ar[dl]^{\tau_X}
     \\
     & \mathrm{Pic}(E)
     \ar@{=>}^o "s"; "t"
   }
   }
 $
 &
 \begin{tabular}{l}
   $E$-orientation of $f$
 \end{tabular}
 \\
 \hline
 \hline
 fiberwise fundamental class with twist $\tau$ &
 $f_! f^\ast \widehat{\tau_X} \simeq \mathbb{D} f_! f^\ast \mathbb{D}(\widehat{\tau_X} \otimes \widehat{\tau})$
 &
 fiberwise twisted Poincar{\'e} duality
 \\
 \hline
\end{tabular}

\newpage

\subsubsection{Examples}

\noindent {\bf a) Particle at the boundary of 2d Poisson-Chern-Simons TQFT.}

$$
  
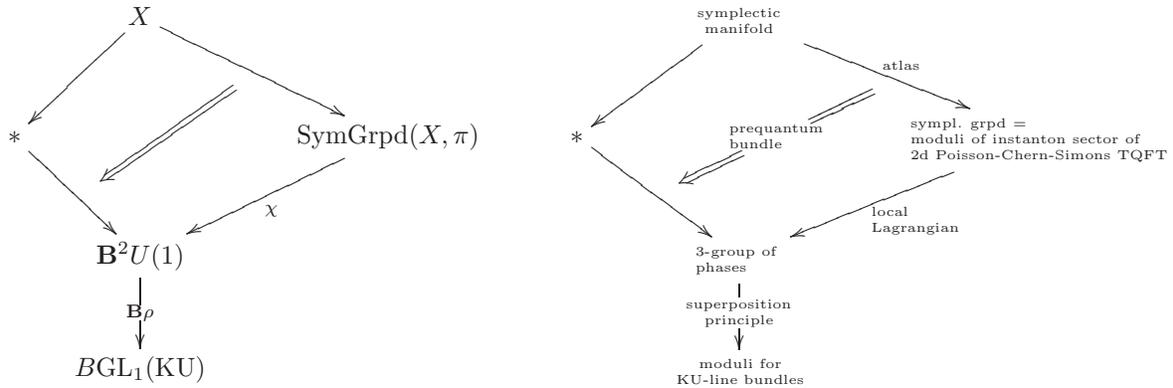
\xymatrix@C=9pt{
    & X
    \ar[dr]_-{\ }="s"
    \ar[dl]
    &
    &&
    \mbox{\tiny \begin{tabular}{c} symplectic \\ manifold\end{tabular}}
    \ar[dr]^{\mbox{\tiny atlas}}_{\ }="s1"
    \ar[dl]
    \\
    \ast \ar[dr]^{\ }="t" &&
    \mathrm{SymGrpd}(X,\pi)\ar[dl]^{\chi}
    &
    \ast
    \ar[dr]^{\ }="t1"
    &&
    \mbox{\tiny
      \begin{tabular}{l}
        sympl. grpd =
        \\
        moduli of instanton sector of
        \\
        2d Poisson-Chern-Simons TQFT
      \end{tabular}
    }
    \ar[dl]^{\mbox{\tiny \begin{tabular}{l}local \\ Lagrangian\end{tabular} }}
    \\
    & \mathbf{B}^2 U(1)
    \ar[d]|{\mathbf{B} \rho}
    &&
    &
    \mbox{\tiny
      \begin{tabular}{l}
        3-group of
        \\
        phases
      \end{tabular}
    }
    \ar[d]|{\mbox{\tiny \begin{tabular}{c} superposition \\ principle\end{tabular}}}
    \\
    & B \mathrm{GL}_1(\mathrm{KU})
    &&
    &
    \mbox{
      \tiny
      \begin{tabular}{c}
        moduli for
        \\
        $\mathrm{KU}$-line bundles
      \end{tabular}
    }
    \ar@{=>} "s"; "t"
    \\
    \mathrm{KU}
    &&
    \mathrm{KU}_{\bullet + \chi}(\mathrm{SymGrpd}(X,\pi))
    \ar[ll]
    &
    &
    \mbox{\tiny \begin{tabular}{c} bundle of Hilbert spaces\\ of quantum states
    \\
    from symplectic leaf-wise
    \\
    geometric quantization
    \end{tabular}}
    \ar@{=>}|{\mbox{\tiny  \begin{tabular}{l}prequantum \\ bundle \end{tabular}}} "s1"; "t1"
  }
$$

\noindent {\bf b) Superstring at boundary of 3d $\mathrm{Spin}$-Chern-Simons TQFT.}
  $$
    \raisebox{20pt}
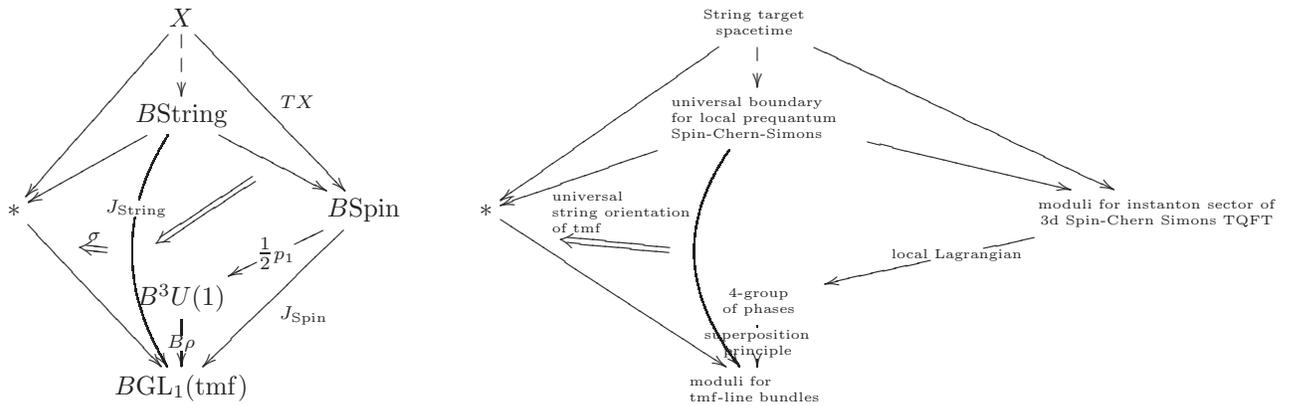
{
    \xymatrix@R=15pt{
      & X
      \ar[ddl]
      \ar[ddr]^{T X}
      \ar@{-->}[d]
      &&
      &\mbox{\tiny
        \begin{tabular}{c}
          String target
          \\
          spacetime
        \end{tabular}
      }
      \ar[ddr]
      \ar[ddl]
      \ar@{-->}[d]
      \\
      & B \mathrm{String}
      \ar[dl]
      \ar[dr]_{\ }="s1"
      \ar@/_1.56pc/[ddd]|<<<<<<<<<<{J_{\mathrm{String}}}_-{\ }="s"^{\ }="t1"
      &&
      &\mbox{\tiny
        \begin{tabular}{l}
          universal boundary
          \\
          for local prequantum
          \\
          $\mathrm{Spin}$-Chern-Simons
        \end{tabular}
      }
      \ar[dl]
      \ar[dr]
      \ar@/_2pc/[ddd]_{\ }="s2"
      \\
      \ast \ar[ddr]^<<<<<<{\ }="t"
      && B \mathrm{Spin} \ar[dl]|-{\tfrac{1}{2}p_1}
      \ar[ddl]^{J_{\mathrm{Spin}}}
      &
      \ast
      \ar[ddr]^<<<<<<<{\ }="t2"
      &
      &
      \mbox{\tiny
      \begin{tabular}{c}
        moduli for instanton sector of
        \\
        3d $\mathrm{Spin}$-Chern Simons TQFT
      \end{tabular}}
      \ar[dl]|{\mbox{\tiny local Lagrangian}}
      \\
      & B^3 U(1)
      \ar[d]|{B\rho}
      &&
      & \mbox{ \tiny
        \begin{tabular}{c}
          4-group
          \\
          of phases
        \end{tabular}
      }
      \ar[d]|{\mbox{\tiny \begin{tabular}{c} superposition \\ principle\end{tabular}}}
      \\
      & B \mathrm{GL}_1(\mathrm{tmf})
      &&
      & \mbox{\tiny
        \begin{tabular}{l}
          moduli for
          \\
          $\mathrm{tmf}$-line bundles
        \end{tabular}
        }
      &&&
      \ar@{=>}_\sigma "s"; "t"
      \ar@{=>} "s1"; "t1"
      \ar@{=>}_{\mbox{\tiny
          \begin{tabular}{l}
            universal
            \\string orientation
            \\
            of $\mathrm{tmf}$
          \end{tabular}
        }} "s2"; "t2"
      \\
      \mathrm{tmf}
      &
      X^{T X}
      \ar[l]
      &
      &&
      \mbox{\tiny
       \begin{tabular}{l}
         integral
         Witten genus =
         \\
         non-perturbative
         string partition function
       \end{tabular}
      }
    }
    }
  $$

\noindent {\bf c) D-Brane Charge and T-Duality.}
$$
  
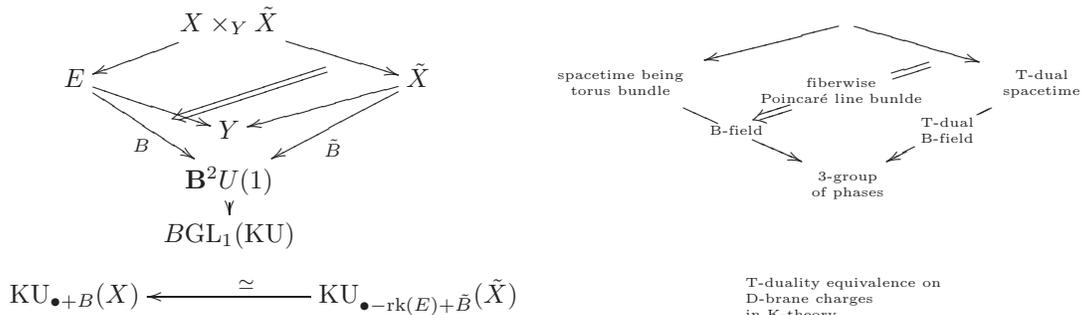
\xymatrix@C=3pt@R=3pt{
     & X \times_Y \tilde X
     \ar[dl]
     \ar[dr]_-{\ }="s"
     &&&
     \ar[dl]
     \ar[dr]_{\ }="s2"
     \\
     E
     \ar[dr]
     \ar[ddr]_{B}^{\ }="t"
     &&
     \tilde X
     \ar[dl]
     \ar[ddl]^{\tilde B}
     &
     \mbox{\tiny
      \begin{tabular}{c}
        spacetime being
        \\ torus bundle
      \end{tabular}
     }
     \ar[ddr]|{     \mbox{
       \tiny
       \begin{tabular}{c}
         B-field
       \end{tabular}
     }}^{\ }="t2"
     &&
     \mbox{\tiny
      \begin{tabular}{c}
        T-dual \\
        spacetime
      \end{tabular}
     }
     \ar[ddl]|{     \mbox{
       \tiny
       \begin{tabular}{c}
         T-dual
         \\
         B-field
       \end{tabular}
     }}
     \\
     & Y
     \\
     & \mathbf{B}^2 U(1)
     \ar[d]
     &&&
     \mbox{
       \tiny
       \begin{tabular}{c}
         3-group
         \\ of phases
       \end{tabular}
     }
     \\
     & B \mathrm{GL}_1(\mathrm{KU})
     \\
     \mathrm{KU}_{\bullet+B}(X)
     &&
     \mathrm{KU}_{\bullet - \mathrm{rk}(E) +  \tilde B}(\tilde X)
     \ar[ll]_\simeq
     &&
     \mbox{\tiny
       \begin{tabular}{l}
         T-duality equivalence on
         \\ D-brane charges
         \\ in K-theory
       \end{tabular}
     }
     \ar@{=>} "s"; "t"
     \ar@{=>}|{\mbox{\tiny \begin{tabular}{c} fiberwise \\ Poincar{\'e} line bunlde \end{tabular}}} "s2"; "t2"
  }
$$

\newpage

\section{Modal homotopy-type theory}

We now discuss the theory in more detail.
Here we begin by reviewing the ambient logical framework. We start in \ref{NotionsJudgementDeduction}
with a lightning review of basics of {(homotopy-)}type theory. Then in \ref{ModalitiesMomentsOpposites}
we state the simple but profound notion of modalities and opposite moments in type theory, following
\cite{Lawvere91, Lawvere94}.
Finally in \ref{AxiomaticMetaphysics} we consider the pattern of such opposite moments
which is called ``cohesion'' in \cite{Lawvere07} and the larger such pattern called
``differential cohesion'' in \cite{dcct}. Then we review from \cite{dcct} the structures induced by these axioms
which we will show to serve as a formal foundation for quantization theory.

Of special interest for quantization is the linear structure that arises,
in \ref{InfinitesimalsAndModules}, from the opposite moments
denoted $(\Re \dashv \oint)$. This we turn to in \ref{LinearTypeTheory} below.

\medskip

We need to assume that the reader is familiar with basics of category theory, topos theory and preferably
with basics of categorical logic, all explained for instance in \cite{LaneMoerdijk}.
Our developments below take place in homotopy topos theory \cite{Rezk} also known
as higher topos theory \cite{HTT} and are naturally expressed
in the internal language of higher toposes ($\infty$-toposes), which is essentially what is being called
``homotopy type theory'' \cite{HoTT}. We will write ``homotopy-type theory'' here to indicate
that we do not work with the formal syntax of \cite{HoTT} but rather speak in terms of its
semantics of geometric homotopy types ($\infty$-stacks) in the homotopy theory of $\infty$-toposes. The type-theoretic perspective for us serves to
make transparent the foundational and essentially formal nature of all assumptions and constructions
such as to lend itself naturally to a fully formal syntactical formalization.

\subsection{Types, Judgements and Deduction}
 \label{NotionsJudgementDeduction}

While there is no room here to give a detailed introduction to homotopy toposes
and homotopy type theory as their internal language (but see \cite{dcct} for review of
all the material that we need), the striking claim of \cite{HoTT}
is that at the bottom of it, homotopy type theory is much more elementary than it might seem,
in fact that it is in a sense more elementary even than traditional elementary formal logic
(since the homotopy-theoretic aspect is obtained not by adding axioms, but by
omitting a traditional axiom, that of uniqueness of identity proofs). Therefore we
can offer here a quick informal introduction from which the inclined reader might
be able to get a working feeling for what is going on.

\medskip

Imagine a computer terminal
of the old sober type: a pitch black screen, empty except for a green prompt
in the top left, expecting your input, like this:

\medskip

\noindent $
  \vdash
$

\medskip

The command line syntax is most elementary: you can enter expressions of the form

\medskip

\noindent$
  \vdash x \colon X
$

\medskip

to be thought of as the judgement that a term $x$ is of type $X$, a representative of some notion.
For instance
entering

\medskip

\noindent$
  \vdash \; \frac{e^2}{4 \pi \epsilon_0 G m_p m_e} \colon \mathrm{RealNumber}
  \,.
$

\medskip

asks the system to judge that Dirac's large number is a representative of the type of real numbers.
The system will accept the input if the judgement is valid, and reject it otherwise.

First of all, for the above example to validate it must at least be true that ``$\mathrm{RealNumber}$''
itself is a type known to the system. So there is a hierarchy of types $\mathrm{Type}$ of all types
pre-installed and the system knows how to validate input of the form

\medskip

\noindent $
  \vdash \; \mathrm{RealNumber} \colon \mathrm{Type}
$

\medskip

Once a judgement $X \colon \mathrm{Type}$ is validated then the system may store that type. To
indicate this it changes the appearance of the prompt to

\medskip

\noindent $
  x \colon X \; \vdash
$

\medskip

Judgements entered after such a prompt are interpreted in this new context, hence may depend on the given type
on the left.
For instance after the system acknowledges the natural numbers, we can consider entering the judgement

\medskip

\noindent $
  n \colon \mathrm{NaturalNumber} \;\vdash \; \mathrm{SU}(n) \colon \mathrm{Group}
$

\medskip

to express that for each natural number $n$, there is a group called $\mathrm{SU}(n)$.

Every judgement in some context may be extended trivially to any other context and we assume that the
system handles this \emph{context extension} implicitly.

One dependent type pre-installed to the system is the collection of (gauge) equivalences
between any two terms of the same type, which might is written

\medskip

\noindent$
  X \colon \mathrm{Type};\; x_1,x_2 \colon X \;\vdash\; (x_1 = x_2) \colon \mathrm{Type}
$

\medskip

expressing that for each type $X$ and any two terms $x_1,x_2$ of that type, there is a type
$(x_1 = x_2)$ of (gauge) equivalence between these.

\medskip

This is to be such that a term of this type

\medskip

\noindent $
  \vdash \; p \colon (x_1 = x_2)
$

\medskip

is a  (gauge) equivalence between $x_1$ and $x_2$.

For example if an $X$-dependent type $x \colon X \vdash P(x) \colon \mathrm{Type}$ is such that
any two terms of it are either distinct or else equivalent in an essentially unique
way, then it may be thought of as a \emph{proposition} about the terms $x$ of type $X$, where a term
$p(x) \colon P(x)$ is a proof of the proposition $P$ about $x$. In this fashion the type theory here
subsumes traditional formal logic; a simple but deep insight known variously as
the \emph{Brouwer-Heyting-Kolmogorov interpretation} or the \emph{Howard isomorphism}
or as the \emph{propositions-as-types paradigm}, see section 1.11 of \cite{HoTT}.

However, crucial for the axiomatization of fundamental
physics is that not all types are propositions in this sense, that type theory is genuinely
more general that predicative logic, in that generally two terms of
a type may be gauge equivalent in more than one essentially distinct ways.

Generally, given a dependent type

\medskip

\noindent$
  x \colon X \;\vdash\; Y(x) \colon \mathrm{Type}
$

\medskip

such as

\medskip

\noindent$
  n \colon \mathrm{NaturalNumber} \;\vdash\; \mathbb{R}^n \colon \mathrm{Type}
$

\medskip

then the system automatically deduces the existence of two further types, the sum (union)
of all the types as $x$ varies in $X$, called the \emph{dependent sum type} and denoted

\medskip

\noindent$
  \vdash \; \underset{{x \colon X}}{\sum} Y(x) \;\colon \mathrm{Type}
$

\medskip

which is the type of terms that are in either of the $Y(x)$ as $x$ varies in $X$, and the
product of all these types, called the \emph{dependent product} and denoted

\medskip

\noindent$
  \vdash \;\colon\; \underset{{x \colon X}}{\prod} Y(x) \;\colon \mathrm{Type}
$

\medskip

which is the type of collections of terms in each $Y(x)$, as $x$ varies in $X$.

For instance if for a given type $X$ the system validates the existence of a term of the form

\medskip

\noindent$
  \vdash \; p \colon \underset{{x_1 \colon X}}{\sum} \; \underset{{x_2 \colon X}}{\prod} (x_1 = x_2)
$

\medskip

then this means a choice of basepoint $x_1 \colon X$ and a choice of $X$-parameterized equivalence
of every other term $x_2$ of type $X$ with $x_1$.

Applied to an $X$-dependent type $P(x)$ which is a proposition in the above sense of the BHK interpretation,
then the dependent sum reduces to the
existential quantifier $\exists_{x \in X}$ in that a term of type $\underset{x \colon X}{\sum} P(x)$
is a proof that there exists at least one $x$ of type $X$ such that $P(x)$ is true; and
the dependent product $\prod_X$ reduces to the universal quantifier $\forall_{x \in X}$ in that
a term of type $\underset{x \colon X}{\prod} P(x)$ is a proof that $P(x)$ holds for all $x$ parameterized
over $X$.

On the other hand, applied to an $X$-dependent type $Y$ whose dependence is trivial, then the dependent sum
reproduces the categorical Cartesian product $X \times Y$ and the dependent product the function space
$X \to Y$.

Imagine we could inspect the computer system's registers and memory. From this we deduce
a category $\mathbf{H}$, called the \emph{type system} or the \emph{category of contexts}, whose objects
are the given types in the plain context, and which for each $X$-dependent type $Y$
has one morphism of the form
$\colon \left(\underset{{x \colon X}}{\sum} Y\left(x\right)\right) \stackrel{p_Y}{\longrightarrow} X$ and which for each $X$-dependent term
$
  x \colon X \;\vdash\; y(x) \colon Y(x)
$
has a commuting diagram of the form
$$
  \raisebox{20pt}{
  \xymatrix{
    & \underset{{x \colon X}}{\sum} Y(x)
    \ar[d]^{p_Y}
    \\
    X
    \ar[ur]^-{y}
    \ar@{=}[r]
    &
    X
  }
  }\,.
$$
For every morphism $f \colon X \longrightarrow Y$ in $\mathbf{H}$ there is the
pullback functor $f^\ast$ between the slice categories and its left and right adjoints,
respectively, are the above operations of dependent sum and dependent product
$$
  (\sum_f \dashv f^\ast \dashv \prod_f)
  \;\colon\;
  \xymatrix{
    \mathbf{H}_{/X}
    \ar@<+11pt>@{->}[rr]|{\sum_f}
    \ar@{<-}[rr]|{f^\ast}
    \ar@<-11pt>@{->}[rr]|{\prod_f}
    &&
    \mathbf{H}_{/Y}
  }
  \,.
$$
This key insight, which is the very basis of ``categorical logic''
and central to our development in
\ref{ComputationalHomotopyTypeTheory} below, goes back to \cite{Lawvere69, Lawvere70}.

One finds that the categories $\mathbf{H}$ appearing this way are equivalently
locally cartesian closed categories
and one says that $\mathbf{H}$ is the \emph{categorical semantics} of the above formal syntax.
More precisely, the above is the sketch of the construction of a 2-functorial equivalence
$$
  \xymatrix@R=2pt{
  \mathrm{syntax}
  &&
  \mathrm{semantics}
  \\
    \mathrm{MLTypeTheories}
      \ar@{<->}[rr]^-\simeq
      &&
    \mathrm{LccCategories}
   }
$$
between a natural 2-category of (Martin-L{\"o}f intuitionistic) type theories and that of locally cartesian closed categories
\cite{Seely84, CD}. In view of this equivalence we here speak about ``type-semantics''.

Finally the homotopy theoretic aspect is reflected by interpreting the $(X \times X)$-dependent type $(x_1 = x_2)$ of gauge equivalences between terms of type $X$ as the diagonal morphism
$$
  \xymatrix{
    X
    \ar[d]^{\Delta_X}
    \\
    X \times X
  }
$$
or rather as its fibrant resolution with respect to a homotopical fibration structure
(essentially as in \cite{Brown}, one of the earliest accounts of what today is called $\infty$-topos theory), hence as a
path space object $X^{I}$ equipped with the two endpoint evaluation maps
$$
  \raisebox{20pt}{
  \xymatrix{
    X^I
    \ar[d]|{(\mathrm{ev}_0, \mathrm{ev}_1)}
    \\
    X \times X
  }
  }\,.
$$
Accordingly, homotopy type theory has semantics in locally Cartesian closed categories that are
in addition equipped with a suitably compatibly homotopical fibration structure. These are such that they
present locally cartesian closed $\infty$-categories such as $\infty$-toposes.
For this see section 2 of \cite{ShulmanInverse}.

\subsection{Modalities, Moments and Opposites}
\label{ModalitiesMomentsOpposites}

A type system $\mathbf{H}$ as specified in \ref{NotionsJudgementDeduction} provides a setting for types to be,
but lacks as yet any determination of qualities these types may have, hence of
modes of being. A central insight
of traditional formal logic is, when generalized from propositions to types,
that such \emph{modalities} are formalized by monads
on the type system \cite{Lawvere70, Goldblatt, Moggi, Kobayashi, Shulman12a},
traditionally called \emph{modalities} or \emph{modal operators}:
\begin{definition}
  A \emph{modality} $\bigcirc$ on a type system $\mathbf{H}$ is a monad (an $\infty$-monad)
  $\bigcirc \;\colon\; \mathbf{H} \longrightarrow \mathbf{H}$. A \emph{co-modality}
  $\Box$ is a co-monad ($\infty$-comonad) $\Box \;\colon\; \mathbf{H} \longrightarrow \mathbf{H}$.
  We say a \emph{$\bigcirc$-modal type} (or $\Box$-\emph{co-modal type}) is a type
  equipped with the structure of a (co-)algebra over this monad.
  \label{modality}
\end{definition}
(In practice we often suppress the ``co-'' terminologically, as it is
determined by the context.)
\begin{remark}
The general theory of $\infty$-monads
on $\infty$-categories is discussed in section 6.2 of \cite{LurieAlg} and in \cite{RiehlVerity}.
By the homotopy monadicity theorem (theorem 6.2.2.5 of \cite{LurieAlg} and def. 6.1.15 with section 7 of
\cite{RiehlVerity}) every $\infty$-monad
$\bigcirc : \mathbf{H} \to \mathbf{H}$ arises as the endomorphism monad
$\bigcirc \simeq R \circ L$ of some $\infty$-adjunction $(L \dashv R) : \mathbf{H} \leftrightarrow \mathcal{D}$
for some $\infty$-category $\mathcal{D}$.
By theorem 5.4.14 in \cite{RiehlVerity} $\infty$-adjunctions have the higher coherence data
of their unit (and counit) uniquely (up to a contractible homotopy type of choices)
induced from the underlying adjunction in the homotopy 2-categories. Therefore a choice of
$\infty$-adjunction $(L \dashv R)$ for $\bigcirc$ re-encodes the coherence data of $\bigcirc$
as a homotopy coherent monoid in the monoidal $\infty$-category $\mathrm{End}(\mathbf{H})$ equivalently as
the choice of $\infty$-category $\mathcal{D}$ and the single datum of an $\infty$-adjunction unit,
see also remark 6.2.0.7 in \cite{LurieAlg}. This allows to present $\infty$-monads
as ordinary monads on the homotopical fibration category underlying $\mathbf{H}$ (as in \ref{NotionsJudgementDeduction}
above), see \cite{Hess} for homotopy monadicity discussed in homotopical (model) categories this way.
All $\infty$-monads that we consider below
arise as endomorphism monads of a given $\infty$-adjunction.
\end{remark}
If such a (co-)monad is \emph{idempotent} in that applying it twice is equivalent to applying
it just once, hence if it behaves as a projection, then it may be thought of as projecting out
from any type one aspect or \emph{moment} that it has. Traditionally this is called
a \emph{closure operator}:
\begin{definition}
  A \emph{moment} $\bigcirc$ is an idempotent modality on $\mathbf{H}$, def. \ref{modality},
  a \emph{co-moment} $\Box$ is an idempotent co-modality.
  Given a moment $\bigcirc$ or co-moment $\Box$ write
$\mathbf{H}_{\bigcirc}, \mathbf{H}_{\Box} \hookrightarrow \mathbf{H}$ for the full subcategory
of its modal types.
\label{MomentAndComoment}
\end{definition}
See \cite{Shulman12a} for moments in homotopy type theory.
Here we focus on these idempotent monads; we encounter non-idempotent monads below
in \ref{InfinitesimalsAndModules}, in \ref{ExponentialConjunction} and in
\ref{ComputationalHomotopyTypeTheory}.
Notice that:
\begin{proposition}
  For $\Box$ a moment ($\bigcirc$ a co-moment), def. \ref{MomentAndComoment}, then
  its modal types $X$, def.\ref{modality}, are equivalently those for which the
  unit $X \to \Box X$ (the co-unit $\bigcirc X \to X$) is an equivalence.
  Moreover,
  these (co-)units exhibit the (co-)modal types as forming a reflective
  subcategory
  $$
    \xymatrix{
      \mathbf{H}_{\bigcirc}
      \ar@<+4pt>@{<-}[r]
      \ar@<-4pt>@{^{(}->}[r]
      &
      \mathbf{H}
    }
    \,,
  $$
  resp. co-reflective subcategory
  $$
    \xymatrix{
      \mathbf{H}_{\Box}
      \ar@<-4pt>@{<-}[r]
      \ar@<+4pt>@{^{(}->}[r]
      &
      \mathbf{H}
    }
    \,.
  $$
  \label{MomentModalTypes}
\end{proposition}
It turns out to be specifically interesting to consider
situations where a reflection and a co-reflection jointly exist in two different ways,
either as an adjoint triple of the form
$$
  \xymatrix{
    \mathbf{H}_{\bigcirc} \simeq \mathbf{H}_{\Box}
    \ar@<+8pt>@{<-}[rr]^-{\bigcirc}
    \ar@{^{(}->}[rr]|-{i_{\Box} \simeq i_{\bigcirc}}
    \ar@<-8pt>@{<-}[rr]_-{\Box}
    &&
    \mathbf{H}
  }
$$
or of the form
$$
  \xymatrix{
    \mathbf{H}_{\bigcirc} \simeq \mathbf{H}_{\Box}
    \ar@<+8pt>@{^{(}->}[rr]^-{i_{\Box}}
    \ar@{<-}[rr]|-{\bigcirc \simeq \Box}
    \ar@<-8pt>@{^{(}->}[rr]_-{i_{\bigcirc}}
    &&
    \mathbf{H}
  }
$$
This is captured by the following
\begin{definition}
We say a moment $\bigcirc$ and co-moment $\Box$ are
\emph{dual} or \emph{opposite} if they are adjoint
$$
  \bigcirc \dashv \Box
  \mbox{\hspace{.7cm} or \hspace{.7cm}}
  \Box \dashv \bigcirc
$$
such that their categories of modal types are canonically equivalent.
\label{OppositeMoments}
\end{definition}
\begin{remark}
The perspective of def. \ref{OppositeMoments}
has been highlighted in  \cite{Lawvere91}, where it is proposed (p. 7)
that adjunctions of this form usefully formalize ``many instances of the \emph{Unity and Identity of Opposites}''
that control Hegelian metaphysics \cite{Hegel}.
 \label{DualMomentsAndHegel}
\end{remark}
When we give such a duality a name $D$, we write
$$
  D \;\colon\; \bigcirc \dashv \Box
  \mbox{\hspace{.7cm} or \hspace{.7cm}}
  D \;\colon\; \Box \dashv \bigcirc
$$
respectively.
Given opposite moments $\bigcirc \dashv \Box$ or $\Box \dashv \bigcirc$, every type $X$
sits naturally in a transformation
$$
  \Box X \longrightarrow X \longrightarrow \bigcirc X
$$
between its two dual moments. This expresses how $X$ is decomposed into these two
moments. For stable homotopy types this further refines to the following much more detailed
decomposition.
\begin{definition}
For pointed $X$ we write $\tilde \Box X$ for the fiber of $\Box X \to X$
and $\tilde \bigcirc X$ for the cofiber of $X \to \bigcirc X$.
\end{definition}
\begin{proposition}
Given opposite moments $\bigcirc \dashv \Box$ as in def. \ref{OppositeMoments},
then every stable homotopy type $X$
sits in a diagram of the form
$$
  \raisebox{30pt}{
  \xymatrix{
    & {\tilde \bigcirc} \Omega X
    \ar[rr]
    \ar[dr]
    & & {\tilde \Box} \Sigma X \ar[dr]
    \\
    \Box {\tilde \bigcirc} \Omega X
    \ar[ur]
    \ar[dr]
    &&
    X
    \ar[ur]
    \ar[dr]
    && \bigcirc {\tilde \Box} \Sigma X
    \\
    & \Box X
    \ar[rr]
    \ar[ur]
    & & \bigcirc X \ar[ur]
  }
  }
  \,,
$$
and the two squares here are homotopy cartesian.
\label{DecompositionIntoMoments}
\end{proposition}
This was highlighted in \cite{BNV}, see section 4.1.2 of \cite{dcct}.
\begin{remark}
 The bottom piece of the diagram in prop. \ref{DecompositionIntoMoments}
 is the basic decomposition of $X$ into its dual moments $\Box X$ and $\bigcirc X$.
 The statement of prop. \ref{DecompositionIntoMoments} is that if $X$ is a stable
 homotopy type then first of all there is a further pair of opposite moments
 of $X$, namely $\tilde \bigcirc \Omega X$ and $\tilde \Box \Sigma X$, and second that
 $X$ may be entirely reconstructed from either gluing $\bigcirc X$ with $\tilde \Box \Sigma X$
 (along $\bigcirc \tilde \Box \Sigma X$) as well as from gluing $\Box X$ with $\tilde \bigcirc \Omega X$
 (along $\Box \tilde \bigcirc \Omega X$).
\end{remark}
There is also a decomposition into moments relative to a base type:
\begin{proposition}
  Let $\bigcirc$ be a moment, def. \ref{MomentAndComoment}, which preserves finite limits. Then for $X \in \mathbf{H}$
  there is a moment $\bigcirc_X$ on $\mathbf{H}_{/X}$ given by sending
  $(E \stackrel{p}{\to} X)$ to the left morphism in the pullback diagram
  $$
    \raisebox{20pt}{
    \xymatrix{
      \bigcirc_X E  \ar[r] \ar[d] & \bigcirc E \ar[d]^{\bigcirc p}
      \\
      X \ar[r] & \bigcirc X
    }}
    \,,
  $$
  where the bottom morphism is the $\bigcirc$-unit.
  Moreover, the universal factorization of $p$ through this pullback
  $$
    \raisebox{20pt}{
    \xymatrix{
      E \ar[dr]_p \ar@{-->}[r] & \bigcirc_X(E)  \ar[r] \ar[d] & \bigcirc E \ar[d]^{\bigcirc p}
      \\
      & X \ar[r] & \bigcirc X
    }}
  $$
  is by a $\bigcirc$-equivalence $E \to \bigcirc_X E$, and this decomposition exhibits an orthogonal factorization system ($\bigcirc$-equivalences / $\bigcirc_X$-modal morphisms) in $\mathbf{H}$.
  \label{RelativeMoments}
\end{proposition}
This is essentially observed in \cite{CJKP}.

\subsection{Axiomatic Metaphysics}
 \label{AxiomaticMetaphysics}

We consider now on a type system as in \ref{NotionsJudgementDeduction}
an iterative pattern of opposite moments, def. \ref{OppositeMoments}.
By inspection we find that in the presence of such a pattern of dual moments the type system
induces a natural formalization of large parts of pre-quantum and quantum physics.

\subsubsection{Determination of the qualities of types}

Notice a very basic example of opposite moments, def. \ref{OppositeMoments}:
\begin{example}
  The dependent sum and dependent product over the empty type
  are adjoint
  $$
    \underset{\emptyset}{\sum} (-)
    \;\dashv\;
    \underset{\emptyset}{\prod} (-)
    \,.
  $$
  This is equivalently the hom-adjunction of the empty type
  $$
    \left(\left(-\right)\times \emptyset \right) \; \dashv \; \left(\emptyset \to \left(-\right)\right)
  $$
  and this is equivalent to the adjunction between the co-monad constant on $\emptyset$ and the
  monad constant on $\ast$
  $$
    \emptyset\dashv \ast
    \,.
  $$
  This adjunction exhibits dual moments.
  The canonical factorization
  $$
    \emptyset \longrightarrow X \longrightarrow \ast
  $$
  appearing in prop. \ref{DecompositionIntoMoments}
  expresses any type $X$ as intermediate between the empty type of no terms
  and the type of an entirely undetermined term.
  \label{OppositeMomentsOfNothingAndBeing}
  \label{InitialTerminalAdjunction}
\end{example}
\begin{remark}
  On p. 7 of \cite{Lawvere91} the adjunction in example \ref{InitialTerminalAdjunction}
  is proposed to be a formalization, in the sense of remark \ref{DualMomentsAndHegel},
  of what \cite{Hegel} calls the unity of the moment of {\it Nichts} (nothing) and
  the moment of {\it reines Sein} (pure being) in {\it Werden} (becoming).
  It is in this sense that
  in \cite{Lawvere91} categories equipped with adjoint moments $\Box \dashv \bigcirc$ are called
  ``categories of being''.
  Indeed, according to \cite{Hegel} I.1.1.C.a.4
  ``there is nothing which is not an intermediate state between being and nothing'',
  which one might feel is well reflected by the fact that the canonical decomposition into moments
  in example \ref{OppositeMomentsOfNothingAndBeing} is the tautological $\emptyset \to X \to \ast$.
  But therefore this is a very unspecific hence un-determinate notion of ``being'', and indeed this un-determination is
  what \cite{Hegel} I.1.1.A takes as the characteristic of {\it reines Sein} (pure being).
  More determinate forms of being equipped with more qualities are to refine this
  (\emph{Dasein}).
  We are hence led, following p. 7 of \cite{Lawvere91}, to consider adding further dual moments in order to equip types
  with more determinate/more qualified being.
\end{remark}

So consider then a further pair of opposite moments $\flat \dashv \sharp$
of the form $\Box \dashv \bigcirc$,
$$
  \xymatrix{
    \emptyset \ar@{-|}[d] \ar@{}[r]|{\subset} & \flat \ar@{-|}[d]
    \\
    \ast \ar@{}[r]|{\subset} & \sharp
  }
$$

Given this we may ask if these in turn are opposite
themselves to pairs of opposites, in that we have
$$
  \xymatrix{
    \int \ar@{-|}[r] \ar@{-|}[d] & \flat \ar@{-|}[d]
    \\
    \flat \ar@{-|}[r] & \sharp
  }
$$
This yields
$$
  \raisebox{20pt}{
  \xymatrix{
    & \int
      \ar@{-|}[d]
    \\
    \emptyset \ar@{-|}[d] \ar@{}[r]|{\subset} & \flat \ar@{-|}[d]
    \\
    \ast \ar@{}[r]|{\subset} & \sharp
  }}
  \,.
$$
We may consider incrementally adding further moments/qualities to the type system this way.
Here we add one more stage of refinement:
\begin{definition}
Given a type system $\mathbf{H}$ as in \ref{NotionsJudgementDeduction} (an $\infty$-topos)
a system of dual moments on $\mathbf{H}$, def. \ref{OppositeMoments}, of the form
$$
  \xymatrix{
    && \Re
    \ar@{-|}[d]
    \\
    &\int \ar@{-|}[d] \ar@{}[r]|{\subset} & \oint \ar@{-|}[d]
    \\
    \emptyset
    \ar@{}[r]|{\subset}
    \ar@{-|}[d]
    & \flat
    \ar@{}[r]|{\subset}
    \ar@{-|}[d]
    &
    \Im
    \\
    \ast
    \ar@{}[r]|{\subset}
    &
    \sharp
  }
$$
we say is a structure of \emph{differential cohesion} in $\mathbf{H}$ \cite{dcct}.
Here ``$\subset$'' denotes inclusion of modal types.
\label{DifferentialCohesion}
\end{definition}
\begin{remark}
  Often we demand in addition that $\int$ preserves finite products and that $\Re$ preserves
finite limits. Neither assumption is needed for the central statements about linear type theory below.
\end{remark}
The system of moments in def. \ref{DifferentialCohesion} determinates
further ``qualities'' carried by all types. In the following we discuss consequences of these
determinate qualities that play a role in the discussion of quantization below.

\subsubsection{Realization of types as moduli stacks}
\label{Realization}

We discuss here how the $(\oint \dashv \Im)$-moments of def. \ref{DifferentialCohesion}
realize each type $X$ as a higher Deligne-Mumford stack (higher Artin stack)
in the sense of \cite{LurieStructured}. This gives an explicit meaning to the statement
that every type with the quality of def. \ref{DifferentialCohesion}
plays the role of a geometric moduli stack (for instance of higher gauge fields,
this we discuss below in \ref{Gauge}).

\begin{claim}
  Given differential cohesion, then for $X$ any type then the
  $\oint_X$-modal $X$-dependent types form an \emph{{\'e}tale topos} over $X$ and
  the $\Im_X$-modality equips this with an $\mathbf{H}$-structure sheaf
  in the sense of \cite{LurieStructured}. Along $\oint$-modal morphisms these base change
  in a module-analog of dependent sum and dependent product.
  \label{UnityOfShapeFlatIsGauge}
\end{claim}
This is discussed in \cite{dcct}.

More in detail:
\begin{proposition}
  For $\mathbf{H}$ equipped with differential cohesion and for $X \in \mathbf{H}$,
  the full subcategory
  $$
    \xymatrix{
      \mathrm{Sh}(X) := (\mathbf{H}_{/X})_{\oint_X}
      \ar@<+10pt>@{<-}[rr]^-{\Re_X}
      \ar@{^{(}->}[rr]
      \ar@<-10pt>@{<-}[rr]_-{\Im_X}
      &&
      \mathbf{H}_{/X}
    }
  $$
  on the $\oint_X$-modal types, prop. \ref{RelativeMoments}, is an $\infty$-topos
  and equipped with the map
  $$
    \mathcal{O}_X
      :
    \mathbf{H}_{\Re}
    \hookrightarrow
    \mathbf{H}
    \stackrel{X^\ast}{\longrightarrow}
    \mathbf{H}/_X
    \stackrel{\Im}{\longrightarrow}
    \mathrm{Sh}(X)
  $$
  it is an $\mathbf{H}_{\Re}$-structured $\infty$-topos in the sense of \cite{LurieStructured}
  (with ``admissible'' morphisms in $\mathbf{H}$ the $\oint$-modal morphisms).
  \label{EtaleTopos}
\end{proposition}
We may call $\mathrm{Sh}(X)$ the \emph{{\'e}tale topos} of $X$.
\begin{proposition}
  For $f \colon Y \longrightarrow X$ a morphism in $\mathbf{H}$ which
  is $\oint$-modal as an $X$-dependent type, this canonically induces
  an {\'e}tale morphism of structured $\infty$-toposes, hence an
  {\'e}tale geometric morphism of $\infty$-toposes
  $$
    \xymatrix{
      \mathrm{Sh}(Y)\simeq \mathrm{Sh}(X)/_{Y}
      \ar@<+12pt>@{->}[rr]^-{f_!}
      \ar@{<-}[rr]|-{f^\ast}
      \ar@<-12pt>@{->}[rr]_-{f_\ast}
      &&
      \mathrm{Sh}(X)
    }
  $$
  together with an equivalence of structure sheaves
  $$
    \mathcal{O}_Y \simeq f^\ast \mathcal{O}_X
    \,.
  $$
\end{proposition}

\subsubsection{Gauge fields}
\label{Gauge}

We discuss here how the $(\int \dashv \flat)$-moments of
def. \ref{DifferentialCohesion} exhibit each stable type as having the quality
of a moduli stack of higher gauge fields/cocycles in generalized differential
cohomology.

\begin{claim}
  Given differential cohesion, def. \ref{DifferentialCohesion},
  then the decomposition of types into their $(\int\dashv \flat)$-moments according to
  prop. \ref{DecompositionIntoMoments} exhibits them as types of cocycles in
  generalized differential cohomology, hence as moduli for pre-quantum (``classical'') fields (higher gauge fields).
\end{claim}
This is discussed in \cite{dcct, BNV}.

\subsubsection{Infinitesimals, Linearity and Modules}
 \label{InfinitesimalsAndModules}

We discuss how the $(\Re \dashv \oint)$-moments of def. \ref{DifferentialCohesion}
induce over each type $X$ a closed symmetric monoidal category
$(\mathrm{Mod}(X), \otimes)$
which behaves like a category of bundles of modules over $X$, hence a generalization of a
category of vector bundles over $X$.

\medskip

\begin{definition}
  Given a type system category $\mathcal{C}$ with terminal object $\ast$ and coproducts, then its
  \emph{maybe-modality} $\ast/$ is the modality (monad, def. \ref{modality}) given by
  $$
    */ \;\colon\; X \mapsto X \sqcup \ast
    \,.
  $$
  \label{MaybeMonad}
\end{definition}
Here our notation reflects the following basic fact.
\begin{proposition}
  The maybe-modal types (def.\ref{modality}), hence the algebras in $\mathcal{C}$
  over the maybe-monad, def. \ref{MaybeMonad}, are equivalently the pointed objects;
  the category of algebras over the maybe monad is the co-slice $\mathcal{C}^{\ast/}$
  under the point.
  \label{PointedObjects}
\end{proposition}
\proof
  That these algebras are the
  pointed objects is already equivalent to the statement of the unit axiom for algebras over the
  maybe-monad. Then action axiom is then automatically satisfied.
\endofproof
\begin{definition}
For $X \in \mathbf{H}$ a type, write $\mathbf{H}_{/X}^{X/} \simeq (\mathbf{H}_{/X})^{\ast/}$
for the category of algebras over the maybe-monad on the slice topos $\mathbf{H}_{/X}$, def. \ref{MaybeMonad},
hence by prop. \ref{PointedObjects} for the category of pointed objects in the slice.
\end{definition}
\begin{remark}
An object in $\mathbf{H}_{/X}^{X/}$ may be interpreted a bundle over $X$ which is equipped with a global
section. The existence of such a global section is a property shared in particular by vector bundles
and more generally by fiber bundles of modules, for which the global section is the zero-section.
\end{remark}
We may axiomatically characterize those sectioned bundles over $X$ which are
like bundles of modules in that they are \emph{linear} bundles as follows.
\begin{definition}
  For $X$ a type, an \emph{infinitesimal extension} or \emph{formal extension}
  of $X$ is an object $(E \to X) \in \mathbf{H}_{/X}^{X/}$ such that
  the underlying morphism in $\mathbf{H}$ is an $\Re$-equivalence,
  equivalently an $\oint$-equivalence.
  Write
  $$
    \mathrm{Mod}_{\Re}(X) \hookrightarrow \left(\mathbf{H}_{/X}^{X/}\right)^{\mathrm{op}}
  $$
  for the opposite of the full subcategory on these objects.
  \label{InfinitesimalExtensions}
\end{definition}
\begin{remark}
  By prop. \ref{RelativeMoments} $\mathrm{Mod}_{\Re}(X)$ is in a sense dual to
  $\mathrm{Sh}(X)$ of def. \ref{EtaleTopos}.
\end{remark}
\begin{remark}
  This definition depends on the system of dual moments in that by $(\Re \dashv \oint \dashv \Im)$
  the notion of ``infinitesimal'' in \ref{InfinitesimalExtensions} is the same as
  the ``infinitesimal'' that characterizes the formally {\'e}tale morphisms in
  def. \ref{EtaleTopos}.
\end{remark}
\begin{example}
  In the typical model of differential cohesion (as in section 4 of \cite{dcct}), the objects $E$ are given by
  sheaves over
  formal duals of commutative algebras (structured algebras)
  of the form $C^\infty(E) = C^\infty(X) \oplus N$, where $N$ is a
  $C^\infty(X)$-module and a nilpotent ideal in $C^\infty(E)$.
  If already $N \cdot N = 0$ then $C^\infty(E)$ is the square-0 extension
  of $C^\infty(X)$ induced from the module $N$.
  A morphism between two such objects in $\mathrm{Mod}(X)$ is a commuting diagram of
  algebras of the form
  $$
    \xymatrix{
      & C^\infty(X)
      \ar[dl]
      \ar[dr]
      \\
      C^\infty(X)\oplus N_1
      \ar[rr]
      \ar[dr]
      &&
      C^\infty(X)\oplus N_2
      \ar[dl]
      \\
      & C^\infty(X)
    }
  $$
  where the vertical maps are the canonical inclusions and projections, respectively.
  For square-0 extensions this is equivalently a homomorphism of modules $N_1 \to N_2$.

  In this fashion $\mathrm{Mod}(X)$ faithfully contains a naive category of module bundles
  over $X$, but it may contain also richer module-like objects.
  In fact, the main theorem in \cite{LurieFormal} says that the infinitesimal extensions of $X$,
  def. \ref{InfinitesimalExtensions}, include the
  sheaves of $L_\infty$-algebras over $X$.
  This means that the ``modules'' considered here are to be thought of
  more generally as modules possibly equipped with further homotopy algebraic structure.
  \label{ModelsForInfinitesimalExtensions}
\end{example}
\begin{lemma}
  The inclusions $\mathrm{Mod}(X)^{\mathrm{op}} \hookrightarrow \mathbf{H}_{/X}^{X/}$
  of def. \ref{InfinitesimalExtensions} preserve limits and colimits.
  \label{InclusionOfInfinitesimalExtensionPreservesLimits}
\end{lemma}
\proof
  Limits in an undercategory are computed in the ambient category, and limits in an
  overcategory are computed as the limits of the diagram with a terminal object
  adjoined in the ambient category. Dually for colimits. We have to show that
  if the diagram in the under-overcategory is in the inclusion of the infinitesimal
  extensions, then so is its (co-)limit. Since $\oint$ preserves all these and by
  assumption on infinitesimal extension, applying $\oint$ to the diagrams with terminal
  (initial object) adjoined make them be diagrams of the shape an $\infty$-groupoid
  with a terminal object, hence of a contractible $\infty$-groupoid, hence be
  essentially constant on $\oint(X)$. This shows that the limit of $\oint$-local
  objects in the slice is itself an $\oint$-equivalence.
\endofproof
We need the following general fact.
  Let $\mathcal{C}$ be a closed symmetric monoidal category with finite limits and colimits
  and reflexive coequalizers.
  Write $\ast \in \mathcal{C}$ for its terminal object and write $\mathcal{C}^{\ast/}$
  for the category of maybe-modal types, def. \ref{MaybeMonad}, hence
  by prop. \ref{PointedObjects} of pointed objects, prop. \ref{PointedObjects},
  in $\mathcal{C}$. The maybe monad $\ast/$ is a commutative monoidal monad
  \cite{Seal}
  and hence canonically induces the structure of a monoidal category on $\mathcal{C}^{\ast/}$.\footnote{
  I am grateful to Todd Trimble and Mike Shulman for discussion of this point, and
  to Zhen Lin Low for pointing out the reference \cite{Seal}.}
\begin{proposition}
  The canonical tensor product induced on the maybe-modal types $\mathcal{C}^{\ast/}$
  is the \emph{smash product} ``$\wedge$'' of pointed objects.
  For $E_1, E_2 \in \mathcal{C}^{\ast/}$
  this operation sends these to the following pushout of coproducts and tensor products
  formed in $\mathcal{C}$
  $$
    E_1 \wedge E_2
    :=
    \;
    \ast \underset{(E_1 \otimes \ast) \coprod (\ast \otimes E_2)}{\coprod} (E_1 \otimes E_2)
  $$
  and this makes $(\mathcal{C}^{\ast/}, \wedge, \ast \coprod \ast)$ a closed symmetric monoidal category
  for the internal hom of pointed morphisms.
  \label{SmashProduct}
\end{proposition}
\proof
  The canonically induced monoidal structure on the category of algebras
  of a commutative monad is often said to go back to \cite{Kock}, where
  indeed the closed structure is discussed, from which the tensor may be obtained
  as the adjunct in suitable circumstances. The monoidal structure appears
  in print explicitly in \cite{Seal} (section 2.2 and theorem 2.5.5).
  Inserting the maybe-monad into the coequalizer formula there straightforwardly yields the
  pushout diagram defining the smash product
  as it appears for instance in construction 4.19 and proposition  4.20 of \cite{ElmendorfMandell}.
\endofproof
Since $\oint$ preserves limits and colimits and hence the construction of the smash product
in prop. \ref{SmashProduct}, the smash product in $\mathbf{H}_{/X}^{X/}$
induces a tensor product on infinitesimal extensions over $X$.
On the other hand, the smash tensor unit $X \coprod X$ of $\mathbf{H}_{/X}^{X/}$ is not $\oint$-local, hence is not an infinitesimal extension over $X$. Therefore we have:
\begin{definition}
  For $X$ a type, write $(\mathrm{Mod}(X), \otimes)$ for the symmetric monoidal category without unit
  which is the opposite of the category of infinitesimal extensions of $X$, def. \ref{InfinitesimalExtensions},
  equipped with the restriction of the
  smash product of prop. \ref{SmashProduct}.
  \label{TensorProductOfModules}
\end{definition}
\begin{example}
  Continuing example \ref{ModelsForInfinitesimalExtensions},
  consider for definiteness the case that these sheaves are representable,
  with $C^\infty(E_i) = C^\infty(X) \oplus N_i$ in the notation there.
  Then we have
  $$
    \begin{aligned}
    C^\infty(E_1 \times_X E_2)
    & =
    \left(C^\infty\left(X\right) \oplus N_1\right)
    \coprod_{C^\infty(X)}
    \left(C^\infty\left(X\right) \oplus N_2\right)
    \\
    & \simeq
    C^\infty(X) \oplus (N_1 \oplus N_2) \oplus (N_1 \otimes N_2)
    \end{aligned}
  $$
  and
  $$
    \begin{aligned}
    C^\infty(E_1 \coprod_X E_2)
    & =
    \left(C^\infty\left(X\right) \oplus N_1\right)
    \times_{C^\infty(X)}
    \left(C^\infty\left(X\right) \oplus N_2\right)
    \\
    & \simeq
    C^\infty(X) \oplus (N_1 \oplus N_2)
   \end{aligned}
   \,.
  $$
  Therefore in these models the tensor product in
  def. \ref{TensorProductOfModules} restricts to the ordinary tensor
  product of modules:
  $$
    C^\infty\left(
      E_1 \wedge_X E_2
    \right)
    \simeq
    C^\infty(X) \oplus (N_1 \otimes N_2)
    \,.
  $$
  \label{TensorProductOfModulesFromSmashOfInfinitesimalExtensions}
\end{example}
\begin{remark}
  Example \ref{TensorProductOfModulesFromSmashOfInfinitesimalExtensions}
  shows that while the tensor unit of the smash product of pointed objects
  does not descend to a tensor unit of modules, in the standard models
  there is after all another object which is the tensor unit for $(\mathrm{Mod}(X), \otimes)$.
  This is the incarnation of the structure sheaf $\mathcal{O}_X$ in the present context,
  yields a closed symmetric monoidal category $(\mathrm{Mod}(X), \otimes, \mathcal{O}_X)$.
\end{remark}

Below in example \ref{BaseChangeForModules}
we find that forming categories of modules over types induces a ``linear''
(meaning: symmetric monoidal but not Cartesian monoidal) analog of
the dependent type theory of \ref{NotionsJudgementDeduction}.
This linearity and its relation to quantization we discuss now in
\ref{LinearTypeTheory}.

\section{Linear homotopy-type theory}
\label{LinearTypeTheory}

The two hallmarks of quantum physics are the \emph{superposition principle}
and \emph{quantum interference}. Together they say that prequantum phases in
$S^1 = U(1)$ are to be freely added, and then are to additively interfere in the ring $E = \mathbb{C}$ of complex numbers
$$
  \xymatrix{
    U(1) \ar@{^{(}->}[rrr]^{\mbox{\scriptsize superposition}} &&& \mathbb{Z}[U(1)] \ar@{->>}[rrr]^{\mbox{\scriptsize interference}} &&& \mathbb{C}
  }
  \,.
$$
This makes the outcome of quantization land in $E$-modules. For $E = \mathbb{C}$ these are
complex vector spaces, but we will see that we should allow other choices of $E$, too. Notably
for realizing quantum mechanics as the boundary field theory of the 2d Poisson-Chern-Simons
TQFT we take $E = \mathrm{KU}$ the complex K-theory ring (an $E_\infty$-ring), for which the superposition
principle is given by \emph{Snaith's theorem} \cite{Snaith}:
$$
  \xymatrix{
     B U(1)\ar@{^{(}->}[rrr]^-{\mbox{\scriptsize superposition}} &&& \mathbb{S}[B U(1)]
      \ar@{->>}[rrr]^-{\mbox{\scriptsize interference}} &&& \mathbb{S}[B U(1)][\beta^{-1}] \simeq \mathrm{KU}
  }
  \,.
$$

Generally, for $E$ a commutative ring (or generally $E_\infty$-ring) write $E \mathrm{Mod}$ for its category of modules.
Under the tensor product $\otimes_E$ of $E$-modules, this is a symmetric closed monoidal category.
As such it is much like a cartesian closed category, only that the tensor product lacks a diagonal map.
Therefore functions in $E \mathrm{Mod}$
may depend only on single (linear) copies of their arguments, and for this reason
\cite{Girard} called the internal logic of such non-cartesian closed symmetric monoidal categories
\emph{linear logic}.

\subsection{Linear logic, Quantum logic, Linear type theory}
 \label{LinearLogic}

Here we briefly survey the literature on quantum logic, and in fact quantum mechanics,
realized in linear type theory; and then point out the
need to refine this to dependent linear homotopy-type theory, which is what we discuss in the following sections.

While there are some deep technical theorems contained in this relation between linear logic, monoidal categories
and quantum physics, namely the coherence theorems, we feel that beyond the technicalities there is a
remarkable conceptual fact to take note of
(which the following discussion hopefully conveys): pure logicians did secretly re-discover the basics of
quantum mechanics (namely linear type theory) independently of physicists and from purely foundational
reasoning. This is not unlike how, as discussed above in \ref{NotionsJudgementDeduction}, pure logicians
did also independently re-discover homotopy theory from purely foundational reasoning. Taken together
this yields linear homotopy-type theory below in \ref{DependentLinearTypeTheory}; and when we show that this
naturally captures quantum field theory below in \ref{Quantization} then the natural origin of all this in
the modern foundations of mathematics seems noteworthy.

\medskip

The study of the internal type theory of non-Cartesian (closed) symmetric monoidal categories
such as $E \mathrm{Mod}$ above
dates back at least to \cite{Lambek}, a systematic account following this is \cite{Szabo}.
After \cite{Girard} introduced ``linear logic'' this was soon realized \cite{Seely89} to be the special case
corresponding to those monoidal categories which are called star-autonomous.
For instance the ``proof nets'' of linear logic are equivalently the string diagrams of these
monoidal categories \cite{Mellies06}.
Since the ``linear'' terminology is apt, today the internal logic/internal type theory of general symmetric monoidal categories is called linear logic, or rather \emph{multiplicative intuitionistic linear logic} for definiteness,
see for instance \cite{HylandDePaiva, BPS}. A recent survey of linear type theory is in \cite{Mellies}.

In parallel to this development, it was understood that \emph{quantum logic} is naturally
taken to be the internal linear logic of
such monoidal categories/linear type theories \cite{Yetter, Pratt, AbDu05, Du06}.
When interpreted in the monoidal category of
Hilbert spaces this reproduces the subobject lattices of the original quantum logic introduced
in \cite{BirkhoffVonNeumann}, a fact explicitly noted in \cite{Crown, HeunenJacobs}.
Conversely, that therefore also categories of cobordisms with their natural symmetric monoidal structure,
which by Atiyah-Segal serve as domains for quantum field theories (see \cite{LurieQFT} for a review)
interpret linear logic (hence quantum logic), was amplified in \cite{Sl, Rosetta}.

In this quantum mechanical interpretation the proof nets of linear logic
correspond to quantum circuits of quantum computing (see e.g. section 3.1 of \cite{Kuperberg}), and Girard's
operational semantics of linear logic via ``Geometry of Interaction''
(see def. 2.6 and remark 5.8 in \cite{AHS}) corresponds to
the construction of categories of superoperators (see e.g. sections 1.4, 1.5 in
\cite{Kuperberg}). That the additional abstract structure needed in order to restrict these
to genuine \emph{quantum operations} on spaces of mixed quantum states
(and hence to provide the probabilistic meaning of quantum physics)
is a ``dagger-compact structure'' on the linear type theory (this we come to in \ref{DaggerCompactness} below)
was observed in \cite{Selinger} following \cite{AbramskyCoecke}.
A quick survey of the wealth of constructions in quantum mechanics
abstractly expressible this way is in \cite{Coecke}.

\medskip

But we have to note that all this captures but a small fragment (to borrow that term) of quantum physics, namely
just basic quantum mechanics, but not quantum field theory. Moreover, modern quantum field theory
is characterized by three additional properties; it is
\begin{enumerate}
  \item \emph{Lagrangian}
  \item \emph{local}
  \item \emph{gauge}
\end{enumerate}
quantum field theory,
and we observe that interpreting each of these three items requires to refine the linear type theory of
a single monoidal category, as considered in the above references, to dependent linear homotopy-type theory:
\begin{enumerate}
  \item The quantum field theories of actual interest in nature and in theory are typically
those that arise by a process of quantization from Lagrangian data. The quantization step is supposed to be
essentially a summing operation of linear types: a space of quantum states is a space of sections
of a pre-quantum bundle, hence of ``sums of fibers'' of a prequantum bundle, and a quantum propagator
between such states is a ``path integral'' summing up contributions of linear maps for all physical
trajectories.  This means that we need a linear analog of the dependent sum operation
$\sum$ discussed in \ref{NotionsJudgementDeduction}, hence that we need dependent linear types
to formalize quantization.
  \item Locality in quantum field theory means, in the Schr{\"o}dinger picture \cite{LurieQFT},
  that spaces of states assigned to codimension-1 slices of spacetime/worldvolume arise by
  integrating up higher categorical data assigned to higher codimension strata.
  This means that spaces of states of gauge field theories
  are ``higher linear types'' (we focus on this below in \ref{DirectedHomotopyTypeTheory}).
  \item Specifically, in the presence of non-trivial gauge equivalences, spaces of quantum states are no longer
  plain vector spaces, but are (co)chain complexes, called ``BRST complexes'' \cite{HenneauxTeitelboim},
  whose differentials encode the gauge symmetry. This means spaces of states of gauge field theories
  are linear homotopy-types.
\end{enumerate}

\medskip

In view of all this and our above discussion in \ref{NotionsJudgementDeduction}
we consider now the refinement of linear type theory to \emph{linear homotopy-type theory}
and hence in particular \emph{dependent} linear type theory, whose induced logic
is hence a form of predicate quantum logic. Or rather, as throughout, we
stick to the categorical semantics of all this. A syntactic account of linear homotopy-type theory is being developed
by Mike Shulman.

\subsection{Linear homotopy-type theory}
\label{DependentLinearTypeTheory}

Given a cohesive space $X$, then there is also the category $E\mathrm{Mod}(X)$ of $E$-module bundles over $X$.
For instance if $X$ is a physical phase space, then the prequantum line bundle is an invertible object
in $E \mathrm{Mod}(X)$. It follows that quantization is to take place in
\emph{dependent} linear type theory, parameterized over the cartesian types $X$ of the pre-quantum geometry.
According to Lawvere's notion of categorical logic embodied in the notion of hyperdoctrines
as made precise in \cite{Seely83}, this means, applied to linear logic, the following:

A \emph{dependent linear logic} or \emph{linear hyperdoctrine} is a category of
contexts $\Gamma$, a symmetric closed monoidal category $\mathcal{C}_{\Gamma}$ for each such
context and functorially for each morphism of contexts $f : \Gamma_1 \longrightarrow \Gamma_2$
an adjoint triple of functors
$$
  (\sum_f \dashv f^\ast \dashv \prod_f)
  \;:\;
  \xymatrix{
    \mathcal{C}_{\Gamma_1}
      \ar@<+16pt>[r]|{f_!}
      \ar@<+4pt>@{<-}[r]|{f^\ast}
      \ar@<-4pt>[r]_{f_\ast}
      &
    \mathcal{C}_{\Gamma_2}
  }
$$
such that $f^\ast$ is strong monoidal and satisfies Frobenius reciprocity, hence such that $f^\ast$ is a
strong closed monoidal functor. Typically one would also demand that consecutive such adjoint triples
satisfy the Beck-Chevalley condition.

The categorical semantics for such dependent linear type theory has been studied in \cite{Shulman08, Shulman12}.
But it is noteworthy that in just slightly different guise these axioms are much older: they
are a version of Grothendieck's ``yoga of six functors'' \cite{May05}, which were recognized as the abstract
reason behind Verdier duality. Specifically, an adjoint triple $(f_! \dashv f^\ast \dashv f_\ast)$
with $f^\ast$ strong closed monoidal
is called a \emph{Wirthm{\"u}ller context} in \cite{May05}. (The literature on Grothendieck's six
operations often considers (also) the dual \emph{Grothendieck contexts}, e.g. \cite{Pol08}.)

\medskip
We now state this more formally. We often say just ``category'' for ``$\infty$-category''.

\begin{definition}
  For $\mathcal{C}$, $\mathcal{D}$ two closed symmetric monoidal
  categories, a \emph{Wirthm{\"u}ller context}
  $f : \mathcal{C} \to \mathcal{D}$ between them is a
  strong closed monoidal functor $f^\ast : \mathcal{D} \to \mathcal{C}$
  such that it has a left adjoint and right adjoint $(f_! \dashv f^\ast \dashv f_\ast)$.
  \label{WirthmuellerMorphism}
\end{definition}
Often it is useful to equivalently reformulate closedness of $f^\ast$ in terms of the following
condition.
\begin{definition}
  Given an adjunction $(f_! \dashv f^\ast)$ between symmetric monoidal categories
  such that $f^\ast$ is a strong monoidal functor,
  then the condition that the canonical natural transformation
  $$
    \overline{\pi}
    \;:\;
    f_!((f^\ast B) \otimes A)
    \longrightarrow
    B \otimes f_!(A)
  $$
  is a natural equivalence is called the \emph{projection formula}. The
  existence of the left adjoint $f_!$ and the validity of the projection formula is
  also referred to as \emph{Frobenius reciprocity} in representation theory and in
  categorical logic (``hyperdoctrines''), and often just called \emph{reciprocity},
  for short.
  \label{FrobeniusReciprocity}
\end{definition}
\begin{remark}
  Below in example \ref{ReciprocityIsLinearity} we observe that
  Frobenius reciprocity is a higher categorical linearity condition.
  This perspective is crucial for making transparent the nature of
  quantum anomaly cancellation that we come to below in \ref{CoboundingTheory},
  see remark \ref{quantumAnomaly} there.
\end{remark}
A basic fact is that:
\begin{proposition}
  Given an adjoint pair $(f_! \dashv f^\ast)$ between closed monoidal categories
  with $f^\ast$ a strong monoidal functor,
  then the condition that $f^\ast$ is strong closed is equivalent to
  Frobenius reciprocity, def. \ref{FrobeniusReciprocity}, hence to $f_!$ satisfying
  its projection formula.
  \label{FrobeniusReciprocityEquivalence}
\end{proposition}
\begin{remark}
  If in a Wirthm{\"u}ller context, def. \ref{WirthmuellerMorphism}, not only $f_!$
  but also $f_\ast$ satisfies its projection formula, then \cite{Haugseng}
  speaks of a ``transfer context'' (def. 4.9 there), because this turns out to be
  an abstract context in which Becker-Gottlieb transfer exists (prop. 4.14 there).
  The abstract construction of Becker-Gottlieb transfer is similar to the construction
  of Umkehr maps via fundamental classes in Wirthm{\"u}ller contexts which we consider in
  \ref{FundamentalClasses} below.
  \label{transferContext}
\end{remark}
The central concept of interest here is now the following.
\begin{definition}
  A \emph{model/semantics for linear homotopy-type theory} is
  a locally Cartesian closed $\infty$-category $\mathbf{H}$ (``of non-linear homotopy-types'')
  and a Cartesian fibration
  $$
    \xymatrix{
      \mathrm{Mod}
      \ar[d]
      \\
      \mathbf{H}
    }
  $$
  (``of dependent linear homotopy-types'')
  such that the $\infty$-functor
  $$
    \mathrm{Mod} \;:\; \mathbf{H}^{\mathrm{op}} \to \mathrm{Cat}_\infty
  $$
  that classifies the fibration by the Grothendieck-Lurie construction (\cite{HTT}, section 3.2)
  takes values in Wirthm{\"u}ller contexts, def. \ref{WirthmuellerMorphism}, hence
  sends objects $X \in \mathbf{H}$ to closed symmetric monoidal $\infty$-categories
  $\mathrm{Mod}(X)$ and sends morphism $f : X \to Y$ to $\infty$-functors
  $f^\ast : \mathrm{Mod}(Y)\to \mathrm{Mod}(X)$ which are strong monoidal,
  have a left and right adjoint, and are strong closed, hence,
  by prop. \ref{FrobeniusReciprocityEquivalence}, satisfy Frobenius reciprocity.
  \label{LinearHomotopyTypeTheory}
\end{definition}
\begin{remark}
Definition \ref{LinearHomotopyTypeTheory} is the evident $\infty$-categorical version of the
\emph{closed monoidal fibrations} considered in \cite{Shulman08} (def. 13.1) and \cite{Shulman12} (theorem 2.14).
Mike Shulman is working on developing formal \emph{syntax} for linear homotopy-type theory
similar to the formal syntax for non-linear homotopy-type theory that is laid out in \cite{HoTT}.
This is to be such that def. \ref{LinearHomotopyTypeTheory} provides the corresponding $\infty$-categorical
semantics/models.
\end{remark}
We consider now three classes of examples of semantics for linear homotopy-type theory.
\begin{example}
  For $\mathbf{H}$ an $\infty$-topos and $f \;\colon\; X_1 \longrightarrow X_2$ any morphism
  in $\mathbf{H}$, the induced {\'e}tale geometric morphism on slice toposes
  is a cartesian Wirthm{\"u}ller context between the slice toposes:
  $$
    (\sum_f \dashv f^\ast \dashv \prod_f)
    \;:\;
    (\mathbf{H}_{/X_1},\, \times_{X_1},\, X_1)
    \longrightarrow
    (\mathbf{H}_{/X_2},\, \times_{X_2},\, X_2)
    \,.
  $$
  Hence the self-indexing
  $$
    \xymatrix{
      \mathbf{H}^{\Delta^1}
      \ar[d]^{\mathrm{cod}}
      \\
      \mathbf{H}
    }
  $$
  of an $\infty$-topos $\mathbf{H}$ (the codomain fibration, see \cite{HTT} 2.4.7)
  is a model for linear homotopy-type theory, def. \ref{LinearHomotopyTypeTheory}.
  \label{ToposBaseChangeIsCartesianWirthmueller}
\end{example}
\proof
    The left adjoint $f_! = \sum_f$ (dependent sum) sends slice objects $(A \to X_1)$
    to the composite $(A \to X_1 \stackrel{f}{\to} X_2)$.
    Therefore by prop \ref{FrobeniusReciprocityEquivalence} it is sufficient to exhibit Frobenius reciprocity
    in the form
    $$
      A \times_{X_1} f^\ast B \simeq A \times_{X_2} B
      \,.
    $$
    But this is equivalently the pasting law for pullbacks in $\mathbf{H}$:
    $$
      \raisebox{20pt}{
      \xymatrix{
        A \times_{X_2} B \simeq& A \times_{X_1} f^\ast B
        \ar[d]
        \ar[r]
        &
        f^\ast B
        \ar[d]
        \ar[r]
        &
        B
        \ar[d]
        \\
        & A \ar[r] & X_1 \ar[r]^f & X_2
      }
      }
    $$
\endofproof
\begin{example}
  For $\mathbf{H}$ a topos and $X \in \mathbf{H}$ an object, write
  $\mathbf{H}_{/X}^{X/} \simeq (\mathbf{H}_X)^{\ast/}$ for the category of pointed objects in the
  slice topos over $X$, prop. \ref{PointedObjects}.
  By prop. \ref{SmashProduct} this is a closed symmetric monoidal category
  $(\mathbf{H}_{/X}^{/X}, \wedge_X, X \coprod X)$ under the smash product $\wedge_X$.
  For $f : X \longrightarrow Y$ a morphism in $\mathbf{H}$, the base change inverse image
  of example \ref{ToposBaseChangeIsCartesianWirthmueller} lifts to a functor
  $f^\ast : \mathbf{H}_{/Y}^{Y/} \longrightarrow \mathbf{H}_{/X}^{X/}$. This is strong closed
  monoidal with respect to the smash product structure and has a left and right adjoint, hence
  constitutes a Wirthm{\"u}ller context
  $$
    (\sum_f \dashv f^\ast \dashv \prod_f)
    \;:\;
    (\mathbf{H}_{/X}^{X/},\, \wedge_{X},\, X \coprod X)
    \longrightarrow
    (\mathbf{H}_{/Y}^{Y/},\, \wedge_{Y},\, Y \coprod Y)
    \,.
  $$
  \label{BaseChangeForPointedObjects}
\end{example}
This appears as examples 12.13 and 13.7 in \cite{Shulman08} and example 2.33 in
\cite{Shulman12}.\\
\proof
  For $f : X \longrightarrow Y$ any morphism in $\mathbf{H}$ then the base change inverse image
  $f^\ast : \mathbf{H}_{/Y} \longrightarrow \mathbf{H}_{/X}$ preserves pointedness,
  and the pushout functor $f_! : \mathbf{H}^{X/} \longrightarrow \mathbf{H}^{Y/}$ preserves co-pointedness.
  These two functors hence form an adjoint pair
 $(f_! \dashv f^\ast) : \mathbf{H}_{/X}^{X/} \longrightarrow \mathbf{H}_{/Y}^{Y/}$.
Moreover, since colimits in the under-over category $\mathbf{H}_{/X}^{X/}$ are computed as colimits in $\mathbf{H}$ of diagrams with an initial object adjoined, and since by the Giraud axioms in the topos $\mathbf{H}$ pullback preserves these colimits, it follows that $f^\ast : \mathbf{H}_{/Y}^{Y/} \to \mathbf{H}_{/X}^{X/}$ preserves colimits.
Since by prop. \ref{PointedObjects} $\mathbf{H}_{/X}^{X/}$ is presentably monadic over $\mathbf{H}_{/X}$
(via the maybe-monad $\ast \coprod (-)$) we have
by  2.78 in\cite{AR}  that $\mathbf{H}_{/X}^{X/}$ and $\mathbf{H}_{/Y}^{Y/}$
are locally presentable categories, so that by the adjoint functor theorem it follows that
$f^\ast$ has also a right adjoint $f_\ast \colon \mathbf{H}_{/X}^{X/} \to \mathbf{H}_{/Y}^{Y/}$.

To see that $f^\ast$ is a strong monoidal functor observe that the smash product is, by prop. \ref{SmashProduct},
given by a pushout over coproducts and products in the slice topos. As above these are all preserved by pullback. Finally to see that $f^\ast$ is also a strong closed functor observe that the internal hom on pointed objects is a fiber product of cartesian internal ohms. These are preserved by example \ref{ToposBaseChangeIsCartesianWirthmueller},
and the fiber product is preserved since $f^\ast$ preserves all limits.
Hence $f^\ast$ preserves also the internal homs of pointed objects.
\endofproof
\begin{example}
  Given an $\infty$-topos $\mathbf{H}$ equipped with moments $(\Re \dashv \oint)$, def. \ref{DifferentialCohesion}, then
  for $f : X \longrightarrow Y$ a morphism in $\mathbf{H}$ the
  Wirthm{\"u}ller context on pointed objects of def. \ref{BaseChangeForPointedObjects}
  restricts to infinitesimal extensions, def. \ref{InfinitesimalExtensions},
  $$
    (\sum_f \dashv f^\ast \dashv \prod_f)
    :
    \mathrm{Mod}_{\Re}(X)^{\mathrm{op}}
    \longrightarrow
    \mathrm{Mod}_{\Re}(Y)^{\mathrm{op}}
    \,.
  $$
  Hence
  $$
    \xymatrix{
      \mathrm{Mod}_{\Re}
      \ar[d]
      \\
      \mathbf{H}
    }
  $$
  is a model for linear homotopy-type theory, def. \ref{LinearHomotopyTypeTheory}.
  \label{BaseChangeForModules}
\end{example}
\proof
  By lemma \ref{InclusionOfInfinitesimalExtensionPreservesLimits} it follows
  that $f_!$ and $f^\ast$ preserve infinitesimal extensions and that
  the restriction of $f^\ast$
  still preserves colimits. Therefore to see that $f_\ast$ restricts it is
  sufficient to see that $\mathrm{Mod}(X)^{\mathrm{op}}$ is locally presentable.
  Since it is the essential fiber of $\oint_X : \mathbf{H}_{/X}^{X/} \to \mathbf{H}_{/\oint(X)}$
  over the singleton subcategory on $\mathrm{id}_{\oint(X)}$
  this statement follows by corollary A.2.6.5 in \cite{HTT}.
\endofproof
In applications already a small subclass of the linear homotopy-types in example
\ref{BaseChangeForModules} carry a fair amount of structure: over sites consisting of
formal duals of square-0 extensions of commutative ($E_\infty$-)rings $E$,
$E$-modules $N \in E \mathrm{Mod}$ are modules in the above abstract sense, by example
\ref{ModelsForInfinitesimalExtensions}, and often we want to consider only these
``representable'' modules. The following example shows that at least over geometrically
discrete homotopy-types, these representable modules in themselves constitute a linear
homotopy-type theory
\begin{example}
  Let $\mathcal{V}$ be a closed symmetric monoidal $\infty$-category with all
  small $\infty$-limits and $\infty$-colimits, such as $\mathcal{V}= E \mathrm{Mod}$
  for $E \in \mathrm{CRing}_\infty$. For $X$ a groupoid, write
  $$
    \mathcal{V}(X) :=\mathrm{Func}(X,\mathcal{V})
  $$
  for the $\infty$-category of $\infty$-functors $X \to \mathcal{V}$
  (also called \emph{$\mathcal{V}$-local systems} on the homotopy type $X$).
  For $f\;:\; X \longrightarrow Y$ a morphism of $\infty$-groupoids,
  the pullback (precomposition) $\infty$-functor
  $f^\ast : \mathcal{V}(Y) \to \mathcal{V}(X)$ has a
  left and right
  $\infty$-adjoint $f_!$ and $f_\ast$, given by left and right
  $\infty$-Kan extension (\cite{HTT}  4.3), hence constitutes an
  adjoint triple
  $$
    (\underset{f}{\sum} \dashv f^\ast \dashv \underset{f}{\prod})
    \;:\;
    \xymatrix{
      \mathcal{V}(X)
        \ar@<+10pt>@{->}[rr]^-{f_!}
        \ar@{<-}[rr]|-{f^\ast}
        \ar@<-10pt>@{->}[rr]_-{f_\ast}
      &&
      \mathcal{V}(Y)
    }
    \,.
  $$
  These are Wirthm{\"u}ller contexts and hence make
  $$
    \xymatrix{
      \mathcal{V}(-)
      \ar[d]
      \\
      \mathrm{Grpd}_\infty
    }
  $$
  a model for linear homotopy-type theory, def. \ref{LinearHomotopyTypeTheory}.
  \label{LocalSystems}
\end{example}
\proof
  That $f_!$ and $f_\ast$ are given by $\infty$-Kan extension is
  prop. 4.3.3.7 in \cite{HTT}.
  We need to show that $f^\ast$ is strong closed, hence by prop. \ref{FrobeniusReciprocityEquivalence}
  that $(f_! \dashv f^\ast)$ satisfies Frobenius reciprocity.
  To that end notice that by the very definitions 4.3.2.2 and 4.3.3.2 in \cite{HTT}
  to which prop. 4.3.3.7 there appeals, $\infty$-Kan
  is given \emph{pointwise} at $y \in Y$ given by $\infty$-colimit over the homotopy fiber
  $f^{-1}(y)\hookrightarrow X$:
  $$
    (f_! A)(y) \simeq \underset{\longrightarrow}{\lim}_{x \in f^{-1}(y)} A(x)
    \,.
  $$
  (For $X$ and $Y$ just 1-groupoids and $\mathcal{V}$ locally presentable
  this follows, with \cite{HTT} A.3.3, also from the more traditional fact that homotopy Kan
  extension is pointwise/strong \cite{Cisinski}.)
  Hence for $f: X\to Y$, and for
  $A, B\in \mathcal{V}(Y)$ we have naturally in $x \in X$ the equivalences
  $$
    \begin{aligned}
      f_! ((f^\ast B) \otimes A)
      & =
      \underset{\longrightarrow}{\lim}_{x \in f^{-1}(y)} (f^\ast B(x) \otimes A(x))
      \\
      &\simeq \underset{\longrightarrow}{\lim}_{x \in f^{-1}(y)} (B(y) \otimes A(x))
      \\
      &\simeq B(y)\otimes \underset{\longrightarrow}{\lim}_{x \in f^{-1}(y)} A(x)
      \\
      & = (B\otimes f_! A)(x)
    \end{aligned}
    \,.
  $$
\endofproof
\begin{remark}
  Below we often consider a model for 
  linear homotopy-type theory $\mathrm{Mod}(-)$ as in example \ref{LocalSystems} together with a
  differentially cohesive $\infty$-topos $\mathbf{H}$, def. \ref{DifferentialCohesion}.
  Then for $\mathbf{Fields} \in \mathbf{H}$ an object we write 
  $$
    \mathrm{Mod}(\mathbf{Fields})
    :=
    \mathrm{Mod}(\int \mathbf{Fields})
  $$
  for the $\infty$-category of linear homotopy-types over $\int X \in \infty \mathrm{Grpd}$
  produced by the shape modality (the ``geometric realization'' of the moduli stack $\mathbf{Fields}$).
\end{remark}
For reference notice that example \ref{LocalSystems} reduces to the following
special cases.
\begin{example}
  For $E = \mathbb{S}$ the sphere spectrum, then $E \mathrm{Mod}(-) = \mathbb{S}\mathrm{Mod}(-) \simeq \mathrm{Spectra}(-)$
  is the theory of parameterized spectra.
  This was shown to be a model for linear homotopy-type theory in our sense here in
  \cite{MaySigurdsson} and \cite{ABGHR} (under the translation in \cite{ABG10}).
  \label{ParameterizedSpectra}
\end{example}
\begin{example}
  For $\mathcal{V}$ a closed monoidal category
  with all small limits and colimits,
  such as $\mathcal{V} = E \mathrm{Mod}$ for
  $E \in \mathrm{CRing}$, then the functor
  $$
    [-,\mathcal{V}]
    \;:\;
    \mathrm{Set}^{\mathrm{op}}
    \longrightarrow
    \mathrm{MonCat}
  $$
  takes values in Wirthm{\"u}ller morphisms.
  \label{WirthmuellerMorphsismsBetweenParameterizedModules}
\end{example}
This appears as example 2.2, 2.17 in \cite{Shulman12}.
\begin{remark}
  Examples \ref{BaseChangeForPointedObjects}, \ref{BaseChangeForModules}
  and \ref{LocalSystems} are
  linearized variants of example \ref{ToposBaseChangeIsCartesianWirthmueller}.
  Accordingly we may think of the notion of systems of Wirthm{\"u}ller morphisms as being a generalization
  of categorical logic (``hyperdoctrines'') to non-cartesian and in particular to linear contexts.
  Notice that in a cartesian Wirthm{\"u}ller context duality becomes trivial, in that
  $\mathbb{D}A \simeq 1$ for all objects $A$. To the extent that all the key constructions
  that we consider here in the following involve duality, they are interesting the ``less cartesian''
  (``less classical'') the ambient Wirthm{\"u}ller context is.
\end{remark}

We now consider some basic constructions in dependent linear type theory/Wirthm{\"u}ller contexts.
The key notion for the discussion to follow is
def. \ref{MeasureInducedByOrientation} below, which gives an abstract formalization of a notion of \emph{measure} against which one can perform an abstract kind of integration.

\subsection{Continuation modality and Dependent linear De Morgan duality}

In general linear logic there is no notion of negation, but its role is played by
duality. Just as negation intertwines ordinary logical conjunction and disjunction,
a basic fact called \emph{de Morgan duality}, so duality in linear logic intertwines
linear dependent sum with linear dependent product. This is prop. \ref{DualityIntertwinesLeftAndRigthtPushforward}
below. Since it is useful to freely pass back and forth along this linear de Morgan duality,
we here collect some basic constructions and facts.

\begin{definition}
  For $(\mathcal{C}, \otimes, [-,-], 1)$ a closed symmetric monoidal category,
  write
  $$
    \mathbb{D} := [-,1] : \mathcal{C}^{\mathrm{op}} \longrightarrow \mathcal{C}
  $$
  for the weak dualization functor.
\end{definition}
\begin{remark}
As usual, we say an object $X \in \mathcal{C}$ is dualizable if it has a (``strong'') dual
with respect to the tensor product $\otimes$, and we say $X$ is \emph{invertible}
if the unit and counit of the duality map are equivalences.
Generally $\mathbb{D}X$ is usually called the \emph{weak} dual of $X$. There is a canonical
natural morphism
$$
  X \longrightarrow \mathbb{D}\mathbb{D} X
$$
which exhibits the unit of the modality (monad) $\mathbb{D}^2(-)$,
also called the \emph{continuation monad} \cite{Mellies08}.
\end{remark}

\begin{proposition}
  For $f$ a Wirthm{\"u}ller morphism, def. \ref{WirthmuellerMorphism},
  the left and right adjoints are intertwined by weak duality in that
  there is a natural equivalence
  $$
    f_\ast \circ \mathbb{D} \simeq \mathbb{D} \circ f_!
    \,.
  $$
  hence
  $$
    \prod_f \circ \mathbb{D} \simeq \mathbb{D} \circ \sum_f
    \,.
  $$
  \label{DualityIntertwinesLeftAndRigthtPushforward}
\end{proposition}
This is a special case of a more general consequence of the axioms of Wirthm{\"u}ller morphisms
which appears in \cite{May05} as prop. 2.8, prop. 2.11. (In fact the ``sixth operation'', the internal hom
$[-,-]$, for us here only ever appears in its specialization to weak duality $\mathbb{D} = [-,1]$.)
\begin{remark}
  Prop. \ref{DualityIntertwinesLeftAndRigthtPushforward}
  is an incarnation of \emph{de Morgan duality} in linear logic. This involves in particular that
  if $A \in \mathcal{D}$ is dualizable, then the $(f^\ast \dashv f_\ast)$-unit $\eta^{f_\ast}$ on $A$ is the dual
  of the  $(f_! \dashv f^\ast)$-counit $\epsilon^{f_!}$ on the dual of $A$.
  $$
    \eta^{f_\ast}_A
    \simeq
    \mathbb{D}(\epsilon^{f_!}_{\mathbb{D}A})
    \,.
  $$
  \label{DualityIntertwines}
\end{remark}

\subsection{Linear polynomial functors and Primary integral transforms}

A central notion in type-semantics is the following:
\begin{definition}
  Given a model of linear homotopy-type theory $\mathrm{Mod} \to \mathbf{H}$,
  def. \ref{LinearHomotopyTypeTheory}, then
  a \emph{multivariate polynomial functor}
  $P : \mathrm{Mod}(X_1) \to \mathrm{Mod}(X_2)$ is
  a functor of the form
  $$
    P \simeq \sum_{f_2} \prod_g f_1^\ast
  $$
  for a given diagram in $\mathbf{H}$ the form
  $$
    \raisebox{20pt}{
    \xymatrix{
      & Y
      \ar[dl]_{f_1} \ar[r]^g
      &
      Z
      \ar[dr]^{f_2}
      \\
      X_1 && & X_2
    }}
    \,.
  $$
  If here $g\simeq \mathrm{id}$, hence if the diagram is a \emph{correspondence}
  $$
    \raisebox{20pt}{
    \xymatrix{
      & Z
      \ar[dl]_{f_1}
      \ar[dr]^{f_2}
      \\
      X_1 && X_2
    }
    }
    \,,
  $$
  then its polynomial functor $\underset{f_2}{\sum} f_1^\ast$
  is called a \emph{linear polynomial endofunctor}.
  \label{PolynomialFunctor}
\end{definition}
\begin{remark}
  In the existing type-theoretic literature the focus is on polynomial \emph{endo}functors, since the initial algebras
  over such endofunctors embody a useful notion of inductive types (``W-types'').
  Polynomial endofunctors in non-linear homotopy-type theory
(our example \ref{ToposBaseChangeIsCartesianWirthmueller})
have been considered in \cite{Kock12, vdBMo13}.
\end{remark}
\begin{remark}
  In the existing representation-theoretic literature, linear polynomial functors are
  known as \emph{categorified integral transforms}, as for instance in
  ``Fourier-Mukai transform'', ``Penrose transform'', ``Harish-Chandra transform'',
  see e.g. \cite{BZ08, BZ09}.
  Here we will call these \emph{primary} integral transforms for emphasis, since below
  in \ref{IntegralTransformations} our focus is on another concept of ``secondary''
  integral transform that will turn out to
  be ``boundaries'' for the primary transforms.
\end{remark}
Categorified integral transforms are often understood to have as correspondence space
the product space (fiber-product in the relative case) and given there not just by
pull-push but by pull-tensor-push. This relates to the above via the following
basic fact (see for instance also p. 10 of \cite{BZ09}).
\begin{proposition}
  Given a correspondence and its universal factorization through the
  product-space correspondence
  $$
    
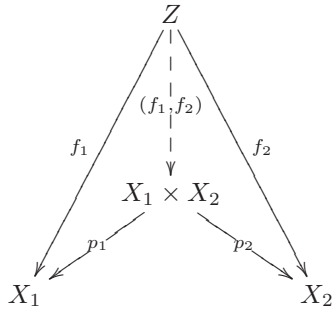
\xymatrix{
      & Z
      \ar[dddl]_{f_1}
      \ar[dddr]^{f_2}
      \ar@{-->}[dd]|{(f_1,f_2)}
      \\
      \\
      & X_1 \times X_2
      \ar[dl]|{p_1}
      \ar[dr]|{p_2}
      \\
      X_1
      &&
      X_2
    }
  $$
  then pull-push through the given correspondence is equivalent pull-tensor-push through the
  product-correspondence, for integral kernel given by
  $K := \underset{(f_1,f_2)}{\sum} 1_Z$:
  $$
    \underset{f_2}{\sum} \circ f_1^\ast
    \;\;
    \simeq
    \;\;
    \underset{p_2}{\sum}\circ (K \otimes(-)) \circ p_1^\ast
    \,.
  $$
\end{proposition}
\proof
  By Frobenius reciprocity, def. \ref{FrobeniusReciprocity}:
  $$
    \begin{aligned}
        (f_2)_! (f_1)^\ast A
        & \simeq
       (p_2)_! (f_1, f_2)_! (f_1,f_2)^\ast p_1^\ast A
       \\
       & \simeq
       (p_2)_! (f_1, f_2)_! (((f_1,f_2)^\ast (p_1^\ast A)) \otimes 1_{Z})
       \\
       &\simeq
       (p_2)_! (p_1^\ast A \otimes ((f_1, f_2)_! 1_Z))
       \\
       & =: (p_2)_! (p_1^\ast A \otimes K )
       \,.
    \end{aligned}
  $$
\endofproof
The terminology ``polynomial functor'' in def. \ref{PolynomialFunctor} is motivated from the following basic example.
\begin{example}
  Consider $\mathbf{H}$ the topos of sets and
  $\mathrm{Set}^{\Delta^1} \stackrel{\mathrm{cod}}{\to} \mathrm{Set}$ the associated
  dependent type structure, example \ref{ToposBaseChangeIsCartesianWirthmueller}.
  If we think of finite sets
  under their cardinality as representing natural numbers, then a polynomial functor
  with $X \simeq \ast$ the singleton acts indeed as a polynomial function under cardinality,
  summing up powers as given by the cardinalities of the fibers of $g$.
  In the other extreme, for general $X$ but with $g = \mathrm{id}$ then a polynomial
  functor is analogously given by multiplication by a matrix with entries in natural numbers.

  Similarly for $\mathrm{Vect}(-) \to \mathrm{Set}$ the linear type theory of
  vector spaces over sets, example \ref{WirthmuellerMorphsismsBetweenParameterizedModules},
  then for finite dimensional vector spaces over finite sets, a polynomial functor
  acts as a polynomial function on the \emph{dimensions} of these vector spaces.
  \label{PolynomialFunctorOnSets}
\end{example}
\begin{remark}
Example \ref{PolynomialFunctorOnSets} shows that the concept of polynomial functor is
a \emph{categorification} of that of polynomial function, hence a kind of
``higher dimensional'' polynomial function.
\end{remark}
In parallel to this, below in \ref{CoboundingTheory} we find that linear polynomial functors
in linear homotopy-type theory constitute the propagators
of $(d+1)$-dimensional topological quantum field theories between ``categorified''
spaces of states, and that they encode (the quantum anomaly
cancellation of) $d$-dimensional topological quantum field theories with propagators
acting between uncategorified spaces of states. Further discussion of this
phenomenon of similar structures appearing in different dimensions we give below in
\ref{Holography}.

\section{Computational homotopy-type theory}
 \label{ComputationalHomotopyTypeTheory}

We had seen that a modality $\bigcirc$ (a monad on the type system, def. \ref{modality}),
which is idempotent in that the monad operation $\bigcirc \bigcirc \stackrel{}{\to} \bigcirc$
is an equivalence, serves to encode qualities and ``moments'' of types, def. \ref{MomentAndComoment}.
If however the monad is not idempotent  then its successive applications equip types with
successive operations, hence with an ``evolution'' or ``computation''. That indeed logically
this captures computation with side effects in functional programming is the celebrated insight of
\cite{Moggi, Kobayashi}. Accordingly one sometimes refers to type systems equipped with a
(non-idempotent) monad $\bigcirc$ as ``computational type theory'' \cite{BBP, Mendler}.\footnote{
In the introduction of \cite{BBP} it is observed that this type-theoretic interpretation of
``modality'' in fact nicely harmonizes with the default interpretation of modal logic as being
about the modality $T$ of ``possibility'':
\begin{quote}
The starting point for Moggi's work is an explicit semantic distinction between {\it computations} and
{\it values}. If $A$ is an object which interprets the values of a particular type, then $T(A)$ is the object which models computation of that type $A$.
[...] For a wide variety of notions of computation, the unary operator $T(-)$  turns out to have the categorical structure of a {\it strong monad} on an underlying cartesian closed category of values.
[...] On a purely intuitive level and particularly if one thinks about non-termination, there is certainly something appealing about the idea that a computation of type $A$ represents the possibility of a value of type $A$.
\end{quote}
}

On the other hand, in the Schr{\"o}dinger picture of quantum physics, evolution is controlled by linear
functionals that arise under quantization as
(secondary) s induced by integral kernels. We now discuss a general abstract formalization
of such integral kernels and of their integral transformations in computational homotopy-type theory.
The non-idempotent monads that appear are those induced by adjoint triples of linear homotopy-types
as in \ref{LinearTypeTheory}.

Taken together, this means that the following may also be thought of as formalizing aspects
of quantum computation. Indeed, there is a vague similarity to the proposal in \cite{LagoFaggian}
of identifying the language for quantum computation to be (linear) type theory equipped with a suitable
monad.

\subsection{Motivation}

To motivate our abstract formalization to follow,
it is helpful to notice that the operation of pull-push of quasi-coherent sheaves of modules through correspondences
famously embodies a higher analog of integral transformations via integral kernels \cite{Pol08, BZ08};
the seminal historical example being the Fourier-Mukai transform.
On the other hand, another higher analog of integral transform theory is given by
pull-push in generalized twisted cohomology (twisted Umkehr maps \cite{ABG11}), as it appears for instance in the
construction of string topology operations (see example \ref{StringTopologyOperations} below)
and in the discussion of T-duality. What has apparently been lacking is the identification of a core general abstract
framework for integral transforms that would make both these applications be special cases
in a way that their main properties follow formally from general statements.
We will show that in a precise sense the latter notion is a ``boundary theory'' of the former,
hence may be called a ``secondary'' integral transform in analogy with the concept of secondary
characteristic classes.

At least a central ingredient of such a formalization should be a ``yoga of six operations''
that Grothendieck famously identified as the abstract mechanism underlying Grothendieck-Verdier duality.
A clean analysis of the possible flavors of this six-operations axiomatics and their corresponding implications
has been given in \cite{May05}. The central formal consequence highlighted there is the
``Wirthm{\"u}ller isomorphism'' which says, when it applies, that pull-push respects dualization,
up to, possibly, a certain natural twist.

We observe below, in example \ref{PullPushInGeneralizedCohomology} following \cite{Nuiten13},
that it is precisely this statement which also controls the twisted Umkehr maps in generalized
cohomology theory \cite{ABG11},
using that these Umkehr maps are built by first dualizing (Poincar{\'e} duality), then pushing, then
dualizing back, possibly picking up a twist thereby.

Motivated by this example, in this note we are after identifying a general abstract (higher) algebraic
formalization of (secondary) integral kernels and integral transformations.

Specifically we are after a formalization
that may mesh well with axiomatic cohesion. In the context of axiomatic cohesion
one regards the ambient category of spaces and (principal) bundles over them as forming an
indexed cartesian closed monoidal category
equipped with a system (a ``yoga'') of base change morphisms (hyperdoctrine) and equipped with some monads used to express
the intended modalities characterizing such spaces. Below it turns out that we are working
essentially in a linear monoidal (``tensor'') or
``quantum'' variant of this setup, where instead of an indexed topos we have an
indexed closed monoidal linear category (as in \cite{Shulman12}, theorem 2.14). Analogous to axiomatic cohesion
we find that the yoga of adjoint base change functors
becomes most natural and transparent when expressed in terms of (just) the (co-)monads induced by these
adjunctions.

\subsection{Exponential modality, Linear spaces of states and Fock space}
\label{ExponentialConjunction}

The secondary integral transforms that we discuss below in \ref{IntegralTransformations}
act on what in the typical model of linear homotopy-type theory are linear spaces of sections
of bundles of modules. In the application to quantization in \ref{Quantization} these are
going to be thought of as spaces of quantum states. Here we discuss
how the concept of forming spaces of sections is naturally captured by the
\emph{exponential modality} of linear logic, lifted to
linear homotopy-type theory.

\medskip

The full set of axioms for linear logic as introduced in \cite{Girard}
contains -- on top of the ``multiplicative fragment'' discussed above in \ref{LinearTypeTheory}
which is interpreted in
($\ast$-autonomous/closed) symmetric monoidal categories -- a co-modality (def. \ref{modality})
denoted ``$!$'' and called the \emph{exponential modality}.
The axioms on $!$ are roughly meant to be such that if $A$ is a linear type, then $!A$ is the linear type
obtained from it by universally equipping it with properties of a non-linear type.
More precisely,
by the general categorical semantics of (co-)modalities recalled above in def. \ref{modality},
the exponential co-modality is to be interpreted as a some co-monad on the type system \cite{BBHdP},
and the axioms on $!$ are such as to make its co-Kleisli category of co-free co-algebras
be cartesian monoidal
\cite{Seely89} (see around prop. 17 of \cite{Biermann} for more discussion).

Based on this in \cite{Benton,Biermann} it is observed that generally the exponential modality
is naturally interpreted as a comonad induced specifically from a strong monoidal adjunction
between the given symmetric monoidal category $\mathrm{Mod}(\ast)$ of linear types and some
given cartesian
closed monoidal category $\mathbf{H}$ (representing a non-linear type system):
$$
  (L \dashv R)
  \;:\;
  \xymatrix{
    \mathrm{Mod}(\ast)^\otimes
    \ar@{<-}@<+4pt>[rr]^-{L}
    \ar@<-4pt>[rr]_-{R}
    &&
    \mathbf{H}^\times
  }
  \;\;\;\;
  ! := L \circ R
  \,.
$$
(If only a strong monoidal functor $L$ like this exists without necessarily a strong monoidal right
adjoint, then \cite{Barber} speaks of the ``structural fragment'' of linear logic.)

Finally in \cite{PontoShulman} (4.3) it is observed that if $\mathrm{Mod}(\ast)$ here is the fiber over the
point of a linear type system $\mathrm{Mod}$ dependent over $\mathbf{H}$, as in \ref{DependentLinearTypeTheory},
then, as our notation already suggests, there is a canonical choice for $L$
induced from the dependent type structure, namely the map that sends $Y \in \mathbf{H}$ to the
$Y$-dependent sum of the unit linear type $1_Y \in \mathrm{Mod}(Y)$:
$$
  L : Y \mapsto \underset{Y}{\sum} 1_Y
  \,.
$$
(On morphisms this $L$ is given by the adjunction counit, we see this below in example
\ref{PontoShulmanOperation} as a special case of the general secondary integral transform formula).
More generally, the dependent linear homotopy-type theory induces for each $X \in \mathbf{H}$
a functor
$$
  L_X \;:\; \mathbf{H}_{/X} \longrightarrow \mathrm{Mod}(X)
$$
given by summing the unit linear type along the fibers of $f$:
$$
  L_X \;:\; (Y \stackrel{f}{\to} X) \mapsto \underset{f}{\sum} 1_Y
  \,.
$$
In this dependent form one recognizes this as the linear version of the operation considered in
section 2 (p. 12) of \cite{LawvereComp}. There the condition that these functors have right adjoints
$R_X$ is found to be the categorical semantics of the foundational \emph{axiom of comprehension}
(\emph{axiom of separation}). Therefore one may say that dependent linear type theory
carries a canonical $!$-modality precisely if it satisfies the linear version of the comprehension axiom.
Since the comprehension axiom in foundations is typically taken for granted, this shows that
the existence of the exponential modality is a rather fundamental phenomenon.

\medskip

In \cite{BPS} it had been found that in familiar models of (multiplicative) linear type theory such
as in the category of vector spaces, the exponential sends a vector space to its
\emph{Fock space}, the vector space underlying the free symmetric algebra on the given space.
This construction is manifestly a categorified exponential.
Now in the simplistic 1-categorical model given by vector spaces over sets, example \ref{WirthmuellerMorphsismsBetweenParameterizedModules},
the left adjoint $L= \sum 1_{(-)}$ canonically induced as above from the dependent type
structure
sends a set to the vector space it spans, $R$ sends a vector space
to its underlying set of vectors, and $! = L \circ R$ hence sends a vector space not quite to
its Fock space, but to the space freely spanned by all the original vectors.
(Another adjunction for vector spaces that does make $!$ produce exactly the Fock space is discussed in \cite{Vicary}.)

More interestingly;
\begin{example}
In the genuinely homotopy-theoretic model of linear homotopy-type theory
given by $E$-module spectra over $\infty$-groupoids, example \ref{LocalSystems},
then $L = E \wedge \Sigma_+^\infty$
sends a homotopy-type to its suspension spectrum, and $R = \Omega^\infty$ sends a spectrum
to its underlying infinite-loop space.
\label{SuspensionSpectrumFromL}
\end{example}
There is a deep sense in which stable homotopy
theory is analogous to linear algebra, namely Goodwillie's calculus of functors
(see section 7 of \cite{LurieAlg}), and
it has been argued in \cite{Arone} that from this point of view at least $\Omega^\infty \circ \Sigma^\infty_+$
is indeed the analogue of the exponential function.\footnote{
I am grateful to Mike Shulman and to David Corfield for highlighting this point and
in fact for driving it home with some patience.}

In conclusion we find that the exponential modality $!$ in linear type theory, when
implemented in linear homotopy-type theory naturally decomposes into an
adjunction who left adjoint is the process of forming spaces of sections, hence quantum states
(and whose induced comonad encodes free second quantization).

\medskip

Below in \ref{Quantization} we find that the process of quantization of an
action functional $\exp(\tfrac{i}{\hbar}S)$ subject to a consistent (anomaly free) choice of
path integral measure $d\mu$ is given by a twisted variant of the canonical left adjoint $L$ as above:
we find there that a choice of action functional comes with a choice of an assignment of
invertible linear homotopy-types $A_X \in \mathrm{Mod}(X)$
(thought of as ``dual prequantum line bundles'') to homotopy-types $X \in \mathbf{H}$,
and the quantization process sends these to their space of dual sections:
$$
  X \mapsto A_X \mapsto \underset{X}{\sum} A_X \in \mathrm{Mod}(\ast)
  \,.
$$
For later reference we note at this point that
\begin{remark}
If we think of $A_X$ as a linear bundle over $X$
(as in example \ref{BaseChangeForModules} via example \ref{ModelsForInfinitesimalExtensions})
then its $X$-dependent product
is to be thought of as its linear space of sections
$$
  \Gamma_X(A) := \underset{X}\prod A
  \,.
$$
If $A$ here is dualizable with dual $\mathbf{L} := \mathbb{D}A$, then by dependent linear de Morgan duality,
prop. \ref{DualityIntertwinesLeftAndRigthtPushforward} this is equivalently
the linear dual
$$
  \Gamma_X(\mathbf{L}^\ast) \simeq \mathbb{D} \underset{X}{\sum} \mathbf{L}
  \,.
$$
In the standard models of linear homotopy-type theory such as that of
definition \ref{LocalSystems}, the dependent sum $\underset{X}{\sum}\mathbf{L}$ has
the interpretation of the \emph{compactly supported} sections of $\mathbf{L}$
(this is of course the default interpretation of $\sum$ in the traditional yoga of six functors,
see the citations in \cite{May05}).
Therefore the linear dual $\mathbb{D} \underset{X}{\sum} \mathbf{L}$ is interpreted as
the space of distributional sections. In a general Verdier-Grothendieck context of six functors these may be
different from the genuine sections (of the dual bundle), but here in the Wirthm{\"u}ller context
they coincide.
\label{LinearSpaceOfSections}
\end{remark}

In the terminology of \cite{Lawvere86} the type
of states $\underset{X}{\prod} \mathbf{L}^\ast$
would be ``intensive'', while $\underset{X}{\sum}\mathbf{L}$ would be ``extensive''.

\medskip

We now turn to the definition of fundamental classes that allow to integrate such sections against
a measure and then we use this to define secondary integral transforms acting on spaces of sections.

\subsection{Fundamental classes and Measures}
 \label{FundamentalClasses}

We discuss here how to axiomatize basic measure theory in dependent homotopy-type theory,
such as to define integrals of sections of linear types (remark \ref{LinearSpaceOfSections})
along maps of contexts that are equipped with a measure.
In terms of linear logic and linear type theory the construction here is a variant and generalization of the
duality structure reflected by the orthocomplementarity in the original BvN quantum logic
\cite{BirkhoffVonNeumann} and more generally in ``dagger-structure'' in linear type theory.
We postpone discussion of this to \ref{DaggerCompactness}. Let throughout
$$
  \xymatrix{
    \mathrm{Mod}(-)
    \ar[d]
    \\
    \mathbf{H}
  }
$$
be a model for linear homotopy-type theory, def. \ref{LinearHomotopyTypeTheory}.

\begin{definition}
 A \emph{fiberwise twisted fundamental class} on a morphism $f : X \longrightarrow Y$
 in $\mathbf{H}$
 is (if it exists) a choice of dualizable object $\tau \in \mathrm{Mod}(Y)$ (the twist) such that
 $f_! f^\ast \mathbb{D}\tau$ is dualizable, together with a  choice of
 equivalence of the form
 $$
   f_! f^\ast (1_Y) \stackrel{\simeq}{\longrightarrow}
   \mathbb{D}(f_! f^\ast (\mathbb{D}\tau))
   \simeq
   f_\ast f^\ast (\tau)
   \,.
 $$
 \label{TwistedOrientationOnWirthmuellerMorphism}
\end{definition}
\begin{remark}
  This is a specialization of the assumption in (4.3) of \cite{May05}.
  There it is emphasized that for many constructions
  the assumption that $f_! f^\ast 1_X$ be dualizable is, while typically
  verified in applications, not necessary. However, it
  is necessary for the definition of secondary integral transforms in def. \ref{IntegralKernelAndTransform}
  below, and therefore we do require it.
  \label{SpecializationOfTwists}
\end{remark}
\begin{remark}
  If here $Y \simeq \ast$ is the terminal object, so that $f : X \to \ast$ is the essentially
  unique terminal morphism, then we usually denote a fundamental class on this morphism
  by $[X]$.
\end{remark}
\begin{proposition}[Wirthm{\"u}ller isomorphism]
  Given a fiberwise fundamental class, def. \ref{TwistedOrientationOnWirthmuellerMorphism},
  on a morphism $f : X \to Y$, then
  for dualizable $A \in \mathrm{Mod}(Y)$ there is a canonical natural equivalence
  $$
    f_\ast f^\ast \mathbb{D}A \simeq \mathbb{D}(f_! f^\ast A)
       \stackrel{\simeq}{\longrightarrow}
    f_! f^\ast( (\mathbb{D}A) \otimes (\mathbb{D}\tau) )
  $$
  and hence a canonical natural transformation
  $$
    f_! f^\ast A
       \longrightarrow
    \mathbb{D}(f_! f^\ast( (\mathbb{D}A) \otimes (\mathbb{D}\tau) ))
    \simeq f_\ast f^\ast \mathbb{D}((\mathbb{D}A) \otimes (\mathbb{D}\tau) ))
  $$
  which is an equivalence if $f_! f^\ast A$ is dualizable.
  \label{TwistedOrientationMakesComonadTwistedCommutedWithDuality}
\end{proposition}
\proof
  The canonical map itself is the composite
  $$
    \xymatrix{
    f_\ast f^\ast \mathbb{D}A
    \ar[d]^\simeq
    \\
    (f_\ast f^\ast \mathbb{D}A) \otimes (\mathbb{D} 1_Y)
    \ar[r]
    &
    (f_\ast f^\ast \mathbb{D}A) \otimes (\mathbb{D} f_! f^\ast 1_Y)
    \ar[d]^\simeq_{}
    \\
    &
    ( f_\ast f^\ast \mathbb{D}A) \otimes ( f_! f^\ast \mathbb{D} \tau)
    \ar[d]^\simeq
    \\
    &
    f_!  (( f^\ast f_\ast f^\ast \mathbb{D}A) \otimes ( f^\ast \mathbb{D} \tau))
    \ar[r]
    &
    f_!  (( f^\ast \mathbb{D}A) \otimes ( f^\ast \mathbb{D} \tau))
    \ar[d]^\simeq
    \\
    &&
    f_! f^\ast  ( \mathbb{D}A \otimes \mathbb{D} \tau)
    }
    \,,
  $$
  where the first is the unit equivalence for $1_Y \simeq \mathbb{D}1_Y$,
  the second is tensoring with the dual of the $(f_! \dashv f^\ast)$-counit,
  the third is tensoring with the dual of the defining equivalence of a fundamental class,
  the fourth is the projection formula of Frobenius reciprocity, def. \ref{FrobeniusReciprocity},
  the fifth comes from the $(f^\ast \dashv f_\ast)$-counit and the last one finally
  is the strong monoidalness of $f^\ast$.

  That this total composite is an equivalence is prop. 4.13 in \cite{May05},
  specialised to twists of the form as in def. \ref{TwistedOrientationOnWirthmuellerMorphism},
following remark \ref{SpecializationOfTwists}.
\endofproof
\begin{remark}
  If the twist in prop. \ref{TwistedOrientationMakesComonadTwistedCommutedWithDuality} vanishes
  (is the tensor unit) then a Wirthm{\"u}ller isomorphism means that $f_!$ coincides with $f_\ast$
  on objects in the image of $f^\ast$. Since $f_!$ is the left adjoint and $f_\ast$ the right
  adjoint of $f^\ast$, this means that in this situation $f^\ast$ has a two-sided adjoint, hence an
  ``ambidextrous'' adjoint. The condition $f_! \simeq f_\ast$ together with a further coherence condition is called
  ``ambidexterity'' in Construction 4.1.8 of \cite{HopkinsLurie}.
\end{remark}

The central construction obtained from a Wirthm{\"u}ller-type six-operations context that we need
below is now the following.\footnote{While this text was being composed,
essentially def. \ref{MeasureInducedByOrientation} for the special case of vanishing twist and in the specific model of $\infty$-module bundles
over homotopy types appeared as Construction 4.0.7 and Notation 4.1.6 of \cite{HopkinsLurie}.}
  Let $f \;:\; X \longrightarrow Y$ be a morphism of contexts in
  a linear homotopy-type theory
  with associated base change
  Wirthm{\"u}ller morphism $(f_! \dashv f^\ast \dashv f_\ast) = (\sum_X \dashv f^\ast \dashv \prod_X)$, def. \ref{WirthmuellerMorphism},
\begin{definition}
  Given a fiberwise fundamental class, def. \ref{TwistedOrientationOnWirthmuellerMorphism},
  on $f$
  and given $A \in \mathrm{Mod}(Y)$ a dualizable object
  such that also $f_! f^\ast A$ is dualizable,
  write
  $$
    [f] \;\colon\; (-)\otimes \tau \longrightarrow f_! f^\ast (-)
  $$
  for the natural transformation given as the composite
  $$
    [f]_A
    \;:\;
    \xymatrix{
      A \otimes \tau
      \ar[rr]^-{\mathbb{D}(\epsilon_{\mathbb{D}(A \otimes \tau)})}
      &&
      \mathbb{D} f_! f^\ast ( \mathbb{D}(A \otimes \tau) )
      \ar[r]^-{\simeq}
      &
      f_! f^\ast A
    }
    \,,
  $$
  where the first morphism is the dual of the $(f_!\dashv f^\ast)$-counit
  on $\mathbb{D}(A \otimes \tau)$, and where the second morphism
  is the equivalence of prop. \ref{TwistedOrientationMakesComonadTwistedCommutedWithDuality}.

  We will usually also refer to $[f]$ as the fundamental class, whence the notation.

  The dual of the $Y$-dependent sum of the fundamental class we call the induced \emph{measure} on $f$ and write
  $$
    d\mu_f(A) := \mathbb{D}\left(\underset{Y}{\sum}[f]_A\right)
    \;:\;
    \mathbb{D}\left(\underset{Y}{\sum} f_!f^\ast A\right)
    \longrightarrow
    \mathbb{D}\left(\underset{Y}{\sum} A\otimes \tau\right)
    \,.
  $$
 \label{MeasureInducedByOrientation}
 \end{definition}
 \begin{remark}
     By remark \ref{DualityIntertwines} the fundamental class
     in def. \ref{MeasureInducedByOrientation} is equivalently
     the composite
  $$
    [f]_A
    \;:\;
    \xymatrix{
      A \otimes \tau
      \ar[rr]^{\eta_{A \otimes \tau}}
      &&
      f_\ast f^\ast (A \otimes \tau)
      \ar[r]^-\simeq
      &
      f_! f^\ast A
    }
    \,.
  $$
 In this form the fundamental class here is manifestly related to
 what appears in remark 4.1.7 of \cite{HopkinsLurie}, we come back to this below in
 example \ref{Ambidexterity}.
 \label{FundamentalClassIsDualUnit}
 \end{remark}

\begin{remark}
 The fundamental class morphism in def. \ref{MeasureInducedByOrientation}
 is reverse to the $(f_! \dashv f^\ast)$-counit.
 We may think of it as the ``Umkehr map'' of the counit
 and find in the following that this naturally induces Umkehr maps for
 other morphisms.
\end{remark}
\begin{remark}
  A measure on a map $f : X \to Y$ in def. \ref{MeasureInducedByOrientation} is to be thought
  of as a $Y$-parameterized collection of measures on all of the (homotopy-)fibers of the map.
  This is explicitly so in the internal linear logic, in which the fundamental class
  reads
  $$
    y : Y,\; A(y) : \mathrm{Type}
      \;\;\vdash\;\;
    [f](y) : A(y)\otimes \tau(y) \to \!\!\!\underset{x \in f^{-1}(y)}{\sum} A(f(x))
    \,.
  $$
  Externally this fiberwise property is directly visible in the model of dependent homotopy type theory given by
  bundles of spectra over $\infty$-groupoids, this is prop. 4.3.5 in \cite{HopkinsLurie}.
\end{remark}

\subsection{Correspondences and Secondary integral transforms}
 \label{IntegralTransformations}

We discuss now how a correspondence of contexts in linear homotopy-type theory
which is equipped with a fiberwise fundamental class on its right leg and with a linear
map between linear homotopy-types pulled back to its corrrespondence space
(a secondary integral kernel)
naturally induces a secondary integral transform.

Examples include ordinary matrices in linear algebra, example \ref{MatrixAsExampleOfAbstractIntegralKernel},
pull-push in twisted generalized
cohomology by twisted Pontryagin-Thom Umkehr maps, example \ref{PullPushInGeneralizedCohomology}
and in particular the ``ambidexterity'' in stable homotopy theory of \cite{HopkinsLurie},
example \ref{Ambidexterity}.

\medskip

Throughout, let $\mathrm{Mod}$ be a linear homotopy-type theory.
A secondary linear integral transform is supposed to be a linear function
between linear spaces of sections, remark \ref{LinearSpaceOfSections}, which is induced from an integral kernel
or matrix given by a linear map between linear bundles $L$ over some correspondence space.
\begin{definition}
  Given a dependent linear homotopy-type theory $\mathrm{Mod}$, then
  a \emph{prequantum integral kernel} is a correspondence
  $$
    \xymatrix{
      & Z
       \ar[dl]_{i_1}
       \ar[dr]^{i_2}
      \\
      X_1 && X_2
    }
  $$
  of contexts -- the \emph{arity} --
  together with linear types $A_1 \in \mathrm{Mod}(X_1)$ and $A_2 \in \mathrm{Mod}(X_2)$
  -- the \emph{coefficients} --
  and a linear function of the form
  $$
    \xi \;:\; i_1^\ast A_1 \longleftarrow i_2^\ast A_2
    \,,
  $$
  the \emph{integral kernel} itself.
  A \emph{quantum integral kernel} or \emph{amplimorphism} is a correspondence
  and linear types as above and a morphism
  $$
    \Xi \;:\; \underset{Z}{\sum} i_1^\ast A_1 \longleftarrow \underset{Z}{\sum} i_2^\ast A_2
    \,.
  $$
  \label{IntegralKernel}
\end{definition}
\begin{remark}
So if $\xi$ is a pre-quantum integral kernel then $\Xi := \underset{Z}{\sum} \xi$
is the corresponding quantum integral kernel.
\end{remark}
Further below in prop. \ref{IntegralKernelAsTransformation} we find a more abstract,
more conceptual origin of prequantum integral kernels. For the moment
we are content with pointing out
that a typical source of prequantum integral kernels are correspondences
dependent on a context of moduli for certain linear types \cite{dcct, Nuiten13}:
\begin{example}
  Let $B$ be some base context and $V \in \mathrm{Mod}(V)$ a $B$-dependent linear type.
  Then every correspondence in $\mathbf{H}_{/B}$ canonically induces an
  invertible prequantum integral kernel, def. \ref{IntegralKernel} as follows.
  In $\mathbf{H}$ (under dependent sum) the correspondence in $\mathbf{H}_{/B}$
  is a diagram of the form
  $$
    \raisebox{20pt}
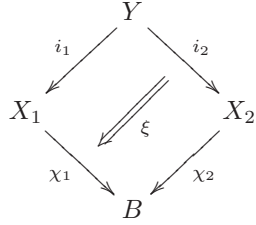
{
    \xymatrix{
      & Y
      \ar[dr]^{i_2}_{\ }="s"
      \ar[dl]_{i_1}
      \\
      X_1
      \ar[dr]_{\chi_1}^{\ }="t"
      && X_2 \ar[dl]^{\chi_2}
      \\
      & B
      \ar@{=>}^\xi "s"; "t"
    }
    }
    \,.
  $$
  Hence putting
  $$
    A_i := \chi_i^\ast V
  $$
  gives the prequantum integral kernel
  $$
    \xi : (i_1)^\ast A_1 \stackrel{\simeq}{\longleftarrow} (i_2)^\ast A_2
    \,.
  $$
    \label{PrequantumIntegralKernelFromPrequantumCorrespondence}
\end{example}
\begin{example}
  For
  $$
    \xymatrix{
      & Z
       \ar[dl]_{i_1}
       \ar[dr]^{i_2}
      \\
      X_1 && X_2
    }
  $$
  a correspondence in $\mathbf{H}$ and $A_2 \in \mathrm{Mod}(X_1)$
  any linear type, then setting
  $$
    A_1 := \underset{i_1}{\sum}(i_2)^\ast A_1
  $$
  yields a non-invertible prequantum integral kernel, where
  $$
    \xi = \eta_{(i_2)^\ast A_1} \;:\; (i_1)^\ast \underset{i_1}{\sum}(i_2)^\ast A_1  \longleftarrow (i_2)^\ast A_1
  $$
  is the unit of the $(\underset{i_2}{\sum} \dashv (i_2)^\ast)$-adjunction. This we may
  call the universal (non-invertible) pre-quantization of the original correspondence
  and the given $A_1$.
\end{example}
\begin{example}
  In the model of linear homotopy-type theory given by an $E_\infty$-ring $E$
  as in example \ref{LocalSystems}
  $$
    \xymatrix{
      E\mathrm{Mod}(-)
      \ar[d]
      \\
      \mathrm{Grpd}_\infty
    }
  $$
  a function between two linear types is a cocycle in bivariant generalized $E$-cohomology.
  Therefore in this case a prequantum integral kernel as in def. \ref{IntegralKernel} is
  a correspondence equipped with a cocycle on its correspondence space. This is, broadly, the
  structure of \emph{motives}. Indeed, we see below in \ref{PullPushInTwistedGeneralizedCohomology}
  that the secondary integral transform in $E\mathrm{Mod}$ for $E = \mathrm{KU}$ may be given by
  KK-theory classes which were argued by Alain Connes to be the K-theoretic analog of
  motives, a point of view that has been made precise in \cite{Snig}.
  We discuss this analogy a bite more below in \ref{Motives}.
\end{example}

In order to apply an integral kernel as a linear map to the spaces of sections of its
coefficient bundles, the idea is to ``pull'' these sections up along one of the two legs of
the correspondence, apply there the map that defines the integral kernel, and then ``push''
the result down along the other leg. Notice that:
\begin{remark}
  For $\mathbb{S} \in \mathrm{CRing}_\infty$ the sphere spectrum regarded as an $E_\infty$-ring
  let $\mathbb{S}\mathrm{Mod}(-)$ be the corresponding model of linear homotopy-type theory
  from example \ref{LocalSystems}.
  Given $f : X \longrightarrow Y$ a morphism in $\infty \mathrm{Grpd}$ then forming the
  suspension spectra yields a morphism of the form
  $$
    \Sigma_+^\infty f  : \Sigma_+^\infty X \longrightarrow \Sigma_+Y
    \,.
  $$
  As in example \ref{SuspensionSpectrumFromL} we have that
  $$
    \Sigma_+^\infty X
    \simeq
    \underset{X}{\sum} 1_X
    \,,
  $$
  where now $1_X$ is the trivial spherical fibration (trivial $\mathbb{S}$-line bundle)
  over $X$.
  Under this identification the above morphism is given by the $(\underset{X}{\sum} \dashv X^\ast)$-counit
  $\epsilon$:
  $$
    \Sigma_+^\infty X
    \simeq
    \underset{X}{\sum} 1_X
    \simeq
    \underset{Y}{\sum} \underset{f}{\sum} f^\ast 1_Y
    \stackrel{\underset{Y}{\sum} \epsilon_{1_Y}}{\longrightarrow}
    \underset{Y}{\sum} 1_Y
    \simeq
    \Sigma_+^\infty Y
    \,.
  $$
  Generally for $E \in \mathrm{CRing}_\infty$ any $E_\infty$-ring, then this construction
  yields the map that is called ``pushforward in generalized $E$-homology'' along $f$
  $$
    E_\bullet(X) \longrightarrow E_\bullet(Y)
    \,.
  $$
  The image of this under dualization $\mathbb{D}$ is the ``pullback in generalized $E$-cohomology''
  along $f$
  $$
    E^\bullet(Y) \longrightarrow E^\bullet(X)
    \,.
  $$
  Beware that these operations are often denoted by ``$f_\ast$'' and ``$f^\ast$'', respectively,
  but that for us these symbols denote push/pull not of sections but of the $E$-module bundles themselves,
  and that sections are pull/pushed instead via the (dual of) the counit, as above.
\end{remark}
Our central notion is now the following, which generalizes the above
to general linear homotopy-type theory, general twists and combines it with Umkehr maps
in order to produce a secondary ``pull-tensor-push integral transform'' on cohomology.
\footnote{While this text was being composed, essentially
def. \ref{IntegralKernelAndTransform} for the special case of vanishing twist and in the specific model of $\infty$-module bundles over homotopy types appeared as Notation 4.1.6 of \cite{HopkinsLurie}.}
\begin{definition}
  Given a prequantum integral kernel $\xi$ or quantum kernel $\Xi$
  as in def. \ref{IntegralKernel} and a
  fiberwise fundamental class,
  def. \ref{TwistedOrientationOnWirthmuellerMorphism}, on the right leg $i_2$,
  with induced fundamental class $[i_2]$, def. \ref{MeasureInducedByOrientation}, then
  we say that the morphism
$$
  \mathbb{D}
  \int_{Z} \Xi\, d\mu_{i_2}
  \;\colon\;
  \xymatrix{
    \underset{X_1}{\sum} A_1
    \ar@{<-}[r]^-{\underset{X_1}{\sum} \epsilon_{A_1}}
    &
    \underset{X_1}{\sum} (i_1)_! (i_1)^\ast A_1
    \ar@{<-}[r]^-{\simeq}
    &
    \underset{Z}{\sum}(i_1)^\ast A_1
    \ar@{<-}[r]^-{\Xi}
    &
    \underset{Z}{\sum} (i_2)^\ast A_2
    \ar@{<-}[r]^-{\simeq}
    &
    \underset{X_2}{\sum} (i_2)_! (i_2)^\ast A_2
    \ar@{<-}[r]^-{\underset{X_2}{\sum} [i_2]_{A_2}}
    &
    \underset{X_2}{\sum} A_2 \otimes \tau
  }
$$
is the induced \emph{dual secondary integral transform}.
The dual morphism
$$
  \int_{Z} \Xi \, d\mu_{i_2}
  \;:\;
  \mathbb{D}\underset{X_1}{\sum}A_1
  \longrightarrow
  \mathbb{D}\underset{X_2}{\sum}(A_2 \otimes \tau)
$$
we call the corresponding \emph{secondary integral transform}.
 \label{IntegralKernelAndTransform}
\end{definition}
\begin{example}
  Consider the simple case of a prequantum integral kernel
  whose underlying correspondence has as right leg an identity
  $$
    \raisebox{20pt}{
    \xymatrix{
      & X
      \ar[dr]^{\mathrm{id}}
      \ar[dl]_{d}
      \\
      Y && X
    }
    }
    \,,
  $$
  where the linear types on the base spaces are the unit types and the integral kernel itself
  is the identity $\xi = \mathrm{id} : f^\ast 1_Y = 1_X \to 1_X = \mathrm{id}^\ast 1_X$.
  Then the right leg $\mathrm{id}_X$ is trivially oriented with vanishing twist and with this choice
  the secondary integral transform formula in def. \ref{IntegralKernelAndTransform} reduces to
  to being map
  $$
    \mathbb{D}\int_X d\mu_{\mathrm{id}}
    \;:\;
    \underset{Y}{\sum} 1_Y
     \stackrel{\underset{Y}{\sum} \epsilon}{\longleftarrow}
    \underset{Y}{\sum} f_! f^\ast 1_X
     \stackrel{\sim}{\longleftarrow}
    \underset{X}{\sum} 1_X
    \,.
  $$
  This we recognize as the operation considered in (4.3) of \cite{PontoShulman}.
  We had discussed the meaning of this operation in \ref{ExponentialConjunction}
  above.
  \label{PontoShulmanOperation}
\end{example}
\begin{remark}
  If the coefficients $A_1$ and $A_2$ in def. \ref{IntegralKernelAndTransform} are dualizable
  with duals $L_1$ and $L_2$, respectively,
  then by linear de Morgan duality, prop. \ref{DualityIntertwinesLeftAndRigthtPushforward},
  and by remark \ref{LinearSpaceOfSections}, the secondary integral transform
  of def. \ref{IntegralKernelAndTransform}
  is a linear function between the linear
  spaces of sections of the dual coefficients:
  $$
    \int_{Z} \Xi \, d\mu_{i_2}
    \;:\;
    \Gamma_{X}(L_1)
    \longrightarrow
    \Gamma_Y(L_2 \otimes \mathbb{D}\tau)
    \,.
  $$
\end{remark}
Example \ref{MatrixAsExampleOfAbstractIntegralKernel} below
shows how basic linear algebra is a special case of
def. \ref{IntegralKernelAndTransform}. This is elementary in itself, but turns out to be
directly the blueprint for the more sophisticated example \ref{PullPushInGeneralizedCohomology} to follow,
which in turn is the context in which one finds genuine quantum physics by example \ref{KUBoundaryQuantization}.
In view of these examples, we make the following observation on the conceptual
interpretation of the construction in def. \ref{IntegralKernelAndTransform},
which the reader with no tolerance for more philosophical considerations is
urged to skip and ignore.
\begin{remark}[logical interpretation of the secondary integral transform]
  \label{SuperpositionAsLinearDependentSum}
  These examples show that we may think of $X$ and $Y$ in
  def. \ref{IntegralKernelAndTransform} as phase spaces
  and, if $A_1$ and $A_2$ are dualizable, think of the linear types $L_1$ and $L_2$ as
  pre-quantum line bundles on these
  (see also section 1.2.10 in \cite{dcct}, surveyed in \cite{Schreiber13}).
  Hence by the BHK correspondence (as reviewed in \ref{NotionsJudgementDeduction}), in the underlying linear logic
  $L_1$ represents a proposition about elements of $X$ (and $L_2$ about elements of $Y$):
  $L_1$ may be thought of as the linear proposition that the given system
  is in state $x \in X$ of its phase space. For a proposition in classical
  logic the fiber of $L_1$ over some $x \in X$ would be either empty or inhabited,
  indicating that the system either is in that state or not. Now in linear logic this
  fiber is a linear space, namely what in physics is a space of \emph{phases}.
  In this vein we have the following stages of interpreting the expression
  $\underset{X}{\sum}L_1$:
  \begin{enumerate}
    \item in logic this expression is the existential quantification $\underset{x \in X}{\exists} L_1(x)$
    asserting that ``there is a state $x$ occupied by the physical system'';
    \item in type theory this expression denotes the collection (type) of all states that the system can be in;
    \item in homotopy-type theory this expression denotes the homotopy-type of all such states, hence properly taking
    their gauge equivalences and higher gauge equivalences into account;
    \item finally in linear homotopy-type theory this expression is the \emph{linear} space of
    all states (with gauge equivalence taken into account) obtained not by disjointly collecting them
    all but by linearly adding up their phases.
  \end{enumerate}
  An analogous comment applies to the middle terms in the composite function in
  \ref{IntegralKernelAndTransform}, $\Xi = \underset{Z}{\sum}\xi$. Here now the correspondence space
  $Z$ is to be interpreted as a space of paths (trajectories) from $X$ and $Y$, with $z \in Z$
  being a path going from $p_1(z) \in X$ to $p_2(z) \in Y$. Hence in analogy to the above we have that
  $\underset{Z}{\sum} \xi$ has the following interpretations:
  \begin{enumerate}
    \item in logic it means ``that there is a path'';
    \item in type theory it means ``the collection of all paths'';
    \item in homotopy-type theory it means ``the collection of all paths with gauge transformations accounted for'';
    \item finally in linear homotopy-type theory it means ``the sum of the phases of all possible paths''.
  \end{enumerate}
\end{remark}

\subsection{Dagger-structure, Fiberwise inner products and Quantum operations}
 \label{DaggerCompactness}

Above in \ref{LinearLogic} we discussed how quantum logic is linear logic, the logic
of closed symmetric monoidal categories. For
core constructions in quantum physics and quantum computation, one considers an additional
structure on these categories, namely what is called
a \emph{strongly compact} \cite{AbramskyCoecke}
or \emph{dagger-compact} ($\dagger$-compact) structure \cite{Selinger}.
Here we discuss how the concept of fundamental classes
in dependent linear type theory that we introduced in \ref{FundamentalClasses}
naturally induces $\dagger$-structure in the special case where the
twist vanishes. Conversely, we may hence regard the concept of
fundamental classes in def. \ref{TwistedOrientationOnWirthmuellerMorphism}
as a generalization of $\dagger$-structure.

\begin{remark}
  In the special case that the twist $\tau$ in
  a fiberwise fundamental class on $f$, def. \ref{TwistedOrientationOnWirthmuellerMorphism}
  vanishes, in that $\tau \simeq 1_{\mathcal{D}}$, then this is then equivalent to an identification of the linear type
  $$
    V_f := f_! f^\ast (1_{\mathcal{D}})
  $$
  with its dual
  $$
    V_f \stackrel{\simeq}{\longrightarrow} V_f^\ast
    \,.
  $$
  This way an untwisted fiberwise fundamental class on $f$ is equivalently a non-degenerate inner product
  $$
    \langle -,-\rangle \;:\; V_f \otimes V_f \longrightarrow 1_{\mathcal{D}}
    \,.
  $$
\end{remark}

In this spirit we say that:
\begin{definition}
  For $A \in \mathrm{Mod}(X)$ dualizable, a choice of \emph{fiberwise inner product}
  is a choice of equivalence
  $$
    A \stackrel{\simeq}{\longrightarrow} \mathbb{D}A
    \,.
  $$
  If this is the inverse of its dual morphism, we say the inner product is
  \emph{symmetric}
  (axiom (T5) in \cite{Selinger10}).
  \label{FibInnerProduct}
\end{definition}
\noindent The corresponding pairing we write
$$
  \langle -,-\rangle_A \;:\; A \otimes A \stackrel{\simeq}{\longrightarrow} A \otimes \mathbb{D}A
  \stackrel{\mathrm{ev}}{\longrightarrow} 1_{X}
$$
and often we find it convenient to use ``$\langle -,-\rangle_A$'' also for the original equivalence itself.
In this notation the symmetry condition is that $\langle -,-\rangle_A \simeq \mathbb{D}\langle -,-\rangle_A^{-1}$.

If $X \simeq \ast$ we may call a fiberwise inner product over $X$ just an ``inner product''
or ``global inner product'', for emphasis.
The following examples show how a fiberwise inner product induces a global one.
\begin{example}
  If $A \in \mathrm{Mod}(X)$ is equipped with a fiberwise inner product $\langle -,-\rangle_A$,
  def. \ref{FibInnerProduct}, and if $X$ (hence the terminal morphism $X \to \ast$) is equipped with an
  untwisted fundamental class $[X]$, def. \ref{TwistedOrientationOnWirthmuellerMorphism},
  then $\underset{X}{\sum}A \in \mathrm{Mod}(\ast)$ is naturally
  equipped with the inner product given by the composite
  $$
    \langle -,-\rangle_{\underset{X}{\sum}A}
    :
    \underset{X}{\sum} A
     \stackrel{\underset{X}{\sum}\langle -,-\rangle_A}{\longrightarrow}
    \underset{X}{\sum} \mathbb{D}A
     \stackrel{\simeq}{\longrightarrow}
    \underset{X}{\prod} \mathbb{D}A
     \stackrel{\simeq}{\longrightarrow}
    \mathbb{D} \underset{X}{\sum} A
    \,,
  $$
  where the second equivalence is the Wirthm{\"u}ller isomorphism induced by the
  fundamental class
  (by the second clause in prop. 4.13 of \cite{May05}, using that $\underset{X}{\sum}1_X$ is dualizable
  by our assumption on fundamental classes, see remark \ref{SpecializationOfTwists})
  and the last one is parameterized linear De Morgan duality, prop. \ref{DualityIntertwinesLeftAndRigthtPushforward}.
  \label{GlobalInnerProduct}
\end{example}
\begin{example}
  If $A \in \mathrm{Mod}(X)$ is equipped with a fiberwise inner product, def. \ref{FibInnerProduct}, and
  $f : Y \to X$ is equipped with an untwisted fiberwise fundamental class,
  def. \ref{TwistedOrientationOnWirthmuellerMorphism}, then
  this induces on
  $f_! f^\ast A$ a fiberwise inner product given as the composite
  $$
    \langle -,-\rangle_{f_! f^\ast A}
    \;:\;
    \xymatrix{
      f_! f^\ast A
      \ar[r]_-{\simeq}
      &
      \mathbb{D}f_! f^\ast \mathbb{D}A
      \ar[rr]_-\simeq^-{\mathbb{D}f_! f^\ast \langle -,-\rangle_A^{-1}}
      &&
      \mathbb{D} f_! f^\ast A
    }
  $$
  of the induced Wirthm{\"u}ller isomorphism, prop. \ref{TwistedOrientationMakesComonadTwistedCommutedWithDuality},
  and the image of the fiberwise fundamental class under $\mathbb{D}f_! f^\ast(-)$.
  \label{InducedOrientationOnSumPullbackA}
\end{example}
A simple but fundamental fact is that between objects that are equipped with (fiberwise) inner products, every morphism
has a canonical reversal:
\begin{definition}
Given a morphism $f : A \longrightarrow B$ between linear types equipped with
fiberwise inner product, def. \ref{FibInnerProduct}, then we say its \emph{transpose} $f^\dagger$ is the composite
$$
  f^\dagger
  \;:\;
  \xymatrix{
    B
    \ar[r]_-\simeq^-{\langle -,-\rangle_B}
    &
    \mathbb{D}B
    \ar[r]^-{\mathbb{D}f}
    &
    \mathbb{D}A
    \ar[r]_-\simeq^-{\langle -,-\rangle_A^{-1}}
    &
    A
  }
  \,.
$$
 \label{Transpose}
\end{definition}
Some comments on this basic abstract construction of $\dagger$-structure are in section 4 of \cite{Selinger10}.

We may now relate the choice of a fiberwise fundamental class to the transpose of the pushforward
along the map.
\begin{proposition}
  Let $A \in \mathrm{Mod}(X)$ be dualizable and equipped with a fiberwise symmetric inner product
  $\langle -,-\rangle_A$,
  def. \ref{FibInnerProduct},
  and let $f : Y \longrightarrow X$ be a morphism of contexts equipped with an untwisted
  fiberwise fundamental class,
  def. \ref{TwistedOrientationOnWirthmuellerMorphism}.
  Then the respective morphism $[f]$, def. \ref{MeasureInducedByOrientation},
  is the transpose, def. \ref{Transpose}, of the $(\sum_f \dashv f^\ast)$-counit:
  $$
    [f] \simeq \epsilon_f^\dagger
    \,,
  $$
  hence, by remark \ref{FundamentalClassIsDualUnit}
  $$
    \epsilon_f^\dagger \simeq \eta_f
    \,.
  $$
  \label{FundamentalClassAsDagger}
\end{proposition}
\proof
  By naturality of the counit we have
  $$
  \xymatrix{
    A \ar[rr]^-{\mathbb{D}\epsilon_{\mathbb{D}A}}
    \ar[d]|{\mathbb{D}\langle -,-\rangle_{A}^{-1}}
    \ar@/^2pc/[rrr]^-{[f]_A}
      &&
    \mathbb{D} f_! f^\ast \mathbb{D}A
     \ar[d]|{\mathbb{D}f_! f^\ast \langle -,-\rangle_{A}^{-1}}
     \ar[r]^-\simeq & f_! f^\ast A
    \\
    \mathbb{D}A
    \ar[rr]^-{\mathbb{D}\epsilon_A}
    &&
    \mathbb{D} f_! f^\ast A
    \ar[ur]_{\langle -,-\rangle_{f_! f^\ast A}^{-1}}
  }
  \,,
  $$
  where the square on the left is the image under $\mathbb{D}$
  of the naturality square of the $(f_!\dashv f^\ast)$-counit on the fiberwise inner
  product $\langle -,-\rangle_A^{-1} : \mathbb{D}A \stackrel{\simeq}{\to} A$, and where the diagonal equivalence
  on the right is the inverse of the map in example \ref{InducedOrientationOnSumPullbackA}.
  By symmetry of the fiberwise inner product on $X$  the left vertical map is equivalent to $\langle -,-\rangle_A$
  and hence the bottom composite of the diagram exhibits $[f]_A$ as the transpose of $\mathbb{D}\epsilon_A$.
\endofproof
\begin{corollary}
  If $X$ itself (hence $X \to \ast$) is equipped with an untwisted fundamental class
  $[X]$  then
  $$
    \underset{X}{\sum}[f] \simeq \left(\underset{X}{\sum}\epsilon\right)^\dagger
  $$
\end{corollary}
\proof
  Combining example \ref{GlobalInnerProduct} and prop. \ref{FundamentalClassAsDagger}.
\endofproof
Therefore:
\begin{remark}
If $A \in \mathrm{Mod}(X)$ is equipped with a fiberwise symmetric inner product
  $\langle -,-\rangle_A$
  and $f : Y \longrightarrow X$ is equipped with untwisted fiberwise fundamental classes,
  def. \ref{TwistedOrientationOnWirthmuellerMorphism},
  then the formula for the secondary integral transform $\mathbb{D}\int_Z \Xi d\mu$ in def.
  \ref{IntegralKernelAndTransform}
  of a prequantum integral kernel on a correspondence
  $$
    
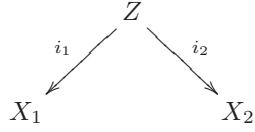
\xymatrix{
      & Z
       \ar[dl]_{i_1}
       \ar[dr]^{i_2}
      \\
      X_1 && X_2
    }
  $$
  becomes
  $$
    \mathbb{D}\int_Z \Xi d\mu
    \;\simeq\;
    \underset{X}{\sum}\epsilon^{i_1}_{A_1} \circ \Xi \circ \underset{X}{\sum} (\epsilon^{i_2}_{A_2})^\dagger
    \,.
  $$
  If moreover $X$ itself is equipped with a fundamental class then this becomes
  $$
    \mathbb{D}\int_Z \Xi d\mu
    \;\simeq\;
    \left(\underset{X}{\sum}\epsilon^{i_1}_{A_1}\right)
       \circ \Xi \circ
    \left(\underset{X}{\sum} \epsilon^{i_2}_{A_2}\right)^\dagger
    \,.
  $$
  \label{pullpushviadagger}
\end{remark}
This kind of operation plays a special role both in abstract quantum physics as
well as in generalized cohomology theory:
\begin{remark}
  In particular for the case that  $i_1 = i_2$ and $A_1 = A_2$ (so that
  in example \ref{MatrixAsExampleOfAbstractIntegralKernel} the integral kernel is a square matrix)
  then the map
  $$
    \Xi\mapsto
     \mathbb{D}\int_Z \Xi d\mu
    \simeq
    \left(\underset{X}{\sum}\epsilon_{A}\right)
       \circ \Xi \circ
    \left(\underset{X}{\sum}\epsilon_{A}\right)^\dagger
  $$
  (which we identify as the path integral quantization map for the integral kernel $\Xi$)
  is what is called a (completely positive) ``quantum operation'', see \cite{Selinger}.
\end{remark}
\begin{remark}
  In the model of linear homotopy-type theory by generalized cohomology theory,
  def. \ref{LocalSystems}, the self-duality of \ref{FibInnerProduct} is
  \emph{Poincar{\'e} duality} (in general with a twist) and the induced transpose maps
  in def. \ref{Transpose} are the ``Umkehr maps'' or ``wrong way maps'' in generalized cohomology.

  Specifically the literature on KK-theory knows that forming Umkehr maps in K-theory is given
  by forming transpose morphisms of the ``right way''-morphisms in the symmetric monoidal category $\mathrm{KK}$, see
  \cite{BMRS}. Definition 2.1 in \cite{BMRS} defines (somewhat implicitly)
  a fundamental class to be a choice of self-duality in $\mathrm{KK}$ (Poincar{\'e} duality in KK)
  and section 3.3 there defines construction of Umkehr maps as the corresponding
  construction of transposes, hence of the dagger-operation as in def. \ref{Transpose}.
  Under this identification the re-formulation of secondary integral transforms via
  dagger operations in remark \ref{pullpushviadagger} corresponds to formula (5.6)
  in \cite{BMRS}.
  \label{UmkehrInKK}
\end{remark}

\section{Directed homotopy-type theory}
\label{DirectedHomotopyTypeTheory}

We have discussed in \ref{LinearLogic} how quantum mechanics is faithfully axiomatized
in linear (homotopy-)type theory; and have indicated that this is not sufficient
for the description of modern quantum physics, as it misses aspects of quantum field theory.
One may think of quantum mechanics as being 1-dimensional quantum field theory, and
conversely of local quantum field theory as being a refinement from quantum propagation
along one dimension (time) to a description of propagation locally in more directions
(space and time) (see \cite{SatiSchreiber} for review and pointers to the literature).
This higher dimensional directionality is formalized, semantically, in
the theory of monoidal $n$-categories (meaning: $(\infty,n)$-categories).
For $n = 1$ these coincide with the categories ($\infty$-categories) that we saw in
\ref{NotionsJudgementDeduction} provide semantics for homotopy-type theory.
Therefore the generalization of this to $(\infty,n)$-categories has been called
\emph{directed homotopy-type theory}.

In this note we do not dwell too much on this except to make here a remark
on what the homotopy-theoretic structure of local quantum field theory is in a little
more detail, just enough so that it becomes clear how the linear homotopy-type theory
which we considered so far sits inside there, and hence how it is that we may
connect to the process of quantization, which is the topic of \ref{Quantization} below.

\subsection{Free directed linear homotopy-types and Quantum field theory}
\label{FreeDirectedLinearHomotopyTypes}

In the traditional Schr{\"o}dinger picture, a $d$-dimensional quantum field theory $Z$ is given
by assigning to each compact space $\Sigma_{d-1}$ a complex vector space
(Hilbert space) $Z(\Sigma_{d-1})$ (of quantum states of fields on $\Sigma_{d-1}$)
and to each cobordism $\Sigma_d \;:\;\Sigma_{d-1}^{\mathrm{in}} \to \Sigma_{d-1}^{\mathrm{out}}$
(``spacetime'' or ``worldvolume'') a linear map
$$
  Z(\Sigma_d) \;:\; Z(\Sigma_{d-1}^{\mathrm{in}}) \longrightarrow Z(\Sigma_{d-1}^{\mathrm{out}})
$$
(propagating quantum states along $\Sigma$) such that this assignment is functorial and
such that it is monoidal, in that the disjoint union of spaces is sent to the tensor product
of vector spaces.
Hence such a non-localized topological quantum field theory is a symmetric monoidal functor of the form
$$
  Z \;:\; \mathrm{Bord}_n^{\coprod} \longrightarrow \mathbb{C} \mathrm{Mod}^{\otimes}
  \,.
$$
The locality of such a quantum field theory is encoded by refining this to an
$(\infty,n)$-functor
$$
  Z \;:\; \mathrm{Bord}_n^{\coprod} \longrightarrow E \mathrm{Mod}_n^{\otimes}
  \,,
$$
where $E \in \mathrm{CRing}_\infty$ is a suitable $E_\infty$-ring and $E \mathrm{Mod}_n$
is an $(\infty,n)$-category of $n$-fold $E$-modules,
hence of higher ``directed'' types of quantum states, this we come to in
\ref{DirectedTypesOfStates}
and \ref{TannakianDescriptionOfDirectedTypesOfQuantumStates} below.

The central theorem of \cite{LurieQFT} says that the $n$-category
$\mathrm{Bord}_n$ is the free symmetric monoidal $(\infty,n)$-category on the collection
of boundary data (branes) and defect data (domain walls). Hence linear-type theoretically
one may think of $\mathrm{Bord}_n$ as a collection of free types in
multiplicative linear directed homotopy type theory.

\subsection{Directed types of quantum states}
\label{DirectedTypesOfStates}

Consider a model
$$
  \xymatrix{
    \mathrm{Mod}
    \ar[d]
    \\
    \mathbf{H}
  }
$$
for linear homotopy-type theory, def. \ref{LinearHomotopyTypeTheory}.
We observe that this naturally comes equipped with a higher directed notion of
linear types, too.
\begin{definition}
For every type $X \in \mathbf{H}$ the symmetric monoidal category
$\mathrm{Mod}(X)$ is canonically a module category over the
symmetric monoidal category $\mathrm{Mod}(\ast)$, via the action
$$
  \mathrm{Mod}(\ast)
  \times
  \mathrm{Mod}(X)
  \longrightarrow
  \mathrm{Mod}(X)
$$
given by
$$
  (\tau, A)
  \mapsto
  (X^\ast \tau) \otimes A
  \,.
$$
  A functor $F : \mathrm{Mod}(X) \longrightarrow \mathrm{Mod}(Y)$
  is called \emph{$\mathrm{Mod}(\ast)$-linear} if it respects this
  action.
  \label{ModBaseLinearity}
\end{definition}
\begin{remark}
That def. \ref{ModBaseLinearity} indeed defines an action is equivalent to the fact that
$$
  X^\ast \;:\; \mathrm{Mod}(\ast)\longrightarrow \mathrm{Mod}(X)
$$
is a strong monoidal functor, by the axioms of linear homotopy-type theory,
def. \ref{LinearHomotopyTypeTheory}.
\end{remark}
\begin{proposition}
  For $f : X\longrightarrow Y$ any map in $\mathbf{H}$, then
  pullback
  $$
    f^\ast : \mathrm{Mod}(X) \longleftarrow \mathrm{Mod}(Y)
  $$
  is a $\mathrm{Mod}(\ast)$-linear functor, def. \ref{ModBaseLinearity},
  as is the sum along the fibers of $f$
  $$
    \underset{f}{\sum}
    \;:\;
    \mathrm{Mod}(X)
     \longrightarrow
    \mathrm{Mod}(Y)
    \,.
  $$
  \label{ReciprocityIsLinearity}
\end{proposition}
\proof
  For $A \in \mathrm{Mod}(X)$
  and $\tau \in \mathrm{Mod}(\ast)$ we naturally have
  $$
    \begin{aligned}
      f^\ast( \tau \cdot A )
      & =
      f^\ast( (Y^\ast \tau) \otimes A )
      \\
      &\simeq
      (f^\ast Y^\ast \tau) \otimes f^\ast A
      \\
      & \simeq
      (Y^\ast \tau) \otimes f^\ast A
      \\
      & =
      \tau \cdot f^\ast A
    \end{aligned}
    \,,
  $$
  where we used that $f^\ast$ is strong monoidal, and
  $$
    \begin{aligned}
      \underset{f}{\sum}(\tau \cdot A)
      &=
      \underset{f}{\sum} ((X^\ast \tau) \otimes A)
      \\
      &\simeq \underset{f}{\sum}( (f^\ast Y^\ast \tau) \otimes A )
      \\
      &\simeq (Y^\ast \tau) \otimes \underset{f}{\sum} A
      \\
      & = \tau \cdot \underset{f}{\sum} A
      \,.
    \end{aligned}
    \,,
  $$
  where the last equivalence is Frobenius reciprocity.
\endofproof
To reflect this we may say:
\begin{definition}
Write
$$
  \mathrm{Mod}_2
  \in
  (\infty,2)\mathrm{Cat}
$$
for the $(\infty,2)$-category
of $\mathrm{Mod}(\ast)$-linear $\infty$-categories of the form $\mathrm{Mod}(X)$
for some $X \in \mathbf{H}$, and $\mathrm{Mod}(\ast)$-linear functors between them.
\label{Mod2}
\end{definition}

In \ref{TannakianDescriptionOfDirectedTypesOfQuantumStates} below we expand a bit more
on this (well-known) concept of higher categorical modules.
In \ref{CoboundingTheory} we consider a kind of quantum field theory that
does have directed spaces of quantum states given by 2-modules of the form
$\mathrm{Mod}(X)$. For this to satisfy the axioms of a TQFT, we will need to require
two extra properties on the ambient model for linear homotopy-type theory.

\begin{definition}
Given a model $\mathrm{Mod}(-) \to \mathbf{H}$
for linear homotopy-type theory, def. \ref{LinearHomotopyTypeTheory},
one says that it satisfies the \emph{Beck-Chevalley condition}
if for all $\infty$-pullback squares in $\mathbf{H}$
$$
  \xymatrix{
    & Z
    \ar[dl]_{h}
    \ar[dr]^{f}
    \\
    X_1 \ar[dr]_k && X_2 \ar[dl]^g
    \\
    & Y
  }
$$
the composition
$$
  f_! h^\ast
  \stackrel{}{\longrightarrow}
  f_1 h^\ast k^\ast k_!
  \stackrel{\simeq}{\longrightarrow}
  f_! f^\ast g^\ast k_!
  \stackrel{}{\longrightarrow}
  g^\ast k_!
$$
is an equivalence
(between pull-push $\mathrm{Mod}(X_1)\to \mathrm{Mod}(X_2)$ along the upper half and push-pull along the lower half).
\label{BeckChevalleyCondition}
\end{definition}
\begin{example}
  The models for linear homotopy-type theory
  $\mathbf{H}^{\Delta^1}\stackrel{\mathrm{cod}}{\to} \mathbf{H}$,
  example \ref{ToposBaseChangeIsCartesianWirthmueller}, and
  $E \mathrm{Mod}(-)\to \infty \mathrm{Grpd}$, example \ref{LocalSystems},
  satisfy the Beck-Chevalley condition, def. \ref{BeckChevalleyCondition}.
\end{example}
\proof
  The first statement is equivalently the pasting law for
  $\infty$-pullbacks in $\mathbf{H}$. The second appears as prop. 4.3.3 in
  \cite{HopkinsLurie}.
\endofproof

\begin{definition}
  We say a model $\mathrm{Mod}(-) \to \mathbf{H}$ for linear homotopy-type theory,
  def. \ref{LinearHomotopyTypeTheory}, is \emph{2-monoidal} if
  for all $X, Y \in \mathbf{H}$ we have
  $$
    \mathrm{Mod}(X \times Y)
    \simeq
    \mathrm{Mod}(X) \otimes_{\mathrm{Mod}(\ast)} \mathrm{Mod}(Y)
  $$
  \label{2MonoidalTypeTheory}
\end{definition}
\begin{example}
  For $E \in \mathrm{CRing}_\infty$ any $E_\infty$-ring, then the model of
  linear homotopy-type theory $E\mathrm{Mod}(-)\to \infty \mathrm{Grpd}$
  is 2-monoidal, def. \ref{2MonoidalTypeTheory}.
\end{example}
\proof
  Since $X \in \infty \mathrm{Grpd} \hookrightarrow (\infty,1)\mathrm{Cat}$ is small
  and $E \mathrm{Mod}(\ast) \in (\infty,1)\mathrm{Cat}$ is locally presentable,
  this follows from basic properties of the symmetric monoidal $\infty$-category
  of locally presentable $\infty$-categories \cite{LurieAlg}.\footnote{
  Thanks to Thomas Nikolaus for discussion of this point.}
\endofproof

\begin{proposition}
 \label{BCImpliesOtimesPreservesIndexedColimits}
 In a model $\mathrm{Mod}(-) \to \mathbf{H}$ for linear homotopy-type theory,
 def. \ref{LinearHomotopyTypeTheory},
 consider $f_1 :X_1 \to Y_1$ and $f_2 : X_2\to Y_2$ in $\mathbf{H}$ and
 $A_i \in \mathrm{Mod}(X_i)$ and $B_i\in \mathrm{Mod}(Y_i)$.
 Then
 $$
   (f_1 \times f_2)^\ast ( p_1^\ast B_1 ) \otimes (p_2^\ast B_2)
   \simeq
   (p_1^\ast f_1^\ast B_1) \otimes (p_2^\ast f_2^\ast B_2)
   \,.
 $$
 If the Beck-Chevalley condition, def. \ref{BeckChevalleyCondition}, holds then also
 $$
   \underset{f_1 \times f_2}{\sum} (p_1^\ast A_1) \otimes (p_2^\ast A_2)
   \simeq
   (p_1^\ast \underset{f_1}{\sum} A_1) \otimes (p_2^\ast \underset{f_2}{\sum} A_2)
   \,.
 $$
\end{proposition}
\proof
The first one follows immediately from the fact that pullback is required to
be strong monoidal. The second one follows using
Frobenius reciprocity and the Beck-Chevalley, as is shown in lemma 3.2 of \cite{PontoShulman}.
\endofproof

\begin{corollary}
  If the given model for linear homotopy-type theory, def. \ref{ToposBaseChangeIsCartesianWirthmueller},
  satisfies the Beck-Chevalley condition, def. \ref{BeckChevalleyCondition} and is 2-monoidal,
  def. \ref{2MonoidalTypeTheory}, then the $(\infty,2)$-functor
  $$
    \mathrm{TQFT}_{d+1} :  \mathrm{Corr}_1(\mathbf{H}) \longrightarrow \mathrm{Mod}_2
  $$
  given by sending correspondence to their linear polynomial functors,
  def. \ref{PolynomialFunctor},
  is monoidal.
  \label{LinearPolynomialFunctorAssignmentIsMonoidal}
\end{corollary}

\subsection{Tannaka duality for directed types of quantum states}
\label{TannakianDescriptionOfDirectedTypesOfQuantumStates}

We briefly recall here a definition of $(E \mathrm{Mod}_n, \otimes)$, the $n$-fold directed higher refinement
of categories of modules serving as higher directed analogs of spaces of quantum states,
and explain the meaning of this definition via Tannaka duality.

\medskip

For $A$ a ring, consider the category $A \mathrm{Mod}$ of its (left) modules.
Tannaka duality (see for instance \cite{Vercruysse} for review and pointers to the literature)
describes the relation between properties of categories such as $A \mathrm{Mod}$
and the underlying ring $A$. For instance if $A$ is a commutative algebra, then
$A \mathrm{Mod}$ naturally inherits the structure of a monoidal category from the
natural tensor product of $A$-modules. As such then $A \mathrm{Mod}$ is itself a higher ``directed''
analog of a (semi-)ring. Therefore one may ask for suitable categories which are
equipped with an action, in the suitable sense, of $(A \mathrm{Mod}, \otimes_A)$.
For instance if $E$ is a commutative ring and $A$ is equipped with the structure of an
$E$-algebra, then $A \mathrm{Mod}$ is naturally a module over the monoidal category
$(E \mathrm{Mod}, \otimes_E)$, with the action given by tensoring an $E$-module with an $A$-module
over $E$.
Such a module category is hence a higher ``directed'' analog of an ordinary module over a ring.
We are going to take $n$-directed $A$-modules to be the result of iterating this reasoning
$n$ times.
But in order to determine what ``suitable'' means in the above and how
to encode these structures in a useful way, we invoke more Tannaka duality.

Being commutative is a very strong condition on a ring, and so one may ask what
the necessary and sufficient structure on $A$ is that makes its category $A \mathrm{Mod}$ of
modules be monoidal. The traditional textbook answer is that this is the structure
of a \emph{bialgebra} on $A$ -- however we need to be careful with the
fine print of the Tannaka duality theorem for bialgebras: the theorem asserts an equivalence
between bialgebras $A$ over $E$ and monoidal category $A \mathrm{Mod}$ that are \emph{equipped} with
their forgetful functor (``fiber functor'') $A \mathrm{Mod} \to E \mathrm{Mod}$.
Possibly more famous is the further refinement of this theorem to those bialgebras which
are Hopf algebras, where Tannaka duality says that Hopf algebra structures on $A$ over $E$ are equivalent
to autonomous monoidal category structure plus fiber functor on $A \mathrm{Mod}$ (here ``autonomous'' means all objects
have duals).

Only more recently did the full Tannaka duality for monoidal categories appear in the literature
(see \cite{Vercruysse} for a review),
where the extra structure of a fiber functor is not assumed: say that a \emph{sesquialgebra}
\cite{TWZ} is an
associative algebra $A$ equipped with one more structure of a (co-)algebra, but not in the category
$E \mathrm{Mod}$ (which would make it an ordinary bialgebra), but in the more flexible 2-category
$E \mathrm{Mod}^{b}$ whose objects are associative $E$-algebras, but whose morphisms are
bimodules between these (and whose 2-morphisms are bimodule homomorphisms). With this definition
the full duality statement is obtained: the structure of a sesquialgebra on the associative $E$-algebra
$A$ is equivalent to the structure of a monoidal category on $A \mathrm{Mod}$.

Observe that the 2-category of algebras with bimodules between then appearing also has another
interpretation: the Eilenberg-Watts theorem says that this is equivalently the 2-category
whose objects are categories of modules of the form $A \mathrm{Mod}$, and whose morphisms
are right exact functors between these that preserve colimits. Since colimits are the
category-theoretic analog of addition, these may be thought of as the \emph{linear}
functors between the categories $A \mathrm{Mod}$ regarded as $E \mathrm{Mod}$-modules.
For this reason we may think of $E \mathrm{Mod}^b$ as being $E \mathrm{Mod}_2$, the
2-category of \emph{directed $E$-modules} or \emph{2-modules} over $E$
(\cite{AQFTFromFQFT}, appendix).
By the above discussion in \ref{FreeDirectedLinearHomotopyTypes}, 2-modules
in this sense should appear as higher order spaces of quantum states in 2-dimensional
quantum field theory. This is indeed the case, see \cite{AQFTFromFQFT}
for a conceptual discussion in the sense used here and for further pointers to
the literature.

Therefore if $A$ is a sesquialgebra, then, by the above, we may say that it is an algebra object
not just in the category $E \mathrm{Mod}$ of $E$-modules, but in fact in the 2-category
$E \mathrm{Mod}_2$ of 2-modules over $E$. Since the sesquialgebra structure on $A$ makes
$A \mathrm{Mod}$ be a monoidal category, we may apply Tannaka duality on this
higher categorical level and regard $(A \mathrm{Mod}, \otimes)$ as a stand-in for
\emph{its} 2-category of module categories. When this is regarded as having as morphisms
bimodule-categories, as 2-morphisms module homomorphisms between these and as 3-morphisms
natural transformation of these, then this forms a 3-category $E \mathrm{Mod}_2^b$
which along the above lines we may regard as $E \mathrm{Mod}_3$, the 3-category
of 3-directed modules over $E$.
For example this means that the Hopf algebras (quantum groups) that famously encode
3-dimensional Dijkgraaf-Witten and Chern-Simons field theory (see for instance the
review in \cite{Rowell}), being in particular bialgebras
and hence in particular sesquialgebras, are naturally stand-ins for 3-modules of quantum states,
just as befits a 3-dimensional quantum field theory.

This process we may now iterate: we say that an $n$-module over $E$ is a an $(n-1)$-fold
algebra over $E$, which in turn means that it is an algebra object in the category
of bimodules over algebra objects in the category of bimodules over ... over $E$.
In this full generality the definition of $E \mathrm{Mod}_n$ is indicated in \cite{FHLT}.
A special strict version of the next step after sesquialgebra (bialgebra) and
3-modules, namely \emph{trialgebras} Tannaka dual to certain
4-modules, has been considered in \cite{Pfeiffer} following \cite{CraneFrenkel}.

We will not dwell on the higher module structures in this note here. For the purposes
of the following section \ref{Quantization}, the upshot of the above discussion is this:
for $E \in \mathrm{CRing}_\infty$ a commutative ring ($E_\infty$-ring), and for all
$n \in \mathbb{N}$, then an $n$-module $N \in E\mathrm{Mod}_n$ over $E$ is in particular an ordinary
$E$-module, equipped in addition with $(n-1)$ compatible algebra structures, and morphisms
between these include in particular $n$-module homomorphisms respecting the extra structure
(while more generally the morphisms are bimodules, respecting extra structure).

\subsection{Boundaries/branes and defects/domain walls}

We consider in the following the first three stages, d =0, 1, 2 of $(n+1)$-dimensional extended QFT.
Moreover, we consider defects of codimension 1 and boundaries of codimension 2. Therefore,
by the cobordism theorem for cobordisms with singularities, the data in dimension
$0 \leq d \leq 2$ in principle determines everything else by higher traces.
But we use coefficients only as suitable for a 2-dimensional QFT, namely the
$(\infty,2)$-category of 2-modules from def. \ref{Mod2}
$$
  Z
  \;:\;
  (\mathrm{Bord}_n)_{0,1,2}
  \longrightarrow
   (\mathrm{Corr}_n(\mathbf{H}))_{0,1,2}
  \longrightarrow
  \mathrm{Mod}_2
$$
By the above discussion this is necessary but of course not fully sufficient for
constructing the fully extended $(n+1)$-dimensional QFT. In order to do so
we would furthermore need to equip the object $Z(\ast)$ with the structure of an
$n$-fold algebra, etc.

The basic singularity datum that we consider here is an elementary \emph{corner}
$$
  \raisebox{20pt}
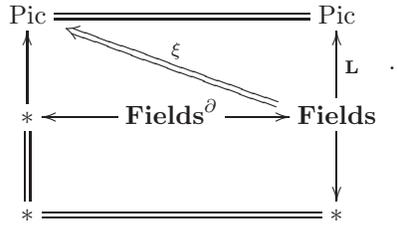
{
  \xymatrix{
    \mathrm{Pic} \ar@{=}[rr]_<{\ }="t" && \mathrm{Pic}
    \\
    \ast \ar[u] \ar@{=}[d] & \mathbf{Fields}^\partial
    \ar[l] \ar[r]^>{\ }="s"
    & \mathbf{Fields} \ar[u]_{\mathbf{L}} \ar[d]
    \\
    \ast \ar@{=}[rr] && \ast
    \ar@{=>}_\xi "s"; "t"
  }
  }
  \,.
$$

Here $\mathbf{Fields}$ are the bulk fields of a $d$-dimensional field theory which we
will denote by $\mathrm{TQFT}_d^\tau$, and which, by this diagram, appears as a defect
that interpolates between the theory constant on $\mathrm{Pic}$ and the
 entirely trivial field theory in dimension $d+1$.
This defect itself has a boundary on the left, giving rise to a $(d-1)$-dimensional
boundary field theory $\mathrm{QFT}_d$ with fields $\mathbf{Fields}^\partial$

Then we also consider what it means to continue this further away from the boundary.
$$
  
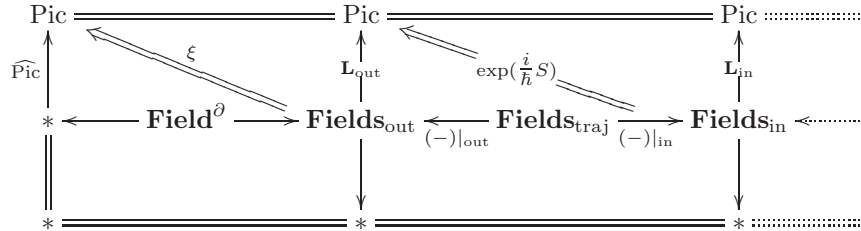
\xymatrix{
    \mathrm{Pic} \ar@{=}[rr]_<{\ }="t1" && \mathrm{Pic} \ar@{=}[rr]_<{\ }="t" && \mathrm{Pic} \ar@{::}[r] &
    \\
    \ast \ar[u]^{\widehat{\mathrm{Pic}}} \ar@{=}[d]
    & \mathbf{Field}^\partial
    \ar[l] \ar[r]^>{\ }="s1" & \mathbf{Fields}_{\mathrm{out}} \ar[u]|{\mathbf{L}_{\mathrm{out}}}
     \ar[d]
    &
    \mathbf{Fields}_{\mathrm{traj}}
    \ar[l]^{(-)|_{\mathrm{out}}}
    \ar[r]_{(-)|_{\mathrm{in}}}^{\ }="s"
    &
    \mathbf{Fields}_{\mathrm{in}}
    \ar[u]|{\mathbf{L}_{\mathrm{in}}}
    \ar[d]
    \ar@{<..}[r]
    &
    \\
    \ast \ar@{=}[rr] && \ast \ar@{=}[rr] && \ast \ar@{::}[r] &
    \ar@{=>}|{\exp(\tfrac{i}{\hbar}S)} "s"; "t"
    \ar@{=>}_\xi "s1"; "t1"
  }
$$
Composition to the right here imposes strong ``anomaly cancellation'' constraints on the
quantization to make the composites functorial. But at the boundary itself, where no further
composition is possible (to the left) these conditions are relaxed and the available choices
are part of what makes the boundary theory itself ``geometric'' (non-topological).

By the above, our truncated coefficients are such that these diagrams in $\mathbf{H}$
are sent first to diagrams in $\mathrm{Mod}_2$ of the form
$$
  \xymatrix{
    \mathrm{Mod}(\mathrm{Pic})
    \ar[d]_{\widehat{\mathrm{Pic}}^\ast}
    \ar@{=}[r]_>{\ }="s1" & \mathrm{Mod}(\mathrm{Pic}) \ar[d]^{\mathbf{L}_{\mathrm{out}}^\ast}
    \\
    \mathrm{Mod}(\ast) \ar@{=}[d] \ar[r]^<{\ }="t1"_>{\ }="s2" & \mathrm{Mod}(\mathbf{Fields})
    \ar[d]^{\underset{\mathbf{Fields}}{\sum}}
    \\
    \mathrm{Mod}(\ast) \ar@{=}[r]^<{\ }="t2" & \mathrm{Mod}(\ast)
    \ar@{=>} "s1"; "t1"
    \ar@{=>} "s2"; "t2"
  }
$$
For every choice of element in $\mathrm{Mod}(\mathrm{Pic})$, hence for every
morphism $\mathrm{Mod}(\ast) \to \mathrm{Mod}(\mathrm{Pic})$ this yields a diagram
$$
  \xymatrix{
    \mathrm{Mod}(\ast)
    \ar@/_2pc/[d]^{\ }="t"
    \ar@/^2pc/[d]_{\ }="s"
    \\
    \mathrm{Mod}(\ast)
    \ar@{=>} "s"; "t"
  }
$$
in $\mathrm{Mod}_2$. By linearity, this is equivalently a morphism in $\mathrm{Mod}(\ast)$
itself. This may hence be regarded as the propagator of the defect $\mathrm{TQFT}_d^\tau$.

\section{Examples}

First, as a simple blueprint for the discussion to follow,
in \ref{matrixCalculus} we spell out how ordinary linear matrix multiplication
arises as a special case of models of the secondary integral transforms in
def. \ref{IntegralKernelAndTransform}. Then
in \ref{PullPushInTwistedGeneralizedCohomology} we
show as our central example how pull-push (Umkehr maps) in
twisted generalized cohomology are another class of examples.

\subsection{Ordinary linear algebra -- Matrix calculus}
 \label{matrixCalculus}

\begin{example}[matrix calculus]
  \label{MatrixAsExampleOfAbstractIntegralKernel}
  Let $k$ be a field, let $\mathbf{H} = \mathrm{Set}$ be the category of sets,
  and for $X \in \mathrm{Set}$ let $\mathrm{Mod}(X) := k \mathrm{Mod}(X) = \mathrm{Vect}_k(X)$
  be the category of $X$-parameterized vector bundles. This is a model for linear homotopy-type theory
  by example \ref{WirthmuellerMorphsismsBetweenParameterizedModules}.
  For $X \in \mathrm{FinSet} \hookrightarrow \mathrm{Set}$ a finite set, then an
  $X$-dependent linear type $A \in \mathrm{Vect}_k(X)$ is an (unordered) $\vert X\vert$-tuple of vector
  spaces, where $\vert X \vert$ is the cardinality of $X$. The dependent sum produces the
  direct sum of these:
  $$
    \underset{X}{\sum} A
    \simeq
    \underset{x \in X}{\oplus} A_x
    \in
    \mathrm{Vect}_k
    \,.
  $$

  Consider then $X_1,X_2 \in \mathrm{FinSet} \hookrightarrow \mathrm{Set}$ two finite sets of cardinality
  $n_1$ and $n_2$, respectively, and consider the projection correspondence
  $$
    \raisebox{20pt}{
    \xymatrix{
      & X_1 \times X_2
      \ar[dl]_{p_1}
      \ar[dr]^{p_2}
      \\
      X_1 && X_2
    }
    }
    \,.
  $$

  Here for $A \in [Y,\mathrm{Vect}_k]$ an $n_2$-tuple of vector spaces, then $(p_2)_! (p_2)^\ast A$
  is the $n_2$-tuple whose value over $y \in X_2$ is $(A_y)^{\oplus^{n_1}} \simeq A_y \otimes k^{n_1}$. The counit
  $(p_2)_! (p_2)^\ast A \to A$ is the morphism that over each $y \in Y$ is given by forming the sum of
  $n_1$ vectors in $A_y$.

  There is an untwisted fiberwise fundamental class on $p_2$, given by the canonical choice of identification
  $k^{n_1} \simeq (k^{n_1})^\ast$ (``regard row-vectors as column vectors'').
  With this choice the equivalence of
  prop. \ref{TwistedOrientationMakesComonadTwistedCommutedWithDuality} is over $y \in Y$ the
  induced isomorphism $A_y \otimes k^{n_1} \simeq A_y \otimes (k^{n_1})^\ast$.
  The induced fundamental class of def. \ref{MeasureInducedByOrientation} is over each $y \in Y$
  the diagonal $A_y \to (A_y)^{\oplus^{n_1}}$. Dually, the induced measure is over each $y \in Y$
  the map $d\mu_{A_y} :  \mathbb{D}(A_y^{\oplus^{n_1}}) \to \mathbb{D}(A_y)$ which is the addition
  operation on $n_1$ covectors. This exhibits the canonical ``counting measure'' on the finite set $X_1$.

  An $n_1 \times n_2$-matrix
  $K \in \mathrm{Mat}_k(n_1, n_2)$
  is equivalently a diagram of functors of the form
  $$
    \xymatrix{
      & X_1 \times X_2
      \ar[dl]_{p_1}
      \ar[dr]^{p_2}_{\ }="s"
      \\
      X_1 \ar[dr]_{k_X}^{\ }="t" && X_2 \ar[dl]^{k_Y}
      \\
      & \mathrm{Vect}_k
      \ar@{=>}^K "s"; "t"
    }
    \,.
  $$
  This defines a (dual) prequantum integral kernel, def. \ref{IntegralKernel}
  between $A_1 = 1_{X_1}$ and $A_2 = 1_{X_2}$ the line bundle on $X_1$ and $X_2$,
  respectively,
  with the morphism
  $$
    \xi
      \;:\;
    (i_2)^\ast 1_{X_2} = 1_{X_1 \times X_2}
      \longrightarrow
    1_{X_1 \times X_2} =  (i_1)^\ast 1_{X_1}
  $$
  given over $(x,y) \in X_1 \times X_2$ by multiplication with the matrix element $K_{x,y}$.

  The induced integral kernel
  $$
    \underset{X_1}{\sum} 1_{X_1}
    \stackrel{\underset{X_1}{\sum}\epsilon}{\longleftarrow}
    \underset{X_1 \times X_2}{\sum} 1_{X_1 \times X_2}
    \stackrel{\Xi}{\longleftarrow}
    \underset{X_1 \times X_2}{\sum} 1_{X_1 \times X_2}
    \stackrel{\underset{X_2}{\sum} [i_2]}{\longleftarrow}
    \underset{X_2}{\sum} 1_{X_2}
  $$
  sends a vector
  $$
    v =
    \left[
      \begin{array}{c}
        v_1
        \\
        v_2
        \\
        \vdots
        \\
        v_{n_2}
      \end{array}
    \right]
    \in
    \underset{X_2}{\sum} 1_{X_2} \simeq k^{n_2}
  $$
  first via the diagonal along $X_1$ to the image under $\underset{X_1 \times X_2}{\sum}$ of
  $$
    \left[
      \begin{array}{cccc}
        v_1 & v_1 & \cdots & v_1
        \\
        v_2 & v_2 & \cdots & v_2
        \\
        \vdots
        \\
        v_{n_2} & v_{n_2} & \cdots & v_{n_2}
      \end{array}
    \right]
    \in
    1_{X_1 \times X_2}
  $$
  then via the integral kernel itself to the image under $\underset{X_1 \times X_2}{\sum}$ of
  $$
    \left[
      \begin{array}{cccc}
        K_{1,1} v_1 & K_{2,1} v_1 & \cdots & K_{n_1,1} v_1
        \\
        K_{1,2} v_2 & K_{2,2} v_2 & \cdots & K_{n_1,2} v_2
        \\
        \vdots
        \\
        K_{1,n_2} v_{n_2} & K_{2,n_2} v_{n_2} & \cdots & K_{n_1, n_2} v_{n_2}
      \end{array}
    \right]
    \in
    1_{X_1 \times X_2}
  $$
  and then via summation over $X_2$ to the image under $\underset{X_1}{\sum}$ of
  $$
    \left[
      \begin{array}{c}
        K_{1,1} v_1 + K_{1,2} v_2 + \cdots + K_{1,n_2} v_{n_2}
        \\
        K_{2,1} v_1 + K_{2,2} v_2 + \cdots + K_{2,n_2} v_{n_2}
        \\
        \vdots
        K_{n_1,1} v_1 + K_{n_1,2} v_2 + \cdots + K_{n_1,n_2} v_{n_2}
      \end{array}
    \right]
    \in
    1_{Y_1}
    \,,
  $$
  hence to the matrix product
  $$
    K \cdot v \in \underset{X_1}{\sum} 1_{X_1} \simeq k^{n_1}
    \,.
  $$
\end{example}

\subsection{Higher linear algebra -- Pull-push in twisted generalized cohomology}
 \label{PullPushInTwistedGeneralizedCohomology}

The previous example \ref{MatrixAsExampleOfAbstractIntegralKernel} considered linear types
given by $k$-modules over sets, for $k$ a commutative ring (a field). This setup has an evident refinement
to (stable) homotopy theory, where sets are refined to $\infty$-groupoids,
commutative rings to $E_\infty$-rings, and modules to module spectra over these.
This homotopy-theoretic refinement of linear algebra used to be advertised as
``brave new algebra'', especially when presented in terms of model categories of
structured ring spectra. In the intrinsic formulation of $\infty$-category theory
it is called ``higher algebra'' in \cite{LurieAlg}.

The following example \ref{PullPushInGeneralizedCohomology} shows how twisted Umkehr maps in generalized cohomology
as in \cite{ABG10} and section 4.1.4  \cite{Nuiten13} are an example of
the general concept of secondary integral transforms in dependent linear homotopy-type theory of def. \ref{IntegralKernelAndTransform}.

\begin{example}[pull-push in twisted generalized cohomology]
  \label{PullPushInGeneralizedCohomology}
  Let $E \in \mathrm{CRing}_\infty$
  be an $E_\infty$-ring spectrum with $\infty$-category of $\infty$-modules denoted $E \mathrm{Mod}$.
  For $X \in \infty \mathrm{Grpd}$ write
  $$
    E \mathrm{Mod}(X) := \mathrm{Func}(X,E \mathrm{Mod})
  $$
  for the $\infty$-category of $\infty$-functors from $X$ to $E \mathrm{Mod}$.
  An object in here is sometimes known as an $X$-parameterized module spectrum,
  and sometimes as a \emph{local system} of $E$-modules on $X$.

  For $f \;: \;X \longrightarrow Y$ a morphism of $\infty$-groupoids, there is an
  induced adjoint triple
  $$
    (\underset{f}{\sum} \dashv f^\ast \dashv \underset{f}{\prod})
    \;:\;
    \xymatrix{
      E\mathrm{Mod}(X)
      \ar@<+10pt>@{->}[rr]^{f_!}
      \ar@{<-}[rr]|-{f^\ast}
      \ar@<-10pt>@{->}[rr]_{f_\ast}
      &&
      E \mathrm{Mod}(Y)
    }
    \,,
  $$
  where $f_!$ and $f_\ast$ are left and right homotopy Kan extension along $f$,
  respectively.
  By example \ref{LocalSystems} this exhibits $E\mathrm{Mod}$ as a linear
  homotopy-type theory.

  For $X,Y, Z \in \infty \mathrm{Grpd}$ three homotopy types,
  consider  a diagram of $\infty$-functors of the form
  $$
    \xymatrix{
      & Z
      \ar[dl]_{g}
      \ar[dr]^{f}_{\ }="s"
       \\
      X \ar[dr]_{\alpha}^{\ }="t" && Y \ar[dl]^{\beta}
      \\
      & E \mathrm{Mod}
      \ar@{=>}^\xi "s"; "t"
    }
    \,.
  $$
  This induces a prequantum integral kernel,
  def. \ref{IntegralKernel}, of the form
  $$
    \xymatrix{
      & E \mathrm{Mod}(Z)
      \ar[dr]^{f_\ast}
      \ar[dl]_{g_\ast}
      \\
      E \mathrm{Mod}(X)
      \ar[dr]_{(p_1)_\ast}
      &&
      E \mathrm{Mod}(Y)
      \ar[dl]^{(p_2)_\ast}
      \\
      & E \mathrm{Mod}
    }
  $$
  with
  $
    \xi : f^\ast \beta \longrightarrow g^\ast \alpha
    \,.
  $
  Comparison with the discussion in \cite{ABG10} shows that
  $
    (p_1)_! \alpha \simeq E_{\bullet + \alpha}(X)
  $
  is the $\alpha$-twisted $E$-homology spectrum of $X$, and
  $
    \mathbb{D}((p_1)_! \alpha) \simeq E^{\bullet + \alpha}(X)
  $
  the $\alpha$-twisted $E$-cohomology spectrum. Similarly for $(Y,\beta)$.

  We may decompose the above slice correspondence $\xi$ as
  $$
    \raisebox{20pt}{
    \xymatrix{
      & Z \ar[dr]^f_{\ }
      \ar[dl]_g
      \ar[dd]^{f^\ast \beta}_{\ }="s"
      \\
      X
      \ar[dr]_\alpha^{\ }="t"
      & & Y \ar[dl]^{\beta}
      \\
      & E\mathrm{Mod}
      \ar@{=>}_\xi "s"; "t"
    }
    }
    \,.
  $$
  Consider then the definition for push-forward along the right leg of this
  diagram the way it appears as def. 4.1.24 in \cite{Nuiten13}. We show that this
  is a special case of the general def. \ref{IntegralKernelAndTransform}.

  To that end, notice that in def. 4.1.24 in \cite{Nuiten13}
  a choice of fundamental class is taken to be a choice of $\gamma \in E \mathrm{Mod}(Y)$
  together with an equivalence
  $$
    f_! f^\ast \beta \stackrel{\simeq}{\longrightarrow} \mathbb{D}(f_! f^\ast \gamma)
    \,.
  $$
  In the language used here this is a Wirthm{\"u}ller isomorphism,
  prop. \ref{TwistedOrientationMakesComonadTwistedCommutedWithDuality},
  for a choice of fiberwise fundamental class, def. \ref{TwistedOrientationOnWirthmuellerMorphism},
  under the identification
  $$
    \gamma = \mathbb{D}(\beta \otimes \tau)
    \,.
  $$
  Indeed, prop. 4.1.27 in \cite{Nuiten13} recovers this identification of $\tau$ for the case
  that $f$ comes from a proper surjective submersion of smooth manifolds; and remark 4.1.28 there
  observes that when such $f$ is $E$-orientable then $\gamma = d$ is the degree shift by the dimension
  of the fibers, as familiar from the classical Poincar{\'e}-Thom collapse map.

  Then further in def. 4.1.24 in \cite{Nuiten13} the corresponding secondary integral transform is
  taken to be the composite
  $$
    \xymatrix{
      E^{\bullet + f^\ast \beta}(Z)
      \ar@{=}[r]
      &
      \mathbb{D}(p_! f_! f^\ast \beta)
      \ar[d]^\simeq
      \ar@{-->}[dr]^{d\mu_f}
      \\
      &
      \mathbb{D}(p_! \mathbb{D}(f_!f^\ast \gamma))
      \ar[r]_-{p_! \mathbb{D} \epsilon_\gamma }
      &
      \mathbb{D}p_!(\mathbb{D}\gamma)
      \ar@{=}[r]
      &
      E^{\bullet - \gamma}(Y)
    }
  $$
  Comparison identifies the dashed diagonal
  composite morphism above indeed as the induced measure $d\mu_f$
  in the sense of def. \ref{MeasureInducedByOrientation}, as indicated.
  By the discussion in \cite{Nuiten13} this identifies the secondary integral transform
  here as given by the twisted Umkehr maps in generalized cohomology due to
  \cite{ABG11}
  $$
    \int_{i_2} \xi \, d\mu_{i_2}
    \;:\;
    E^{\bullet + \alpha}(X)
    \stackrel{}{\longrightarrow}
    E^{\bullet + {\beta + \tau}}(Y)
    \,,
  $$
\end{example}

The following example spells out how the construction considered in \cite{HopkinsLurie}
is a special case of the above.
\begin{example}
  Let $\mathcal{C}$ be a stable $\infty$-category with all limits and colimits,
  for instance the $\infty$-category $E \mathrm{Mod}$ of $\infty$-modules over some
  $E_\infty$-ring, in which case the following is
  a special case of example \ref{PullPushInGeneralizedCohomology}.
  For $X,Y \in \infty \mathrm{Grpd}$
  two homotopy types and $f : X \longrightarrow Y$ a morphism between them, consider
  the prequantum integral kernel, def. \ref{IntegralKernel}, given by
  the correspondence
  $$
    \xymatrix{
      & [X, \mathcal{C}]
      \ar[dr]^{f_\ast}_{\ }="s"
      \ar[dl]_{f_\ast}
      \\
      [Y,\mathcal{C}]
      \ar[dr]_{\mathrm{id}}^{\ }="t"
      &&
      [Y,\mathcal{C}]
      \ar[dl]^{\mathrm{id}}
      \\
      & [Y,\mathcal{C}]
      \ar@{=>}^= "s"; "t"
    }
  $$
  and by a choice of objects $C,D \in \mathrm{Func}(X,\mathcal{C})$ and
  a choice of a morphism
  $$
    \xi := f_! u \;:\; f_! f^\ast C \longrightarrow f_! f^\ast D
    \,.
  $$
  Suppose this $f$ is such that it carries a functorial un-twisted fundamental class,
  hence according to def. \ref{MeasureInducedByOrientation} a natural transformation
  $$
    \mu := [f] : \mathrm{id} \longrightarrow f_! f^\ast
    \,.
  $$
  Then according to def. \ref{IntegralKernelAndTransform} the dual secondary integral transformation induced by this
  data is the morphism
  $$
    \mathbb{D}
    \int_{f} \xi \, d\mu_f
    \;:\;
    D \stackrel{\epsilon}{\longleftarrow} f_! f^\ast D \stackrel{f_!(u)}{\longleftarrow} f_! f^\ast C\stackrel{\mu}{\longleftarrow} C
    \,.
  $$
  This is the notion of integration considered in Notation 4.1.6 of \cite{HopkinsLurie}
  (almost exactly denoted  there by the same symbols as here, only that we call it the \emph{dual}
  integration map, following the interpretation in the examples above).
  \label{Ambidexterity}
\end{example}

\section{Quantization}
\label{Quantization}

We now put everything together and discuss how the secondary integral transforms
axiomatized in dependent linear homotopy-type theory
in \ref{FundamentalClasses}, realized in the model of twisted generalized
cohomology theory in \ref{PullPushInTwistedGeneralizedCohomology}
serve to formalize aspects of quantum field theory.

The way quantum field theory appears in linear homotopy-type theory
turns out to be following a principle which in fundamental physics
has come to be known as the ``holographic principle'', where
a quantum field theory of some dimension carries another quantum field theory
of one dimension less ``on its boundary''. We find this phenomenon
appear in two stages:

First, in \ref{CohomologicalQuantization}, we find non-topological
$(d-1)$-dimensional quantum field theories $\mathrm{QFT}_{d-1}$ on the boundary
of $d$-dimensional topological quantum field theories $\mathrm{TQFT}_{d}^\tau$,
notably we find ordinary quantum mechanics on the boundary of the
non-perturbative 2d Poisson-Chern-Simons theory.
This is as in \cite{Nuiten13, dcct} and we will be brief.
But here the $d$-dimensional topological quantum field theory
is subject to various ``twists'' and in general there is an obstruction for these
twists to consistently orient such as to produce a globally well defined theory.
In \ref{CoboundingTheory} we observe that this consistent global choice of fundamental classes
is equivalent to the $d$-dimensional theory itself being the boundary theory
of an (untwisted) $(d+1)$-dimensional theory $\mathrm{TQFT}_{d+1}$.

Here the obstruction to consistent orientations is also called a \emph{quantum anomaly},
and hence the existence of $\mathrm{TQFT}_{d+1}$ is a quantum anomaly cancellation condition.

\subsection{$\mathrm{TQFT}_d^\tau$ via secondary integral transform quantization}

A fairly well-known example that has been part of the motivation for
a formalization of path integral quantization in terms of pull-push in (generalized) cohomology
is the following.
\begin{example}[string topology operations]
  Let the ground ring $E = \mathbb{S} \in \mathrm{CRing}_{\infty}$ be the sphere spectrum
  and consider the linear homotopy-type theory
  of example \ref{ParameterizedSpectra}
  $$
    \xymatrix{
      \mathrm{Sp}
      \ar[d]
      \\
      \mathrm{Grpd}_\infty
    }
    \,,
  $$
  where
  $\mathbb{S}\mathrm{Mod} \simeq \mathrm{Sp}$ is the $\infty$-category of parameterized
  spectra.
  Let $X \in \mathrm{Grpd}_\infty$ be the homotopy-type
  of a compact manifold and consider correspondences induced from
  2-dimensional cobordisms
  $$
    
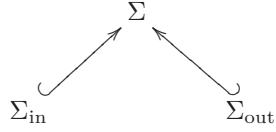
\xymatrix{
      & \Sigma
      \\
      \Sigma_{\mathrm{in}}
      \ar@{^{(}->}[ur]
      &&
      \Sigma_{\mathrm{out}}
      \ar@{_{(}->}[ul]
    }
  $$
  with non-empty outgoing boundary $\Sigma_{\mathrm{out}}$
  as
  $$
    
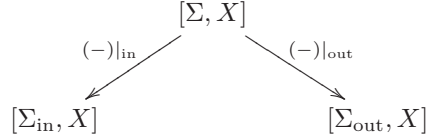
\xymatrix{
      & [\Sigma, X]
      \ar[dl]_{(-)|_{\mathrm{in}}}
      \ar[dr]^{(-)|_{\mathrm{out}}}
      \\
      [\Sigma_{\mathrm{in}},X]
      &&
      [\Sigma_{\mathrm{out}}, X]
    }
    \,.
  $$
  Via example \ref{PullPushInGeneralizedCohomology}
  the secondary integral transforms of def. \ref{IntegralKernel} are
  given by pull-push of ordinary cohomology in this case.
  By \cite{CohenJones}
  these are equivalently the
  \emph{Chas-Sullivan string topology operations} on the
  ordinary homology of $X$ . As a system these
  constitute a 2-dimensional topological quantum field theory $\mathrm{TQFT}_2$
  \cite{CohenGodin}, see also example 4.2.16 and remark 4.2.17 in
  \cite{LurieQFT}.
  \label{StringTopologyOperations}
\end{example}

While example \ref{StringTopologyOperations} stands out as one of the few
examples of extended quantum field theories that have already been explicitly constructed by
a quantization process from geometric prequantum data, in the
traditional formulation this prequantum data is not very explicit.

The following example very briefly recalls how local prequantum field
data is encoded in section 3.9.14 of \cite{dcct}
\begin{example}[local Lagrangian prequantum field theory]
\label{LocalLagrangianFieldTheory}
Let $\mathbf{H}$ be an $\infty$-topos equipped with differential cohesion,
def. \ref{DifferentialCohesion} whose shape modality $\int$
induces localization at a line object $\mathbb{A}^1$.
Write
$$
  U(1) := \mathbb{A}^1 / \mathbb{Z}
  \in
  \mathrm{Grp}(\mathbf{H})
$$
for the induced circle group.
Then for $n \in \mathbb{N}$ there is, by cohesion, canonically a moduli stack
$\mathbf{B}^{n+1} U(1)_{\mathrm{conn}}\in \mathbf{H}$ of $n$-form connections.
The slice $\mathbf{H}_{/\mathbf{B}^n U(1)_{\mathrm{conn}}}$
is the collection of \emph{local Lagrangians}
for $(n+1)$-dimensional local field theory
$$
  \left[
  \raisebox{20pt}{
  \xymatrix{
    \mathbf{Fields}
    \ar[d]^{\mathbf{L}}
    \\
    \mathbf{B}^{n+1} U(1)_{\mathrm{conn}}
  }
  }
  \right]
  \;\;
  \in \mathbf{H}_{/\mathbf{B}^{n+1}U(1)_{\mathrm{conn}}}
  \,,
$$
where the dependent sum
$$
  \mathbf{Fields} := \underset{\mathbf{B}^{n+1}U(1)_{\mathrm{conn}}}{\sum} \mathbf{L}
$$
is interpreted as the moduli stack of fields on which the Lagrangian is defined.

The (concretified) automorphism $n$-groups of such a local Lagrangian in
$\mathbf{H}_{\mathbf{L}}$ encode the Hamilton-de Donder-Weyl equations of motion
of the classical local field theory induced by this Lagrangian
(this is discussed in section 1.2.11 of \cite{dcct}, a lightning review is in \cite{Schreiber13}).

More generally,
given a type of incoming fields $\mathbf{Fields}_{\mathrm{in}} \in \mathbf{H}$
and a type of outgoing fields $\mathbf{Fields}_{\mathrm{out}}$, then
a type of field \emph{trajectories} (i.e. paths, histories) from these incoming to
these outgoing fields
is encoded by a correspondence of the form
$$
  \raisebox{20pt}
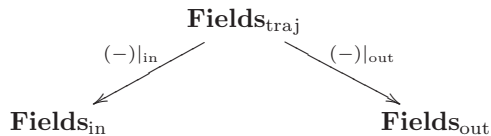
{
  \xymatrix{
    & \mathbf{Fields}_{\mathrm{traj}}
    \ar[dl]_{(-)|_\mathrm{in}}
    \ar[dr]^{(-)|_\mathrm{out}}_{\ }="s"
    \\
    \mathbf{Fields}_{\mathrm{in}}
    &&
    \mathbf{Fields}_{\mathrm{out}}
  }
  }
  \,.
$$
Now given a ground ring ($E_\infty$-ring) $E \in \mathrm{CRing}_\infty$,
then we say that
a choice of \emph{superposition principle} is a choice of representation
$$
  \rho
  \;:\;
  \mathbf{B}^n U(1)
  \longrightarrow
  \mathrm{GL}_1(E)
$$
of the $\infty$-group of phases $\mathbf{B}^n U(1)$ on the invertible elements of $E$.
Given this, the local Lagrangian defines a linear homotopy-type
depending on $\mathbf{Fields}$ which is given by the composite
$$
  \xymatrix{
    \mathbf{Fields}
    \ar[d]^{\mathbf{L}}
    \\
    \mathbf{B} (\mathbf{B}^n U(1))
    \ar[d]^{\mathbf{B}\rho}
    \\
    \mathbf{B}\mathrm{GL}_1(E)
  }
$$
and which by convenient abuse of notation we denote by the same symbol:
$$
  \mathbf{L} \in E \mathrm{Mod}(\mathbf{Fields})
  \,.
$$
In this form one may think of $\mathbf{L}$ equivalently as the local
(fully extended) \emph{prequantum line bundle} of the field theory,
see \cite{hgp}.

Now with a correspondence given by field trajectories as above, and
given local Lagrangians on the incoming and on the outgoing fields,
then an extension of this data to a \emph{prequantized Lagrangian correspondence}
is a completion to diagram in $\mathbf{H}$ of the form
$$
  \raisebox{20pt}{
  \xymatrix{
    & \mathbf{Fields}_{\mathrm{traj}}
    \ar[dl]_{\mathrm{in}}
    \ar[dr]^{\mathrm{out}}_{\ }="s"
    \\
    \mathbf{Fields}_{\mathrm{in}}
    \ar[dr]_{\mathbf{L}_{\mathrm{in}}}^{\ }="t"
    &&
    \mathbf{Fields}_{\mathrm{out}}
    \ar[dl]^{\mathbf{L}_{\mathrm{out}}}
    \\
    & E \mathrm{Mod}
    \ar@{=>}|{\exp(\tfrac{i}{\hbar}S)} "s"; "t"
  }
  },.
$$
Here the homotopy filling this diagram, which we denote
``$\exp(\tfrac{i}{\hbar}S)$'' plays the role of the
\emph{action functional} that sends each trajectory to the
quantum phase that is induces.

The upshot for the present discussion is that in terms
of linear homotopy-type theory such data of a
prequantized Lagrangian correspondence
is equivalently a prequantum integral kernel
in the sense of def. \ref{IntegralKernel},
$$
  \exp(\tfrac{i}{\hbar}S)
  \;:\; \mathrm{in}^\ast \mathbf{L}_{\mathrm{in}} \longleftarrow \mathrm{out}^\ast \mathbf{L}_{\mathrm{out}}
$$
on the given  trajectory correspondence.
\end{example}
Specific examples of this class of example include
example \ref{KUBoundaryQuantization} below. Further below in \ref{CoboundingTheory}
we see a more fundamental origin of these action functionals.

\subsection{$\mathrm{QFT}_{d-1}$ via boundary field theory}
 \label{CohomologicalQuantization}

We now combine all the ingredients and indicate how the cohomological formulation
of boundary field theory quantization in \cite{Nuiten13, dcct} appears in the formal axiomatization in
cohesive linear homotopy-type theory.

\begin{itemize}
  \item example \ref{KUBoundaryQuantization}  -- quantum particle at the boundary of 2d Poisson-Chern-Simons theory;
  \item example \ref{stringattheboundary} -- quantum superstring at the boundary of 3d Spin-Chern-Simons theory;
  \item example \ref{DBraneChargeAndTDuality} -- D-brane charge and T-duality
\end{itemize}

\medskip

The following example indicates how example \ref{LocalLagrangianFieldTheory}
specialises to reproduce traditional non-perturbative geometric quantization
(refined from symplectic manifolds to Poisson manifolds).

\begin{example}[quantum particle at the boundary of 2d Poisson-Chern-Simons theory]
 \label{KUBoundaryQuantization}
Let $(X,\pi)$ be a Poisson manifold (hence a foliation of symplectic manifolds,
hence a foliation of phase spaces of mechanical systems).
By the discussion in \cite{hgp, Bongers} this canonically induces
the local action functional of a 2-dimensional Poisson-Chern-Simons
topological field theory
$$
  \xymatrix{
    \mathrm{SymplGrpd}(X,\pi)_{\mathrm{conn}}
    \ar[d]^{\exp(\tfrac{i}{\hbar} S_{2})}
    \\
    \mathbf{B}^2 U(1)_{\mathrm{conn}}
  }
$$
whose moduli stack of fields is a differential cohomological refinement of
what is called the symplectic groupoid $\mathrm{SymplGrpd}(X,\pi)$ associated with
the Poisson manifold. This is such that perturbatively (on vanishing instanton sectors)
it reproduces the Poisson-sigma-model, whose perturbative quantization has
been shown by \cite{CattaneoFelder} to yield on its boundary Kontsevich's
perturbative deformation quantization of $(X,\pi)$.
Here we are after the full, non-perturbative quantization
(geometric quantization).

The quantization of $\exp(\tfrac{i}{\hbar} S_2)$ is supposed to yield
a 2-dimensional $\mathrm{TQFT}_2^{\tau}$. But here we will concentrate only
on the boundary conditions of this theory, hence of its higher space of states in
codimension 2. For this we may forget the differential refinement and consider
just the action on the underlying instanton sectors
$$
  \xymatrix{
    \mathrm{SymplGrpd}(X,\pi)
    \ar[d]^{\exp(\tfrac{i}{\hbar} S_{2})}
    \\
    \mathbf{B}(\mathbf{B}^1 U(1)_{\mathrm{conn}})
  }
  \,.
$$
This restriction  classifies what is known as the
groupoid pre-quantization of the symplectic groupoid \cite{Bongers}. From this one
observes that the original Poisson manifold is canonical a boundary
condition for the 2d Poisson-Chern-Simons theory that it induces, in that there
is a morphism in $\mathrm{Corr}(\mathbf{H}_{\mathbf{B}^2 U(1)})$ which in
$\mathbf{H}$ is of the form
$$
  \xymatrix{
    & X
    \ar[dr]_-{\ }="s"
    \ar[dl]
    \\
    \ast \ar[dr]^{\ }="t" && \mathrm{SymplGrpd}(X,\pi)\ar[dl]
    \\
    & \mathbf{B}^2 U(1)
    \ar@{=>} "s"; "t"
  }
$$
Now let $E = \mathrm{KU}$ be the complex K-theory $E_\infty$-ring.
By Snaith's theorem mentioned above this prequantum correspondence linearizes
to a prequantum integral kernel represented by
$$
  
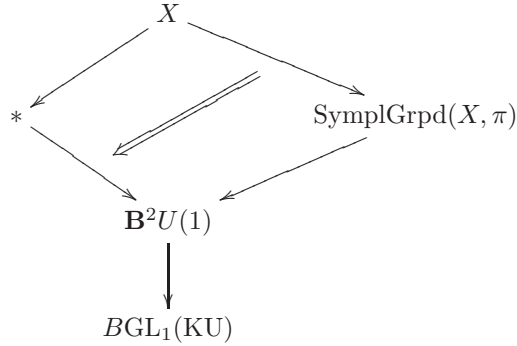
\xymatrix{
    & X
    \ar[dr]_-{\ }="s"
    \ar[dl]
    \\
    \ast \ar[dr]^{\ }="t" && \mathrm{SymplGrpd}(X,\pi)\ar[dl]
    \\
    & \mathbf{B}^2 U(1)
    \ar[d]
    \\
    & B \mathrm{GL}_1(\mathrm{KU})
    \ar@{=>} "s"; "t"
  }
$$
via the construction in example \ref{PrequantumIntegralKernelFromPrequantumCorrespondence}.

Therefore if the right leg can be equipped with a fiberwise fundamental class
(typically requiring $X$ to be compact) this correspondence induces a secondary integral transform, def. \ref{IntegralKernel},
which by example \ref{PullPushInGeneralizedCohomology} is given by
pull-push in K-theory as $\mathrm{KU}$-linear map
$$
  \mathrm{KU} \longleftarrow \mathrm{KU}_{\bullet + \chi}(\mathrm{SymplGrpd(X)})
$$
which exhibits a class in the complex K-theory of the symplectic groupoid induced by $X$.
In view of remark \ref{UmkehrInKK} this may be expressed for instance by a composite of
bivariant classes in KK-theory.
This class may be represented by complex vector spaces associated to the symplectic leaves of
$X$ and one finds that these represent the spaces of quantum states under traditional geometric
quantization. 

More generally, if there is a Lie group $G$ with a Hamiltonian action on $X$ 
(for instance time evolution for $G = \mathbb{R}$) then 
the quantization takes place in $G$-equivariant K-theory and produces a $G$-equivariant
K-theory class of the point, hence a $G$-representation, being the action of the $G$-quantum operators
on the space of quantum states.
For more details see section 5.2 of \cite{Nuiten13}.
\end{example}

The following example lifts the previous one up in dimension, from particles to strings.

\begin{example}[string at the boundary of 3d Chern-Simons theory]
  \label{stringattheboundary}
  The (fully extended) local Lagrangian for 3d Chern-Simons theory is \cite{StackyPerspective}
  $$
    \raisebox{20pt}{
    \xymatrix{
      \mathbf{B}\mathrm{Spin}_{\mathrm{conn}}
      \ar[d]^{\tfrac{1}{2}\widehat{\mathbf{p}_1}}
      \\
      \mathbf{B}^3 U(1)_{\mathrm{conn}}
    }
    }
    \,.
  $$
  The \emph{universal} boundary condition for this \cite{dcct},
  namely the homotopy fiber of $\tfrac{1}{2}\widehat{\mathbf{p}_1}$ is the moduli 2-stack
  of String-principal 2-connections \cite{DiffClasses}
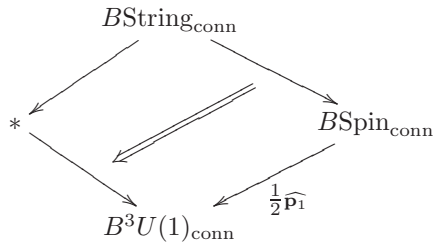
  $$
    \raisebox{20pt}{
    \xymatrix{
      & B \mathrm{String}_{\mathrm{conn}}
      \ar[dl]
      \ar[dr]_{\ }="s"
      \\
      \ast \ar[dr]^{\ }="t"
      && B \mathrm{Spin}_{\mathrm{conn}} \ar[dl]^{\tfrac{1}{2}\widehat{\mathbf{p}_1}}
      \\
      & B^3 U(1)_{\mathrm{conn}}
      \ar@{=>} "s"; "t"
    }
    }
    \,.
  $$
  Notice that this means that for a Spin-manifold
  $\nabla_{\mathrm{Spin}} : X \stackrel{}{\longrightarrow} B \mathrm{Spin}$
  to constitute a boundary condition
  is equivalent to it having (differential) String structure:
  $$
    \raisebox{20pt}
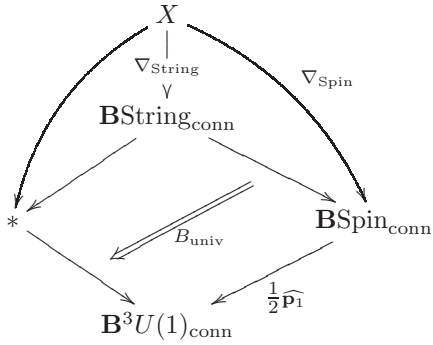
{
    \xymatrix{
      & X
       \ar@{-->}[d]|{\nabla_{\mathrm{String}}}
       \ar@/^1pc/[ddr]^{\nabla_{\mathrm{Spin}}}
       \ar@/_1pc/[ddl]
      \\
      & \mathbf{B} \mathrm{String}_{\mathrm{conn}}
      \ar[dl]
      \ar[dr]_{\ }="s"
      \\
      \ast \ar[dr]^{\ }="t" && \mathbf{B} \mathrm{Spin}_{\mathrm{conn}} \ar[dl]^{\tfrac{1}{2}\widehat{\mathbf{p}_1}}
      \\
      & \mathbf{B}^3 U(1)_{\mathrm{conn}}
      \ar@{=>}^{B_{\mathrm{univ}}} "s"; "t"
    }
    }
    \,.
  $$
  Here the total homotopy filling this diagram is physically the twisted Kalb-Ramond B-field on
  $X$ to which the heterotic string on $X$ couples \cite{TwistedDifferential}.

  The canonical choice for an $E_\infty$-ring with a superposition principle in this degree
  is $\mathrm{tmf}$  (see section 8 of \cite{ABG10}).   $$
    \raisebox{20pt}
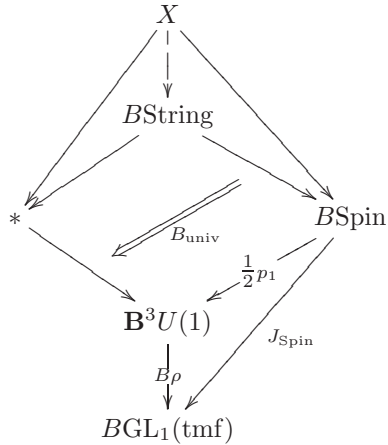
{
    \xymatrix{
      & X
       \ar@{-->}[d]|{}
       \ar[ddr]^{}
       \ar[ddl]
      \\
      & B \mathrm{String}
      \ar[dl]
      \ar[dr]_{\ }="s"
      \\
      \ast \ar[dr]^{\ }="t" && B \mathrm{Spin}
        \ar[dl]|{\tfrac{1}{2}p_1}
        \ar[ddl]^{J_{\mathrm{Spin}}}
      \\
      & \mathbf{B}^3 U(1)
      \ar[d]|{B \rho}
      \\
      & B \mathrm{GL}_1(\mathrm{tmf})
      \ar@{=>}^{B_{\mathrm{univ}}} "s"; "t"
    }
    }
    \,.
  $$
  The corresponding
  prequantum $\mathrm{tmf}$-line bundle is that classified by the Spin J-homomorphism
  $$
    J_{\mathrm{Spin}} \simeq B \rho\circ \tfrac{1}{2}p_1
  $$
  (see diagram (8.1) in \cite{ABG10}).
  Therefore for considering pull-push through this correspondence, it is natural to factor it through
  the pullback of $J_{\mathrm{Spin}}$ to $B \mathrm{String}$, which is the
  universal String J-homomorphism $J_{\mathrm{String}}$.
  Such a factorization
  $$
    \raisebox{20pt}
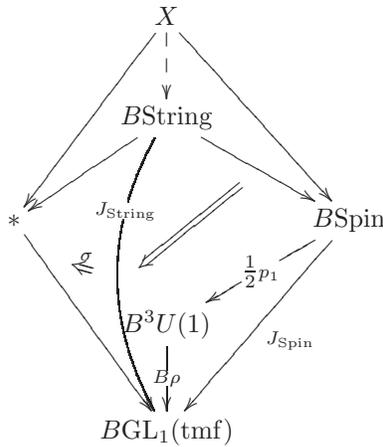
{
    \xymatrix{
      & X
      \ar[ddl]
      \ar[ddr]
      \ar@{-->}[d]
      \\
      & B \mathrm{String}
      \ar[dl]
      \ar[dr]_{\ }="s1"
      \ar@/_1.56pc/[ddd]|<<<<<<<<<<{J_{\mathrm{String}}}_-{\ }="s"^{\ }="t1"
      \\
      \ast \ar[ddr]^<<<<<<{\ }="t"
      && B \mathrm{Spin} \ar[dl]|-{\tfrac{1}{2}p_1}
      \ar[ddl]^{J_{\mathrm{Spin}}}
      \\
      & B^3 U(1)
      \ar[d]|{B\rho}
      \\
      & B \mathrm{GL}_1(\mathrm{tmf})
      \ar@{=>}_\sigma "s"; "t"
      \ar@{=>} "s1"; "t1"
    }
    }
  $$
  is given
  by the homotopy $\sigma$ which is the \emph{String orientation of $\mathrm{tmf}$} \cite{AHR}
  in its moduli incarnation
  (this boundary condition correspondence \emph{is} the lower two third of diagram (8.1) in \cite{ABG10}).

  Under quantization the left part of this decomposition becomes, via example \ref{PullPushInGeneralizedCohomology}, the
  $\mathrm{tmf}$-linear map
  $$
    \xymatrix{
      &&
      X^{T X}\wedge \mathrm{tmf}
      \ar[d]
      \\
      \mathrm{tmf} = \mathrm{tmf}_\bullet(\ast)
      \ar@{<-}[r]^-\sigma
      &
      \mathrm{tmf}_{\bullet + J_{\mathrm{String}}}(B \mathrm{String})
      \ar@{}[r]|-{\simeq}
      &
      M\mathrm{String}\wedge \mathrm{tmf}
    }
  $$
  underlying which is the $\mathbb{S}$-linear map (we denote all these incarnations by the same
  symbol ``$\sigma$'', for convenience)
  $$
    \mathrm{tmf} \stackrel{\sigma}{\longleftarrow} M\mathrm{String}
  $$
  which is the String orientation as usually given.
  In \cite{AHR} it is show that this is on homotopy groups the \emph{Witten genus}.
  This is the  partition function of the heterotic superstring
  \cite{WittenGenus} (with target space the given String-spacetime $X$ above).
\end{example}

\begin{example}[D-brane charge and T-duality]
  \label{DBraneChargeAndTDuality}
  Given a superstring spacetime $X$ equipped with a spin-structure and a Kalb-Ramond B-field
  $\hat B : \to \mathbf{B}^2 U(1)_{\mathrm{conn}}$, then a prequantum D-brane $Q$ in $X$ is a boundary condition
  for this, hence given by a diagram
  $$
    \raisebox{20pt}{
    \xymatrix{
      & Q \ar@{^{(}->}[dr]^{i}_{\ }="s"
      \ar[dl]
      \\
      \ast
      \ar[dr]^{\ }="t"
      && X \ar[dl]^{\chi}
      \\
      & \mathbf{B}^2 U(1)
      \ar@{=>}^\xi "s"; "t"
    }
    }
    \,.
  $$
  Here the homotopy trivializing this diagram is a trivialization of the class of the background B-field
  restricted to the brane $Q$, given by a (twisted) Chan-Paton gauge field $\xi$ on the brane. Via the canonical linearization $\mathbf{B}^2 U(1)\to B \mathrm{GL}_1(\mathrm{KU})$
  already used in example \ref{KUBoundaryQuantization}
  this is naturally quantized over $\mathrm{KU}$-coefficients. For compact $Q$ a fiberwise fundamental class 
  on $i$ exists with twist $\tau$ the third integral Stiefel-Whitney class $\nu_Q$ of the normal bundle of the inclusion $i$. This twist vanishes if the normal bundle admits a $\mathrm{Spin}^c$-structure. Given such,
  the path integral quantization of this boundary is a $\mathrm{KU}$-linear map
  $$
    \mathrm{KU} \longleftarrow \mathrm{KU}_{\bullet + \chi}(X)
  $$
  exhibiting a class in the $\chi$-twisted K-theory of spacetime $X$. This the
  \emph{D-brane} of $(Q, \xi)$ in $X$ and the $\mathrm{Spin}^c$-condition is the
  \emph{Freed-Witten anomaly cancellation} condition for the brane.
  
  More generally a quantum boundary
  condition according to def. \ref{IntegralKernel}
  is any class $\xi$ in the $(i^\ast \chi + \nu_Q)$-twisted K-theory of $Q$ and the 
  $D$-brane charge is the push-forward of that to the $\chi$-twisted K-theory of $Q$. 
  In this case $\xi$ is a non-abelian Chan-Paton gauge bundle on $Q$ and the fact that 
  it is in $(i^\ast \chi + \nu_Q)$-twisted K-theory is the \emph{Freed-Witten-Kapustin anomaly}
  cancellation.   For more see section 5.2 4 of \cite{Nuiten13}.

  Now suppose that $X$ is a torus-fiber bundle over some base $Y$. Then a
  (topological) T-duality transformation is a correspondence of the form
  $$
  \xymatrix{
     & X \times_Y \tilde X
     \ar[dl]
     \ar[dr]_-{\ }="s"
     \\
     X
     \ar[dr]
     \ar[ddr]_{\chi}^{\ }="t"
     &&
     \tilde X
     \ar[dl]
     \ar[ddl]^{\tilde \chi}
     \\
     & Y
     \\
     & \mathbf{B}^2 U(1)
     \ar[d]
     \\
     & B \mathrm{GL}_1(\mathrm{KU})
     \ar@{=>} "s"; "t"
  }
$$
where $\tilde X \to Y$ is another torus bundle spacetime with another B-field $\tilde \chi$,
and the homotopy filling this diagram is a (twisted) line bundle on the torus fiber product
$X \times_Y \tilde X$ with the property that on each fiber it restricts to the canonical Poincar{\'e}-line
bundle on the product of a torus and its dual, up to equivalence.
The fundamental theory of topological T-duality says, in our language here, that the
path integral quantization of this correspondence induces an equivalence
$$
  \xymatrix{
    \mathrm{KU}_{\bullet + \chi}
    \stackrel{\simeq}{\longleftarrow}
    \mathrm{KU}_{\bullet + \tilde \chi}
  }
$$
between the twisted K-theories of the two T-dual string backgrounds, hence between their
D-brane charges. In terms of KK-theory this is discussed in \cite{BMRS}.

According to example \ref{stringattheboundary} the above correspondences should be
regarded as correspondences-of-correspondences between those in example \ref{stringattheboundary},
where the B-field is not a map to $B\mathrm{GL}_1(\mathrm{KU})$ but a homotopy between maps
to $B \mathrm{GL}_1(\mathrm{tmf})$. This would mean that T-duality then induces an 
equivalence of non-perturbative Witten genera, namely of tmf-classes of the String-spacetimes
$X$ and $\tilde X$. This has indeed been found to be the case by Thomas Nikolaus for a fairly
large class of cases\footnote{See Nikolaus' talk at the String Geometry Network meeting.}.
\end{example}

\subsection{$\mathrm{TQFT}_{d+1}$ via quantum anomaly cancellation}
\label{CoboundingTheory}

Given a correspondence
$$
  \xymatrix{
    X_1
    \ar@{<-}[r]^-{i_1}
    &
    Z
    \ar[r]^-{i_2}
    &
    X_2
  }
$$
we defined in def. \ref{IntegralKernelAndTransform} an integral kernel based on this
corrrespondence to be data
of the form
$$
  \xi : i_1^\ast A_1 \longleftarrow  i_2^\ast A_2
  \,.
$$
One may ask where this form of data comes from. In example
\ref{PrequantumIntegralKernelFromPrequantumCorrespondence} and
then more specifically in example \ref{LocalLagrangianFieldTheory} we gave a class of
constructions that occur naturally in practice which do yield this kind of
data. But here we want to go one step further and understand this
data as being in turn the boundary field theory data of a TQFT of
yet one more dimension higher.

Moreover, so far the $\mathrm{TQFT}_d^\tau$ which we obtained correspondence-wise by quantization
via secondary integral transforms may be
``anomalous'' in that its correspondence-wise construction does not actually extend to a
monoidal functor
$$
  \mathrm{TQFT}_d^\tau
    :
  \xymatrix{
   \mathrm{Bord}_n \ar[rr]^-{\exp(\tfrac{i}{\hbar}S)d\mu}
   &&
   \mathrm{Corr}(\mathbf{H})
   \ar[r]^-{\int (-) d\mu }
   &
    \mathrm{Mod}(\ast)
    }
  \,.
$$
Here we show that the condition that $\mathrm{TQFT}_d^\tau$ is quantum anomaly free
means that it is itself the boundary field theory of yet another
$\mathrm{TQFT}_{d+1}$.\footnote{The result here is joint with Joost Nuiten.}

\begin{definition}
\label{coboundingTQFTFunctor}
For $\mathrm{Mod}(-)\to \mathbf{H}$ a model for linear homotopy-type theory,
which satisfies Beck-Chevalley, def. \ref{BeckChevalleyCondition}, and is 2-monoidal,
def. \ref{2MonoidalTypeTheory},
write
$$
  \mathrm{TQFT}_{d+1}
  \;:\;
  \mathrm{Corr}_1(\mathbf{H})
  \longrightarrow
  \mathrm{Mod}_2
$$
for the $(\infty,1)$-functor
from the $(\infty,1)$-category of correspondences in $\mathbf{H}$
to the $(\infty,2)$-category of 2-modules, def.\ref{Mod2},
given by sending homotopy-types $X$ to their
$\infty$-categories $\mathrm{Mod}(X)$ of linear homotopy-types
dependent on them, and sending correspondences $X_1 \stackrel{i_1}{\leftarrow} Z \stackrel{i_2}{\rightarrow} X_2$
as above to
their \emph{linear polynomial functors}
$$
  \xymatrix{
    \mathrm{Mod}(X_1)
    \ar@{<-}[rr]^{\underset{i_1}{\sum} \circ i_2^\ast}
    &&
    \mathrm{Mod}(X_2)
  }
$$
as in def. \ref{PolynomialFunctor}.
\end{definition}
\begin{remark}
  That def. \ref{coboundingTQFTFunctor} indeed gives a monoidal $(\infty,2)$-functor
  is the content of cor. \ref{LinearPolynomialFunctorAssignmentIsMonoidal}.
\end{remark}
\begin{proposition}
  In a linear homotopy-type theory which
  satisfies Beck-Chevalley, def. \ref{BeckChevalleyCondition}, and is 2-monoidal, def. \ref{2MonoidalTypeTheory},
  then the functor $\mathrm{TQFT}_{d+1}$ in def. \ref{coboundingTQFTFunctor} is monoidal.
\end{proposition}
\proof
 By assumption of 2-monoidalness it suffices to see that
 for $X_1 \stackrel{i_1}{\leftarrow} Z\stackrel{i_2}{\to} X_2$
 and $\tilde X_1\stackrel{\tilde i_1}{\leftarrow} \tilde Z \stackrel{\tilde i_2}{\to}\tilde X_2$
 two correspondences in $\mathbf{H}$, and $(p_1^\ast A) \otimes (p_2^\ast \tilde A) \in \mathrm{Mod}(X_2 \times \tilde X_2)$,
 then
 $$
   \underset{i_1\times \tilde i_1}{\sum} (i_2 \times \tilde i_2)^\ast
   \left((p_1^\ast A) \otimes (p_2^\ast \tilde A)\right)
   \simeq
   \left(
      p_1^\ast \underset{i_1}{\sum}i_2^\ast A_1
   \right)
   \otimes
   \left(
      p_2^\ast \underset{i_1}{\sum}\tilde i_2^\ast \tilde A
   \right)
   \,.
 $$
 Given the assumption of Beck-Chevalley, this is the statement of prop.\ref{BCImpliesOtimesPreservesIndexedColimits}.
\endofproof

We now consider boundary conditions for $\mathrm{TQFT}_{d+1}$. For
that purpose write
$$
  1_{d+1}
  \;:\;
  \mathrm{Corr}(\mathbf{H})
  \longrightarrow
  \mathrm{Mod}_2
  \,,
$$
for the $(\infty,2)$-functor which sends every correspondence to the
identity functor on $\mathrm{Mod}(\ast)$.
\begin{proposition}
  A $\mathrm{Mod}(\ast)$-linear natural transformation
  $$
    \exp(\tfrac{i}{\hbar}S)
    \;:\;
    1_{d+1}\longrightarrow \mathrm{TQFT}_{d+1}
  $$
  is over each correspondence $X_1 \leftarrow Z \rightarrow X_2$
  equivalently a prequantum integral kernel, def. \ref{IntegralKernel}.
  \label{IntegralKernelAsTransformation}
\end{proposition}
\proof
Consider the naturality square
$$
  \xymatrix{
    \mathrm{Mod}(\ast)
    \ar@{<-}[rr]^-{\mathrm{id}}_<{\ }="s"
    \ar[d]_{1_\ast \mapsto A_1}
    &&
    \mathrm{Mod}(\ast)
    \ar[d]^{1_\ast \mapsto A_2}
    \\
    \mathrm{Mod}(X_1)
    \ar@{<-}[rr]_{\underset{i_1}{\sum} \circ i_2^\ast}^>{\ }="t"
    &&
    \mathrm{Mod}(X_2)
    \ar@{<=}|{\exp(\tfrac{i}{\hbar}S)} "s"; "t"
  }
  \,.
$$
Here by $\mathrm{Mod}(\ast)$-linearity the vertical functors are fixed
by their image of the tensor unit, which we denote by $A_1$, $A_2$
respectively.
Therefore the unit component of this natural transformation on this $A_2$
has to be a morphism in $\mathrm{Mod}(X_1)$ of the form
$$
  \exp(\tfrac{i}{\hbar}S)_{A_2}
   :
  \underset{i_1}{\sum} i_2^\ast A_2
  \stackrel{}{\longrightarrow}
  A_1
  \,.
$$
On a general object $\tau \in \mathrm{Mod}(\ast)$ the component of the
transformation has to be of the form
$$
  \exp(\tfrac{i}{\hbar}S)_{A_2\otimes \tau}
  :
  \underset{i_1}{\sum} (i_2^\ast A_2 \otimes Z^\ast \tau)
  \simeq
  \underset{i_1}{\sum} (i_2^\ast A_2 \otimes i_1^\ast X_1^\ast \tau)
  \stackrel{}{\longrightarrow}
  A_1 \otimes X_1^\ast \tau
$$
which by Frobenius reciprocity, def. \ref{FrobeniusReciprocity}, is
equivalently of the form
$$
  \exp(\tfrac{i}{\hbar}S)_{A_2\otimes \tau}
  :
  (\underset{i_1}{\sum} i_2^\ast A_2) \otimes X_1^\ast \tau
  \stackrel{}{\longrightarrow}
  A_1 \otimes X_1^\ast \tau
  \,.
$$
By linearity this is fixed to be
$\exp(\tfrac{i}{\hbar}S)_{A_2\otimes \tau} \simeq \exp(\tfrac{i}{\hbar}S)_{A_2}\otimes \mathrm{id}_{\tau}$
and hence the transformation is equivalent to the data consisting of
$A_1$, $A_2$ and $\exp(\tfrac{i}{\hbar}S)_{A_2}$.
Finally observe that by the $(\underset{i_1}{\sum} \dashv i_1^\ast)$-adjunction
the datum $\exp(\tfrac{i}{\hbar}S)_{A_2}$ is equivalently
given by its adjunct
$
  \xi : i_2^\ast A_2 \longrightarrow i_1^\ast A_1
  \,,
$
which is the integral kernel in question.
\endofproof
\begin{definition}
Given a choice of untwisted fiberwise fundamental class on $i_2$, def. \ref{TwistedOrientationOnWirthmuellerMorphism},
consider the transformation
$$
  \mathbb{D}\int(-)d\mu \;:\; \mathrm{TQFT}_{d+1} \longrightarrow 1_{d+1}
$$
restricted to the given correspondence $X_1 \stackrel{i_1}{\leftarrow} Z \stackrel{i_2}{\to} X_2$
whose component there is
$$
  \raisebox{20pt}{
  \xymatrix{
    \mathrm{Mod}(X_1)
    \ar@{<-}[rr]^{\underset{i_1}{\sum} \circ i_2^\ast}_<{\ }="t"
    \ar[d]_{\underset{{X_1}}{\sum}}
    &&
    \mathrm{Mod}(X_2)
    \ar[d]^{\underset{X_2}{\sum}}
    \\
    \mathrm{Mod}(\ast)
    \ar@{<-}[rr]_-{\mathrm{id}}^>{\ }="s"
    &&
    \mathrm{Mod}(\ast)
    \ar@{=>}|{\underset{X_2}{\sum}[i_2]} "s"; "t"
  }
  }
  \,,
$$
where the transformation filling this diagram is the
$X_2$-dependent sum
of the given fundamental class $[i_2]$ on $i_2$, def. \ref{MeasureInducedByOrientation}.
\end{definition}
Combining this we obtain a twisted
$$
  \mathrm{TQFT}_d^\tau
  :=
  \mathbb{D}\int \exp(\tfrac{i}{\hbar}S) \,d\mu
  \;:\;
  \mathrm{Corr}(\mathbf{H})
  \longrightarrow
  \mathrm{Mod}(\ast)
$$
as the unit component of the composite of these two transformations,
hence as a defect from the trivial $(d+1)$-dimensional theory to itself.
It sends a correspondence to the unit component of the pasting composite of natural transformations
as follows
$$
  \xymatrix{
    \mathrm{Mod}(\ast)
    \ar@{<-}[rr]^-{\mathrm{id}}_<{\ }="s1"
    \ar[d]_{1_\ast \mapsto A_1}
    &&
    \mathrm{Mod}(\ast)
    \ar[d]^{1_\ast \mapsto A_2}
    \\
    \mathrm{Mod}(X_1)
    \ar@{<-}[rr]|{\underset{i_1}{\sum} \circ i_2^\ast}_<{\ }="t"^>{\ }="t1"
    \ar[d]_{\underset{X_1}{\sum}}
    &&
    \mathrm{Mod}(X_2)
    \ar[d]^{\underset{X_2}{\sum}}
    \\
    \mathrm{Mod}(\ast)
    \ar@{<-}[rr]_-{\mathrm{id}}^>{\ }="s"
    &&
    \mathrm{Mod}(\ast)
    \ar@{<=}^{\exp(\tfrac{i}{\hbar}S)} "s1"; "t1"
    \ar@{=>}|{\underset{X_2}{\sum}[i_2]} "s"; "t"
  }
  \,.
$$

\begin{proposition}
 The unit component of this pasting composite
 $$
  \mathbb{D}\int \exp(\tfrac{i}{\hbar}S)d\mu
  \;:\; 1_{d+1} \stackrel{\exp(\tfrac{i}{\hbar}S)}{\longrightarrow}
  \mathrm{FQFT}_{d+1}
  \stackrel{\mathbb{D}\int(-)d\mu_{i_2}}{\longrightarrow}
  1_{d+1}
$$
 is the dual secondary integral transform
 $$
   \mathbb{D}\int_{Z} \xi d\mu_{i_2}
   :
   \underset{X_1}{\sum}A_1
     \stackrel{}{\longleftarrow}
   \underset{X_2}{\sum}A_2
 $$
 which is associated by
 def. \ref{IntegralKernelAndTransform} to the integral kernel $\xi$
 corresponding to $\exp(\tfrac{i}{\hbar}S)$ via the proof of
 prop. \ref{IntegralKernelAsTransformation}.
 \label{IntegralTransformFromTransformation}
\end{proposition}
\proof
The pasting natural transformation here has as unit component the map
$$
  \underset{X_1}{\sum} A_1
  \stackrel{\underset{X_1}{\sum}\exp(\tfrac{i}{\hbar}S)}{\longleftarrow}
  \underset{X_1}{\sum}\underset{i_1}{\sum}i_2^\ast A_2 \stackrel{\underset{X_2}{\sum} [i_2]}{\longleftarrow}
  \underset{X_2 }{\sum} A_2
$$
By the general formula for adjuncts we have that $\xi$ and $\exp(\tfrac{i}{\hbar}S)$
are related by
$$
  \exp(\tfrac{i}{\hbar}S) :
    A_1
    \stackrel{\epsilon}{\longleftarrow}
   \underset{i_1}{\sum} i_1^\ast A_1 \stackrel{\underset{i_1}{\sum}\xi}{\longleftarrow} \underset{i_1}{\sum} i_2^{\ast} A_2
   \,.
$$
Inserting this into the first expression manifestly yields the
secondary integral transform formula of
def. \ref{IntegralKernelAndTransform}, up to canonical equivalence.
\endofproof
\begin{remark}(consistent orientations and quantum anomalies)
  Proposition \ref{IntegralTransformFromTransformation} provides a
  succinct formulation of what it takes to choose
  fiberwise fundamental classes, def. \ref{TwistedOrientationOnWirthmuellerMorphism},
  on a system of correspondences \emph{consistently}, namely
  such that the operation of secondary integral transforms is functorial
  in the correspondences: the condition is that
  $\int(-)d\mu : \mathrm{FQFT}_{d+t} \longrightarrow 1_{d+1}$ is indeed a
  natural transformation, hence indeed a boundary condition for the
  tautological $(d+1)$-dimensional theory.
  The existence of such consistent orientations is the central obstruction
  to the existence of the quantization process, and such obstructions
  to quantization are known in the physics literature as
  \emph{quantum anomalies}. A clean account of quantum anomalies
  as traditionally considered is in \cite{Freed86};
  for quantum anomalies from the perspective as considered here
  see also \cite{Freed}.
  Therefore finding consistent orientations
  is \emph{quantum anomaly cancellation}.
  The problem of finding consistent orientations for integral
  transforms given by pull-push had previously
  been highlighted  in \cite{FHT07} for the special case of
  pull-push in equivariant K-theory.
  \label{quantumAnomaly}
\end{remark}

\section{Conclusion and Outlook}

\subsection{Holographic principle}
\label{Holography}

  Ever since the widely (and wildly) cited articles \cite{Maldacena, Witten98}
  it has become common in fundamental physics to conjecture that
  topological/gravitational quantum
  field theories $\mathrm{TQFT}_{d+1}^\tau$ of dimension $d$ carry on their boundary,
  as \emph{boundary field theories} non-topological quantum field theories
  $\mathrm{QFT}_{d-1}$, such that the boundary fields of $\mathrm{TQFT}_d^\tau$
  encode the correlators of $\mathrm{QFT}_{d-1}$. This relation has
  come to be known as the \emph{holographic principle}, and pairs of
  field theories related this way as \emph{holographic duals}.
  With all the ambition that goes into these conjectures, it is often
  underappreciated that there has previously been a seminal example of field theories
  for which such a holographic relation is an established theorem.
  This is the relation between 3-dimensional Chern-Simons theory
  for a compact gauge group $G$, and the 2-dimensional Wess-Zumino-Witten model,
  describing the propagation of a quantum string on $G$
  (see \cite{Gawedzki} for a survey of both systems, and specifically see around
  p. 30 for their ``holographic'' relation).

  Often such holographic relations sit in longer hierarchies of boundary field
  theories see notably \cite{Sati} and \cite{Witten}.
  In particular the $d$-dimensional topological field theory $\mathrm{TQFT}_d^\tau$ may
  be anomalous,
  with its quantum anomaly encoded by it being the (partial) boundary
  of some $\mathrm{TQFT}_{d+1}$.
  At the pre-quantum level a formalization of such hierarchies
  of field theories
  has been discussed in section 3.9.14 of \cite{dcct}.

  Comparison shows that the quantization process
  via secondary integral transforms discussed here
  reflects this holographic form.
  By prop. \ref{IntegralTransformFromTransformation}
  and via the examples in \ref{CohomologicalQuantization} it produces
  $(d-1)$-dimensional quantum field theories $\mathrm{QFT}_{d-1}$ as
  boundary conditions of $d$-dimensional topological quantum field theories
  $\mathrm{TQFT}_d^\tau$, which however in general involve an intricate
  twisting, hence a ``quantum anomaly''. The consistent orientation of the
  twists and hence the ``cancelling of the quantum anomaly'' is equivalent
  to $\mathrm{TQFT}_d^\tau$ in turn being a (two-sided) boundary for
  a $(d+1)$-dimensional (untwisted) $\mathrm{TQFT}_{d+1}$.

\medskip

\begin{center}
\begin{tabular}{|l|c|l|}
  \hline
  {\bf field theory} & {\bf spaces of states} & {\bf propagator}
  \\
  \hline
  $\mathrm{TQFT}_{d+1}$ & $\mathrm{Mod}_2 \in \mathrm{Cat}_2$ & integral transform
  \\
  \hline
  $\mathrm{TQFT}_{d}^\tau$ & $\mathrm{Mod}(\ast) \in \mathrm{Mod}_2$ &
   \begin{tabular}{l} secondary integral transform, \\ path integral \end{tabular}
  \\
  \hline
  $\mathrm{QFT}_{d-1}$ & $\underset{X}{\sum} A_X \in \mathrm{Mod}(\ast)$ &
  \begin{tabular}{l}
    equivariance under
    \\
    Hamiltonian group action
  \end{tabular}
  \\
  \hline
\end{tabular}
\end{center}

\medskip

  We might informally depict this situation as follows
  $$
    \xymatrix{
      \mathrm{QFT}_{d-1}
      \ar@{^{(}->}[r]^-\partial
      &
      \mathrm{TQFT}^\tau_d
      \ar@{^{(}->}[r]^-\partial
      &
      \mathrm{TQFT}_{d+1}
    }
  $$

  One famous example of such a hierarchy of ``holographic'' relations
  is supposed to be the between 3-dimensional Chern-Simons theory $\mathrm{CS}_3$
  and the 2-dimensional Wess-Zumino-Witten model $\mathrm{WZW}_2$
  $$
    \xymatrix{
      \mathrm{WZW}_2
      \ar@{^{(}->}[r]^-\partial
      &
      \mathrm{CS}_3
      \ar@{^{(}->}[r]^-\partial
      &
      \mathrm{TQFT}_{4}
    }
    \,.
  $$
  Another expected example of much current interest is
  the relation between the 7-dimensional Chern-Simons theory $\mathrm{CS}_7$
  that appears as the compactification of the 11-dimensional Chern-Simons
  term in 11-dimensional supergravity \cite{FSS}
  on whose boundary the $\mathrm{AdS}_7/\mathrm{CFT}_6$ correspondence (see \cite{Nastase})
  predicts the 6-dimensional $(2,0)$-supersymmetric field theory
  $\mathrm{WZW}_6$
  $$
    \xymatrix{
      \mathrm{WZW}_6
      \ar@{^{(}->}[r]^-\partial
      &
      \mathrm{CS}_7
      \ar@{^{(}->}[r]^-\partial
      &
      \mathrm{TQFT}_{8}
    }
    \,.
  $$
  These two examples have also been amplified in \cite{Freed}.

  Notice that $\mathrm{WZW}_6$ is thought to become
  4-dimensional(super-)Yang-Mills theory
  after Kaluza-Klein reduction on a torus whose modulus is the combined
  Yang-Mills coupling constant and $\theta$-angle
  \cite{WittenSix}.
  (Under this reduction M{\"o}bius transformations of the torus are supposed
  to yield the S-duality of super-Yang-Mills.)
  Therefore there is a plausible
  route that connects the quantization process described here to
  one of the core problems of quantization theory, that of Yang-Mills theory
  \cite{JaffeWitten}.

\subsection{Motives}
\label{Motives}

We conclude here with some comments revolving around the observation that the
quantization process found in linear homotopy-type theory above is broadly analogous
to the construction of categories of \emph{motives}
(see for instance \cite{Barry} for the general idea of motives and see \cite{Levine}
for a technical introduction).

\medskip

Consider specifically the model of linear homotopy-type theory in example \ref{LocalSystems}
above, given by bundles of $E$-module spectra over $\infty$-groupoids
(``local systems'' with coefficients in $E$-modules), for $E \in \mathrm{CRing}_\infty$ some
$E_\infty$-ring. As we have seen in
example \ref{PullPushInGeneralizedCohomology} in this case a
prequantum integral kernel, def. \ref{IntegralKernel}, consists of
\begin{enumerate}
  \item a correspondence of homotopy types of spaces
   $$
     \raisebox{40pt}{
     \xymatrix{
       & Z
       \ar[d]^{(i_1,i_2)}
       \\
       & X_1 \times X_2
       \ar[dr]^{p_2}
       \ar[dl]_{p_1}
       \\
       X_1
       &&
       X_2
     }
     }\,;
   $$
  \item
   a choice of twists $\chi_i$ of $E$-cohomology on $X_i$;
  \item
    a fiberwise $E$-linear map
    $$
      \xi : i_1^\ast \chi_1 \longrightarrow i_2^\ast \chi_2
    $$
    on the correspondence space $Z$
    inducing an $E$-linear map
   $$
    \Xi := \int_Z \xi d\mu \;:\; E_{\bullet + \chi_1}(X_1) \longrightarrow E_{\bullet + \chi_2}(X_2)
   $$
   between the twisted homology spectra.
\end{enumerate}
First notice that here in the special case that $E_{\bullet + \chi_1}(X_1)$ happens to be equivalent to
$E_\bullet(\ast)$ (for instance because $X_1 \simeq \ast$), then $\Xi \in E_{\bullet +\chi_2}(X_2)$
is equivalently just a cycle in the $\chi_2$-twisted $E$-homology of $X_2$.
Conversely if the left hand spectrum is equivalent to $E_\bullet(\ast)$ then
$\Xi \in E^{\bullet + \chi_1}(X_1)$ is equivalently a cocycle in the $\chi_1$-twisted $E$-cohomology
of $X_1$. In the general case we may speak of $\Xi$ here as being a
$(\chi_1, \chi_2)$-twisted cocycle in \emph{bivariant} $E$-cohomology.
Notice furthermore that in the case that the twists vanish then the $E$-cohomology of a homotopy
type carries the structure of an $E_\infty$-ring and so one may think
in this case of $\Xi$ as exhibiting the cohomology of the correspondence space
as a bimodule over the cohomology of its legs.

This form of a datum given by a correspondence of spaces equipped with cycles
on the correspondence space
is familiar from the definition of (pure or mixed) \emph{motives},
where the spaces considered are (projective) smooth schemes and the cocycles are algebraic cycles
(see section 8 of chapter 1 of \cite{ConnesMotives} for a review).
The literature also considers correspondences of schemes equipped instead with cycles in algebraic K-theory
(see section 5.3 of \cite{TabuadaComparison} for a review). These map to
pure (Chow-)motives via the algebraic Chern character and faithfully embed into
\emph{non-commutative motives} (\cite{TabuadaComparison}, p. 9,10)
where the role of the cycles is played by bimodules over algebraic data
associated with the left and the right leg of the correspondence (see \cite{Tabuada} for a survey).

Indeed, in the special case of linear homotopy-type theory
over $E = \mathrm{KU}$ the
complex K-theory spectrum (as in the examples in \ref{CohomologicalQuantization} above)
and restricting attention to homotopy types $X_i$ and $Z$ of
suitable compact manifolds, then the bivariant cocycles $\Xi$ above may be equivalently expressed as classes in
bivariant operator K-theory, i.e. KK-theory (section 4.2 of \cite{Nuiten13}). It had long been suggested
by Connes (originally in section 3 of \cite{ConnesSkandalis},
for more see section 1 of chapter 4 of \cite{ConnesMotives})
that KK-theory (or its pseudo-abelian envelope given by cyclic cohomology) is to be regarded as the analog in
non-commutative topology ($C^\ast$-algebra theory) of
the theory of motives in algebraic geometry. Recently in \cite{Snig} this analogy
has been substantiated by the construction of a canonical map from
KK-theory to non-commutative motives.

In summary this means that there is a zig-zag of comparison functors relating
motives in algebraic geometry to prequantized integral kernels in linear homotopy-type theory
as considered here:
$$
  
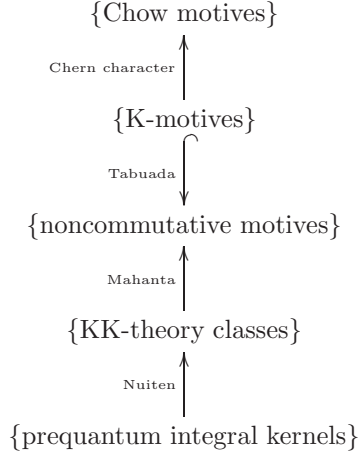
\xymatrix{
    \mbox{\{Chow motives\}}
    \\
    \mbox{\{K-motives\}}
    \ar[u]^{\mbox{\tiny Chern character}}
    \ar@{^{(}->}[d]_{\mbox{\tiny Tabuada}}
    \\
    \mbox{\{noncommutative motives\}}
    \\
    \mbox{\{KK-theory classes\}}
    \ar[u]^{\mbox{\tiny Mahanta}}
    \\
    \mbox{\{prequantum integral kernels\}}
    \ar[u]^{\mbox{\tiny Nuiten}}
  }
$$

So far this is just the pre-quantum level. When we now
consider quantization, then the relation to motivic theory
goes further still. Above in \ref{CohomologicalQuantization}
we discussed the structure of a ``consistent orientation'' or
``anomaly free densitized action $\exp(\tfrac{i}{\hbar}S)d \mu$'' as the necessary
structure to make the secondary integral transform  operation of ``pull-tensor-push'' through such
motivic correspondences, def. \ref{IntegralKernelAndTransform}
well defined, such as to yield a functor
$$
  \mathrm{QFT}_d
  =
  \int \exp(\tfrac{i}{\hbar}S) d\mu
  \;:\;
  \mathrm{Corr}(\mathbf{H})
  \longrightarrow
  E\mathrm{Mod}(\ast)
$$
from correspondences to $E$-modules over the point.
In example \ref{PontoShulmanOperation} we saw that given such a functor,
it is of interest to consider its restriction along the inclusion
$$
  \mathbf{H} \longrightarrow \mathrm{Corr}(\mathbf{H})
$$
which sends morphisms in $\mathbf{H}$ to ``right way''-morphisms.
Conversely therefore one may consider equipping a given functor
$\mathbf{H} \longrightarrow E \mathrm{Mod}$
with the structure of a lift to a functor on correspondences.
(Notice that we are considering homology-valued QFTs, but by postcomposing with $E$-duality they become
cohomologically valued functors on $\mathbf{H}^{\mathrm{op}}$).
Such extensions of (contravariant) functors on the given category of geometric spaces
to functors on correspondences equipped with cycles are known as
\emph{sheaves with transfer} in the theory of motives.
(Notice that, by remark \ref{transferContext}, the concept of ``transfer'' here is
indeed at least closely related to the concept of Umkehr maps via fundamental classes that we consider.)
Those complexes of sheaves with transfer on smooth schemes
that are only sensitive to the $\mathbb{A}^1$-homotopy type of schemes form the
category of ``effective geometric motives''
(see section 3.1 of \cite{Voevodsky} or section 3 of lecture 2 of \cite{Levine}).

Therefore, translated this way to the present setup
(and imposing that the presheaves are monoidal functors)
effective geometric motives are analogous to QFTs obtained by
quantization via secondary integral transforms as discussed here.

\medskip

While we feel that the canonical constructions in linear homotopy-type theory
that we considered here exhibits this analogy between quantization and motives
in a pleasingly holistic way,
at least various aspects of such a relation between motivic structures and quantization
have a long tradition in the literature, often implicitly.
In the context of mechanics it goes back to \cite{Weinstein} and in the context of
field theory it was amplified in \cite{Freed92}.
A fairly comprehensive and commented list of
references does not fit into the present note, but may be found online
at \href{http://ncatlab.org/nlab/show/motivic+quantization#PrevLit}{ncatlab.org/nlab/show/motivic+quantization\#PrevLit}.

\medskip

Notice that in the approximation to quantization
given by perturbation theory (formal algebraic deformation quantization) a deep relation to
motivic structures was suggested in \cite{Kontsevich}, where
the space of choices of formal deformation quantizations is argued to be a torsor over a
quotient of the motivic Galois group equivalent to the Grothendieck-Teichm{\"u}ller group.
A precise version of this statement was proven in \cite{Dolgushev}.

Since our discussion here concerns the full (non-perturbative, geometric)
quantization (as in example \ref{KUBoundaryQuantization} above) it seems natural
to expect that the motivic structures that appear here may be related to these
seen in formal deformation quantization. This remains to be explored.

\medskip
\medskip
\medskip

\noindent {\bf Acknowledgement.}
The discussion of pull/push in cohomology and its relation to quantization is joint
and ongoing work with Joost Nuiten.
While writing this text I received very helpful comments from Mike Shulman
on aspects of linear homotopy-type theory. 
Thanks to Adeel Khan Yusufzai
for discussion of aspects of geometric motives and for pointing out Tabuada's
proof of the relation to noncommutative motives; and thanks to Thomas Nikolaus
for pointing out Haugseng's work on transfer.
Thanks to David Corfield for discussion of formalized metaphysics and thanks to
Todd Trimble for discussion of linear type theory and
detailed first-hand information on the history of their categorical semantics.
The text also profited from technical discussion with Yonatan Harpaz and Marc Hoyois.
Finally, thanks again to David Corfield for plenty of
comments and profound thoughts on the text and its content.

\end{document}